\documentclass[fleqn,usenatbib]{mnras}

\usepackage{newtxtext,newtxmath}

\usepackage[T1]{fontenc}

\DeclareRobustCommand{\VAN}[3]{#2}
\let\VANthebibliography\thebibliography
\def\thebibliography{\DeclareRobustCommand{\VAN}[3]{##3}\VANthebibliography}

\usepackage{graphicx}	%
\graphicspath{{figures/}}
\usepackage{amsmath}	%
\usepackage{hyperref}
\usepackage[flushleft]{threeparttable}
\usepackage[scientific-notation=true]{siunitx}
\usepackage{physics}
\usepackage{siunitx}
\usepackage{xcolor}

\newcommand{\Msun}{\mathrm{M}_{\odot}}

\title[Cooling Flows in FIRE]{
Cooling Flows as a Reference Solution for the Hot Circumgalactic Medium
}

\author[I. Sultan et al.]{Imran Sultan,$^{1}$\thanks{E-mail: sultan@u.northwestern.edu}
Claude-André Faucher-Giguère,$^{1}$
Jonathan Stern,$^{2}$
Shaked Rotshtein,$^{2}$
Lindsey Byrne,$^{1}$
\newauthor
Nastasha Wijers$^{1}$
\\
$^{1}$Center for Interdisciplinary Exploration and Research in Astrophysics (CIERA) and Department of Physics and Astronomy, Northwestern University,\\1800 Sherman Ave, Evanston, IL 60201, USA\\
$^{2}$School of Physics \& Astronomy, Tel Aviv University, Tel Aviv 69978, Israel\\
}

\date{Accepted XXX. Received YYY; in original form ZZZ}

\pubyear{2023}

\begin{document}
\label{firstpage}
\pagerange{\pageref{firstpage}--\pageref{lastpage}}
\maketitle

\begin{abstract}
The circumgalactic medium (CGM) in $\gtrsim 10^{12}\ \Msun$ halos is dominated by a hot phase ($T \gtrsim 10^{6}$ K). While many models exist for the hot gas structure, there is as yet no consensus. We compare cooling flow models, in which the hot CGM flows inward due to radiative cooling, to the CGM of $\sim 10^{12}-10^{13}\ \Msun$ halos in galaxy formation simulations from the FIRE project at $z\sim0$. The simulations include realistic cosmological evolution and feedback from stars but neglect AGN feedback. At both mass scales, CGM inflows are typically dominated by the hot phase rather than by the `precipitation’ of cold gas. Despite being highly idealized, we find that cooling flows describe $\sim 10^{13}\ \Msun$ halos very well, with median agreement in the density and temperature profiles of $\sim 20\%$ and $\sim 10\%$, respectively. This indicates that stellar feedback has little impact on CGM scales in those halos. For $\sim 10^{12}\ \Msun$ halos, the thermodynamic profiles are also accurately reproduced in the outer CGM. For some of these lower-mass halos, cooling flows significantly over-predict the hot gas density in the inner CGM. This could be due to multidimensional angular momentum effects not well captured by our 1D cooling flow models and/or to the larger cold gas fractions in these regions. Turbulence, which contributes $\sim 10-40\%$ of the total pressure, must be included to accurately reproduce the temperature profiles. Overall, cooling flows predict entropy profiles in better agreement with the FIRE simulations than other idealized models in the literature.

\end{abstract}


\section{Introduction}
The role of the circumgalactic medium in galaxy formation, although not fully understood, is likely crucial.
The circumgalactic medium (CGM) is the gas and dust contained inside the dark matter halo (extending to $\sim R_{\mathrm{vir}}$, although there is no strict boundary), and outside the galaxy.
We distinguish this medium from the gas contained within the galaxy, i.e. the interstellar medium (ISM), although the end of the ISM and beginning of the CGM is not always clear.
Ample observational evidence of the CGM includes UV absorption lines (e.g. Mg II, C IV, O VI, and other absorption lines), soft X-ray absorption lines (e.g. O VII and O VIII absorption lines in the Milky Way), and emission, also in rest UV and X-rays.
Note that the CGM is multiphase, with phases existing at different temperatures traced by different metal absorption and emission lines.
For reviews of observational and theoretical work in the field, see reviews by, e.g., \cite{tumlinsonCircumgalacticMedium2017} and \cite{faucher-giguereKeyPhysicalProcesses2023}.

The CGM is the intermediary between gas in the galaxy and gas in the intergalactic medium (IGM).
Thus the CGM may play a key role in shaping gas flowing between the IGM and ISM.
Since this gas fuels star formation and black hole growth, the CGM may have major implications for the evolution of the galaxy. 

In the classical picture of halo virialization, gas that is gravitationally collapsing into the CGM from the intergalactic medium reaches supersonic velocities.
The resulting shock heats the gas to temperatures similar to the virial temperature of the halo, which at $z=0$ is given by
\begin{equation}\label{eq:Tvir}
    T_{\mathrm{vir}}=\frac{\mu m_p}{2 k_B} v_c(R_{\mathrm{vir}})^2=\SI{6e5}{\kelvin} \left( \frac{M_{\mathrm{halo}}}{10^{12}\ \Msun} \right)^{2/3}.
\end{equation}
Here $\mu m_p$ is the mean mass of a molecule in the gas, and $v_c(R_{\mathrm{vir}})$ is the circular velocity in the gravitational potential of the halo at its virial radius (of order the escape velocity of the halo).

In this picture, gas with a cooling time ($t_{\mathrm{cool}}$: the time to radiate away all internal energy) longer than the free-fall time ($t_{\mathrm{ff}}$: the time for gravitational collapse to the potential center) is able to maintain a hot virialized steady state at temperatures $T \sim T_{\mathrm{vir}}$. 
On the other hand, gas with a cooling time shorter than the free-fall time quickly cools and accretes onto the galaxy in cold clouds or streams.
The halo mass threshold that marks the transition from cold to hot accretion mode is $M_{\mathrm{halo}} \sim 10^{11} -10^{12}\ \Msun$, with the exact value depending on CGM mass, metallicity, and radius (e.g., \citealt{reesCoolingDynamicsFragmentation1977, whiteCoreCondensationHeavy1978, birnboimVirialShocksGalactic2003, keresHowGalaxiesGet2005, keresGalaxiesSimulatedLCDM2009, faucher-giguereSmallCoveringFactor2011, sternMaximumAccretionRate2020}).

Interestingly, this transition (the characteristic mass for halo virialization) is comparable to the mass scale above which galaxies are quenched. 
Recent work has linked the virialization of the inner CGM (and resulting rotating cooling flows) with the emergence of large, thin galactic disks \citep{sternVirializationInnerCGM2021, hafenHotmodeAccretionPhysics2022}, further emphasizing the connection between the hot CGM and galaxy formation.
In this interpretation, the formation of large disks precedes quenching, presumably by AGN feedback (e.g., \citealt{byrneEffectsMultichannelActive2024, byrneStellarFeedbackregulatedBlack2023}).
 
In this paper we will focus on the hot phase of the CGM in virialized halos, with temperatures of $T_{\mathrm{vir}} \sim 10^6$ K for the halos we study at or above the characteristic mass for halo virialization ($M_{\mathrm{halo}} \sim 10^{12}$--$10^{13}\ \Msun$ at low redshift).
Observational evidence for this hot component includes X-ray absorption and emission lines detected in our own galaxy (e.g., \citealt{henleyORIGINHOTGAS2010, guptaHUGERESERVOIRIONIZED2012, fangXMMNEWTONSURVEYLOCAL2015, bregmanExtendedDistributionBaryons2018}), as well as far UV absorption lines, including O VI and Ne VIII detected in extragalactic systems (e.g., \citealt{tumlinsonLargeOxygenRichHalos2011, burchettCOSAbsorptionSurvey2019, quCosmicUltravioletBaryon2024}).
Our main goal is to identify the physical processes that shape the hot gas profiles in these halos.
Our mass range is selected such that the CGM is expected to be fully virialized, but also such that the cooling radius is a significant fraction of the virial radius (e.g., see review by \citealt{donahueBaryonCyclesBiggest2022}).

Analytic modeling of the hot phase generally starts with the assumption of hydrostatic equilibrium (HSE), under which thermal pressure balances the gravitational force on the gas.
However, HSE alone is not sufficient to fully constrain the thermodynamic properties of a spherically symmetric ideal gas; a constraint on the pressure profile of the halo leads to a degeneracy in temperature and density profiles since $P \propto n T$.
Thus, modelers have made various assumptions to constrain the thermodynamics of the hot phase. 
These include assuming constant temperature (isothermal; e.g. \citealt{faermanMASSIVEWARMHOT2017}), constant entropy (isentropic; e.g., \citealt{faermanMassiveWarmHot2020}), and power-law entropy profiles (e.g., the power law found by \citealt{voitBaselineIntraclusterEntropy2005} for non-radiative cosmological hydrodynamical simulations of clusters).
Other models assume a constant $t_{\mathrm{cool}}/t_{\mathrm{ff}}$ in the hot phase, with cold clouds forming in regions where the ratio falls below the constant ratio (`precipitation' model; e.g., \citealt{sharmaStructureHotGas2012, voitAmbientColumnDensities2019}).
See \cite{singhComparisonModelsWarmHot2024} for a comparison of analytic models for a Milky Way-like halo (see also, \citealt{orenSunyaevZeldovichSignals$L^$2024}).

In addition to the aforementioned HSE models, there is the cooling flow model. 
Cooling flow models were developed in previous studies to characterize the flow of gas into galaxies and clusters of galaxies.
The physics modeled by a cooling flow is the inflow of gas driven by radiative cooling in a gravitational potential (for a review, see e.g. \citealt{fabianCoolingFlowsClusters1984}).
Crucially, this idealized model assumes ongoing heating feedback is negligible compared to radiative losses.
The minimal model assumes the gas in the cooling flow is spherically symmetric; the result is a single-phase flow, with a temperature of approximately the virial temperature since heating due to compression approximately balances radiative losses.
The implied inflow time approximately equals the $t_{\mathrm{cool}}$, which is also the time for thermal instabilities to grow; this ensures that initially small density fluctuations do not have time to become large, preventing cold structures from spontaneously forming out of the hot gas and maintaining a single-phase flow (\citealt{balbusTheoryLocalThermal1989}; \citealt{sternCoolingFlowSolutions2019}).

Cooling flow solutions were first studied in detail for cluster-mass halos.
However, observations of clusters showed a lack of cooling of low temperature gas below $\sim 1$ keV, and the star formation rates of central galaxies in clusters were observed to be only $\sim 1-10$\% of the predicted mass flow rates of cooling gas, giving rise to the `cooling flow problem' (see e.g. \citealt{mcdonaldRevisitingCoolingFlow2018}).
Feedback from supermassive black holes (i.e. active galactic nuclei or AGN feedback) and thermal conduction were among the possible mechanisms proposed to counteract cooling in clusters (see \citealt{mcnamaraHeatingHotAtmospheres2007} for a review). 
The cooling flow problem may not be an issue for galaxy-scale halos, however.
For example, \cite{sternCoolingFlowSolutions2019} demonstrated that O VII and O VIII absorption in the Milky Way, and the density profile predicted by O VII and O VIII emission in our galaxy, can both be explained by a cooling flow with a mass flow rate of order the star formation rate in our galaxy.
\cite{wijersNeVIIIWarmhot2024} similarly showed that cooling flow models can explain the median Ne VIII profiles around external $\sim L^*$ galaxies.

The application of cooling flow models was initially limited to the cluster scale due to the fact that X-ray emission, a primary probe of hot gas, is much weaker for lower mass halos. 
Recent observations of ionic tracers of gas in the spectra of the Milky Way and other galaxies have enabled more direct measurements of the hot halo gas.
These observations include soft X-ray absorption lines (e.g. O VII and O VIII detected in Milky Way observations with the Chandra and XMM-Newton space telescopes; \citealt{guptaHUGERESERVOIRIONIZED2012, fangXMMNEWTONSURVEYLOCAL2015}) and far ultraviolet absorption lines (e.g. Ne VIII detected in the CGM of nine $z \gtrsim 0.5$ galaxies with the Hubble Cosmic Origins Spectrograph instrument; \citealt{burchettCOSAbsorptionSurvey2019}) observed in the spectra of background quasars.
Motivated by the new observations, \cite{sternCoolingFlowSolutions2019, sternMaximumAccretionRate2020} extended the previous cooling flow models down to Milky Way-mass scale halos (several orders of magnitude lower than cluster-mass halos).
Most recently, \cite{sternAccretionDiscGalaxies2024} further extended cooling flow models to include rotation more explicitly. 

Thus there is a lack of consensus in modeling hot halos, and the state of the art in the field is a collection of distinct models that differ from one another and make different predictions for various observables.
To test the predictions of cooling flow models against our current understanding of galaxy formation and cosmology, and assess their effectiveness relative to other models, we use cosmological simulations of galaxy formation.
Simulations are a powerful tool to break the degeneracy in modeling hot halos, since they let us measure realized thermodynamic profiles of the CGM that arise in a self-consistent model of galaxy formation.

In this paper we analyze cosmological zoom-in simulations carried out by the Feedback in Realistic Environments (FIRE) project\footnote{\url{https://fire.northwestern.edu}}.
Cosmological zoom-ins combine the large-scale dark matter information with high-resolution hydrodynamical simulations focused on selected galaxies, enabling the modeling of galaxies and their CGM in their cosmological context.
As a result of the detailed models of star formation and stellar feedback, galaxies in FIRE simulations up to the mass of the Milky Way have excellent agreement with observations when exploring a wide range of measurements including stellar masses \citep{hopkinsGalaxiesFIREFeedback2014, hopkinsFIRE2SimulationsPhysics2018, feldmannFIREboxSimulatingGalaxies2023}, mass-metallicity relations (e.g., \citealt{maOriginEvolutionGalaxy2016, bassiniInflowOutflowProperties2024, marszewskiHighRedshiftGasPhaseMass2024}), and galaxy structural properties (e.g., \citealt{el-badryGasKinematicsMorphology2018}).

\begin{table*}
\begin{center}
\caption{FIRE simulation set used in our analysis, including m12- and m13-mass halos. Our set includes both FIRE-2 and FIRE-3 simulations. Columns 1-5 show the halo name, lowest redshift $z$ to which the simulation was run, virial mass at $z$, virial radius at $z$, and the baryonic mass resolution, respectively. Column 6 indicates whether the simulation included magnetohydrodynamics.}
\begin{tabular}{c|c|c|c|c|c}
Halo Name & $z$ & $M_{\mathrm{vir}}(z)/(10^{12}\Msun)$ & $R_{\mathrm{vir}}(z)/\mathrm{pkpc}$ & Baryonic Mass Resolution ($\Msun$) & MHD? \\ \hline
m12b (FIRE-2)    & 0   & 1.3  & 286 & \num{7e3} & no \\
m12c (FIRE-2)    & 0   & 1.3  & 283 & \num{7e3} & no \\
m12f (FIRE-2)    & 0   & 1.5  & 302 & \num{7e3} & no \\
m12i (FIRE-2)    & 0   & 1.1  & 268 & \num{7e3} & no \\
m12m (FIRE-2)    & 0   & 1.5  & 296 & \num{7e3} & no \\
m12r (FIRE-2)    & 0   & 1.0  & 266 & \num{7e3} & no \\
m12w (FIRE-2)    & 0   & 1.0  & 261 & \num{7e3} & no \\
m12z (FIRE-2)    & 0   & 0.8  & 244 & \num{4e3} & no \\ \hline
m12a (FIRE-3)    & 0   & 2.0  & 329 & \num{6e4} & yes \\
m12d (FIRE-3)    & 0   & 1.3  & 289 & \num{6e4} & yes \\
m12e (FIRE-3)    & 0   & 2.2  & 341 & \num{6e4} & yes \\
m12f (FIRE-3)    & 0   & 1.4  & 290 & \num{7e3}   & yes \\
m12g (FIRE-3)    & 0   & 2.6  & 360 & \num{7e3} & yes \\
m12j (FIRE-3)    & 0   & 0.9  & 255 & \num{7e3} & yes \\
m12k (FIRE-3)    & 0   & 2.2  & 342 & \num{6e4} & yes \\
m12n (FIRE-3)    & 0   & 1.5  & 302 & \num{7e3} & yes \\
m12q (FIRE-3)    & 0   & 1.5  & 297 & \num{7e3}   & yes \\
m12r (FIRE-3)    & 0   & 0.9  & 257 & \num{7e3}   & yes \\
m12u (FIRE-3)    & 0   & 0.6  & 227 & \num{3e4}   & yes \\
m12w (FIRE-3)    & 0   & 0.9  & 257 & \num{7e3}   & yes \\
m12x (FIRE-3)    & 0   & 0.6  & 215 & \num{4e3} & yes \\
m12z (FIRE-3)    & 0   & 0.7  & 232 & \num{4e3}   & yes \\ \hline
m13h2 (FIRE-3)   & 0.48 & 21.8 & 560 & \num{3e5}   & yes \\
m13h7 (FIRE-3)   & 0   & 25.0 & 771 & \num{3e5}   & yes \\
m13h29 (FIRE-3)  & 0.05 & 13.9 & 615 & \num{3e5}   & yes \\
m13h113 (FIRE-3) & 0   & 9.4  & 557 & \num{3e5}   & yes \\
m13h206 (FIRE-3) & 0   & 6.7  & 496 & \num{3e5} & yes \\
m13h223 (FIRE-3) & 0.06 & 46.0 & 909 & \num{3e5} & yes \\
m13h236 (FIRE-3) & 0.20 & 9.7  & 498 & \num{3e5} & yes
\end{tabular}\label{tab:sims}
\end{center}
\end{table*}

The primary question we ask in this work is whether cooling flows are a useful baseline model for the hot CGM of massive halos in the absence of AGN feedback.
We analyze a large set of FIRE simulations evolved to $z \approx 0$, including both Milky Way-mass ($M_{\mathrm{halo}} \sim 10^{12}\ \Msun$) and more massive ($M_{\mathrm{halo}} \sim 10^{13}\ \Msun$) galaxies.
This halo mass range allows us to probe the characteristic mass for thin-disk formation and, later, halo virialization and quenching.
To model cooling flows in FIRE, we start with the derivation of \cite{sternCoolingFlowSolutions2019, sternMaximumAccretionRate2020} for cooling flows with an approximate 1D angular momentum treatment; we extend the model to include non-thermal pressure support in the form of turbulence, which we find can be a significant source of pressure in FIRE halos. 

We show that cooling flows effectively model the hot phase of the halos we study in FIRE simulations without AGN feedback, and produce much better agreement with FIRE halos than other analytic models.
The agreement of the cooling flows is excellent for $\sim 10^{13}\ \Msun$ mass halos within the cooling radius.
The $\sim 10^{12}\ \Msun$ halos are well described by cooling flows in the outer CGM while they deviate from the idealized model in the inner CGM, but nevertheless cooling flows more effectively model these halos than other analytic models.
We thus propose the use of cooling flows as a benchmark model that captures the physics of radiative cooling and gravity, which can be used as a baseline in future work when analyzing simulations with the additional physics of supermassive black holes (i.e. AGN feedback).

This paper is organized as follows.
We begin in Section \ref{sec:Methods} with a description of the FIRE simulations analyzed in this paper and our halo analysis methods, including our method for selecting the hot, virialized phase of the CGM.
In Section \ref{sec:AnalyticModels} we summarize the cooling flow model and our procedure for fitting cooling flows to simulations.
In Section \ref{sec:Results}, we present our results on how well the hot phase in FIRE is modeled by cooling flows, the significance of non-thermal pressure support (turbulence and magnetic fields) in FIRE halos, and a comparison to other analytic models of hot halos.
Finally, we provide a discussion of our results in Section \ref{sec:Discussion} and summarize the main findings of this study in Section \ref{sec:Conclusions}.

Throughout this paper we use $\log$ to refer to the base-10 logarithm.
We use a standard flat $\Lambda$CDM cosmology with $h \approx 0.68$ and $\Omega_{m} \approx 0.3$ \citep{aghanimPlanck2018Results2020}.
We assume the solar metal mass fraction $Z_{\odot}=0.0142$ of \cite{asplundChemicalCompositionSun2009} in our results.

\section{Simulations and Methods}\label{sec:Methods}
In this section we describe the FIRE simulations used in this work, as well as our halo analysis methods.

\begin{figure*}
    \includegraphics[width=\textwidth]{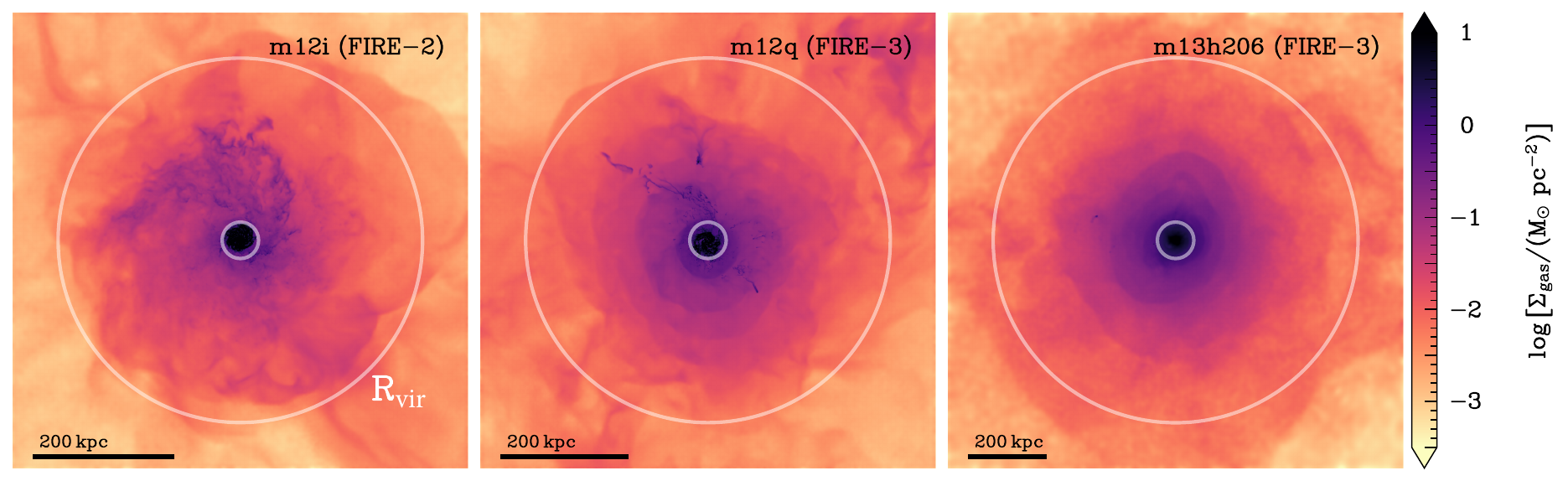}
    \caption{Gas surface density maps at $z=0$ for three of the halos in our analysis set (see Table \ref{tab:sims}).
    Maps of $\log \Sigma_{\mathrm{gas}}$ are shown for m12i (FIRE-2), m12q (FIRE-3), and m13h206 (FIRE-3).
    The inner and outer circles in each panel indicate $0.1 R_{\mathrm{vir}}$ and $R_{\mathrm{vir}}$, respectively.
    The maps show a 30 kpc slice of the galactic plane, viewed face-on.
    }
    \label{fig:Sigmamap}
\end{figure*}

\begin{figure*}
    \includegraphics[width=\textwidth]{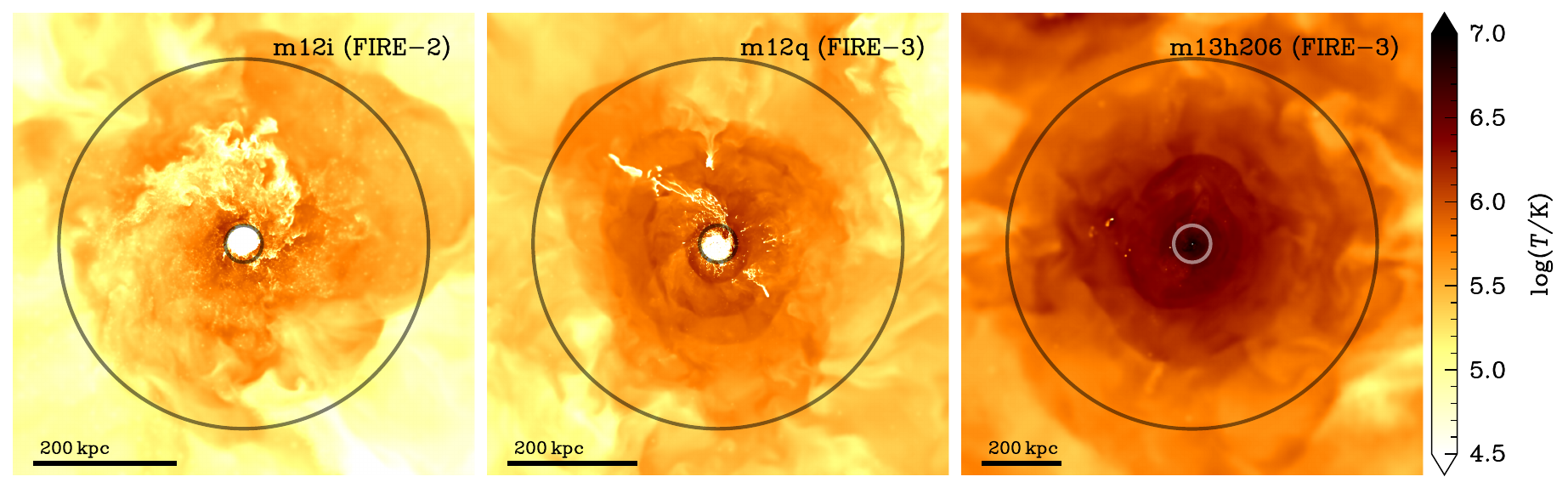}
    \caption{Similar to Figure \ref{fig:Sigmamap}, but for gas temperature maps at $z=0$ for three of the halos in our analysis set.
    Maps of $\log T$ weighted by mass are shown in a 30 kpc slice.
    Significant cold gas structures are present in the m12 halos, while the higher-mass m13 halo is more fully dominated by hot gas throughout the halo.
    }
    \label{fig:Tmap}
\end{figure*}

\subsection{FIRE Simulations}
The cosmological zoom-in simulations used in this study are from the FIRE project.
The simulations were run with the GIZMO\footnote{\url{http://www.tapir.caltech.edu/~phopkins/Site/GIZMO.html}} gravity+hydrodynamics code \cite{hopkinsNewClassAccurate2015} using a meshless finite-mass hydrodynamics method.
Considerable progress has been made by the FIRE project in modeling the physics of stellar feedback and the multi-phase interstellar medium (ISM). 
For example, the rates of Type-Ia and Type-II supernovae explosions are modeled on an individual basis per star particle, and feedback including mass, metals, energy, and momentum are ejected into the nearby ISM. 
Star particles also lose mass by stellar winds for both OB and AGB stars.
Radiative feedback models that take into account multiple wavelengths are utilized to simulate the photoionization and photoelectric heating effects, and the corresponding radiation pressure is also modeled.
A complete description of the FIRE-2 methods is given in \cite{hopkinsFIRE2SimulationsPhysics2018}. 

In addition to FIRE-2 simulations, in this study we analyze simulations carried out as part of FIRE-3, which builds on the improved numerical methods of the previous version of the project.
FIRE-3 has kept the core FIRE-2 physical processes unchanged, while improving some aspects of the microphysics (including stellar evolution and gas cooling) and numerical algorithms for star formation and supernova feedback, among others.
For example, there are updates to the meta-galactic ultraviolet background (the model of \citealt{faucher-giguereCosmicUVXray2020} is now used, in contrast with the \citealt{faucher-giguereNewCalculationIonizing2009} model used in FIRE-2), supernovae rates, mass-loss rates of OB and AGB stars, stellar luminosities, and supernovae and stellar mass-loss yields. 
Magnetic fields are also now included by default.
We refer the reader to \cite{hopkinsFIRE3UpdatedStellar2023} for a full list of updates made in the FIRE-3 version of the FIRE code.

The FIRE simulation set we analyze in this work is given by Table \ref{tab:sims}.
We focus our analysis on Milky Way-mass ($M_{\mathrm{halo}} \sim 10^{12}\ \Msun$, which we call our `m12' halos) and more massive ($M_{\mathrm{halo}} \sim 10^{13}\ \Msun$; our `m13' halos) galaxies.
Our simulation set consists of 22 m12 halos and 7 m13 halos, where each halo represents a different initial condition for the zoom-in simulation.
We analyze the latest snapshot in time to which each simulation was evolved: this corresponds to redshift $z=0$ for all of the m12 halos in our set, and redshifts in the range $z=0.48$ to $z=0$ for the m13 halos in our set.\footnote{The m13 halos sometimes become prohibitively expensive to evolve, e.g. due to the build up of very dense stellar cores, so some runs were stopped at $z>0$.}
Our m12 halo set includes both FIRE-2 and FIRE-3 simulations.

The FIRE-2 m12 halos in our set are part of the `core' FIRE-2 simulation suite (see \citealt{wetzelPublicDataRelease2023} for the FIRE-2 public data release, and additionally \citealt{hopkinsFIRE2SimulationsPhysics2018, garrison-kimmelLocalGroupFIRE2019, samuelProfileFIREResolving2020}).
Analyses of several FIRE-3 m12 and m13 simulations in our set have recently been carried out, exploring galaxy properties \citep{byrneEffectsMultichannelActive2024} and Ne VIII absorption in the CGM \citep{wijersNeVIIIWarmhot2024}.
For more details on the FIRE-3 halos, readers can refer to the aforementioned studies, in addition to the FIRE-3 code paper \citep{hopkinsFIRE3UpdatedStellar2023} and Gandhi et al., in prep. 
The FIRE-3 simulations we analyze here use the supernova feedback implementation described in \cite{hopkinsFIRE3UpdatedStellar2023} \citep[variants, corresponding to different assumptions for the terminal momentum, are discussed in][]{hopkinsImportanceSubtletiesScaling2024}. 
This implementation tends to produce lower stellar masses at the halo mass scales studied here than FIRE-2. 
Since we do not study galaxy properties in detail, this difference between our FIRE-2 and FIRE-3 runs is not directly significant for our results. 
The broadly consistent results we obtain regarding cooling flows for the m12 halos from the FIRE-2 and FIRE-3 suites suggest that our main results are not very sensitive to the supernova feedback algorithm.

We analyze simulations run with the default FIRE physics models, i.e., simulations with no black holes and no cosmic rays.
All simulations were run with a subgrid model for the turbulent diffusion of metals in gas \citep{colbrookScalingLawsPassivescalar2017, escalaModellingChemicalAbundance2018}.

The initial mass of baryonic resolution elements (gas and stars) ranges from $\num{4e3}\ \Msun$ to $\num{6e4}\ \Msun$ for the m12 simulations, and is $\num{3e5}\ \Msun$ for the m13 simulations (i.e. the baryonic mass resolution).
Baryonic particles can gain and lose mass due to stellar mass loss and SNe; particles that stray more than a factor of three away from the median mass are split or merged.
Dark matter particles have a mass that is a factor of $(\Omega_m - \Omega_b)/\Omega_b \approx 5$ larger than the baryonic resolution elements in the high-resolution region.

For gas resolution elements, gravitational softening is treated in an adaptive manner, where it is set equal to the smoothing length of the gas.
The minimum (Plummer equivalent) force softening length for gas ranges from 0.1 pc to 1 pc for the m12 simulations, and is 0.2 pc for the m13 simulations (physical units).
The gravitational softening is fixed for star and dark matter particles.
The Plummer equivalent force softening length for stars ranges from 3 pc to 8 pc for the m12 simulations, and is 18 pc for m13 simulation.
Dark matter in the high-resolution region has a Plummer equivalent force softening length ranging from 30 pc to 80 pc for the m12 simulations, and is 190 pc for the m13 simulations. 

\subsection{Halo centering and definition}
We find the center of each halo following the iterative center of mass of a ``shrinking sphere'' method of \cite{powerInnerStructureLCDM2003}, centering on the high-resolution dark matter particles at the simulation output snapshots presented in this study.

Consistent with \cite{bryanStatisticalPropertiesXRay1998}, we define the virial radius $R_{\mathrm{vir}}$ of the halo as the radius of a sphere centered on the halo with density $\Delta_{\mathrm{vir}} \rho_{\mathrm{crit}}$, where $\rho_{\mathrm{crit}}(z)$ is the critical density of the universe at redshift $z$ and we use their fitting function for the virial overdensity: $\Delta_{\mathrm{vir}}(z) = 18 \pi^2 +82 x - 39 x^2,$ where
$x(z) = \Omega_m(z) - 1$.
The virial mass $M_{\mathrm{vir}}$ is the total mass enclosed within the sphere.
Note that in this paper, we use particle velocities in the rest-frame of the halo by subtracting the velocity of the center of mass of all particles within $R_{\mathrm{vir}}$ of the halo center.

\subsection{Selecting the hot, virialized phase of the CGM}\label{sec:selectinghotphase}

The CGM is multiphase, comprised of gas that occupies different regions in temperature-density phase space.
In Figures \ref{fig:Sigmamap} and \ref{fig:Tmap}, we show maps of surface density and temperature for three halos in our analysis.
The visualizations were produced with FIRE Studio \citep{gurvichFIREStudioMovie2022}.
The density maps show a 30 kpc slice (i.e., a projection of all gas mass within $|z|<15$ kpc on the xy-plane), where the halo has been rotated such that the total angular momentum vector of the galaxy (summed over all particles--- dark matter, gas, and stars--- within $0.1 R_{\mathrm{vir}}$) lies along the z-axis.
The CGM is contained roughly between the two circles in each panel, indicating $0.1 R_{\mathrm{vir}}$ and $R_{\mathrm{vir}}$.

As shown by the density and temperature maps, within the inner boundary is the cold, dense interstellar medium of the galaxy.
The volume-filling phase of the CGM outside the inner boundary is hot and more diffuse.
Outside the outer boundary is the warm, diffuse intergalactic medium.
The visualizations highlight the key point of the multiphase nature of the CGM, since coexisting alongside the volume-filling hot phase, there are cold, dense regions possibly corresponding to gas flows, satellites, etc.
The cold streams embedded in the medium are more prominent for the $\sim 10^{12}\ \Msun$ halos, while the $\sim 10^{13}\ \Msun$ halo is almost entirely filled by the hot phase gas.
As we will show, the $\sim 10^{13}\ \Msun$ halos are better modeled by a steady state cooling flow than the lower mass halos.

The purpose of this study is to model the hot, virialized phase of the CGM that makes up the bulk of the gas at our halo mass range.
To that end, we carry out a selection procedure to isolate the hot phase in our simulations, and we analyze only the gas particles that belong to this phase.
We additionally exclude radii at which the majority of the gas does not belong to the hot phase from our analysis (i.e. we exclude radial shells where $<50$\% of the total gas belongs to the hot phase).
The details of our hot phase selection procedure are given in Appendix \ref{app:virialbranch}.

In the rest of this work, unless otherwise specified our results will show only the gas we identified as belonging to the hot, virialized phase of the CGM.

\begin{figure*}
	\includegraphics[width=6.5in]{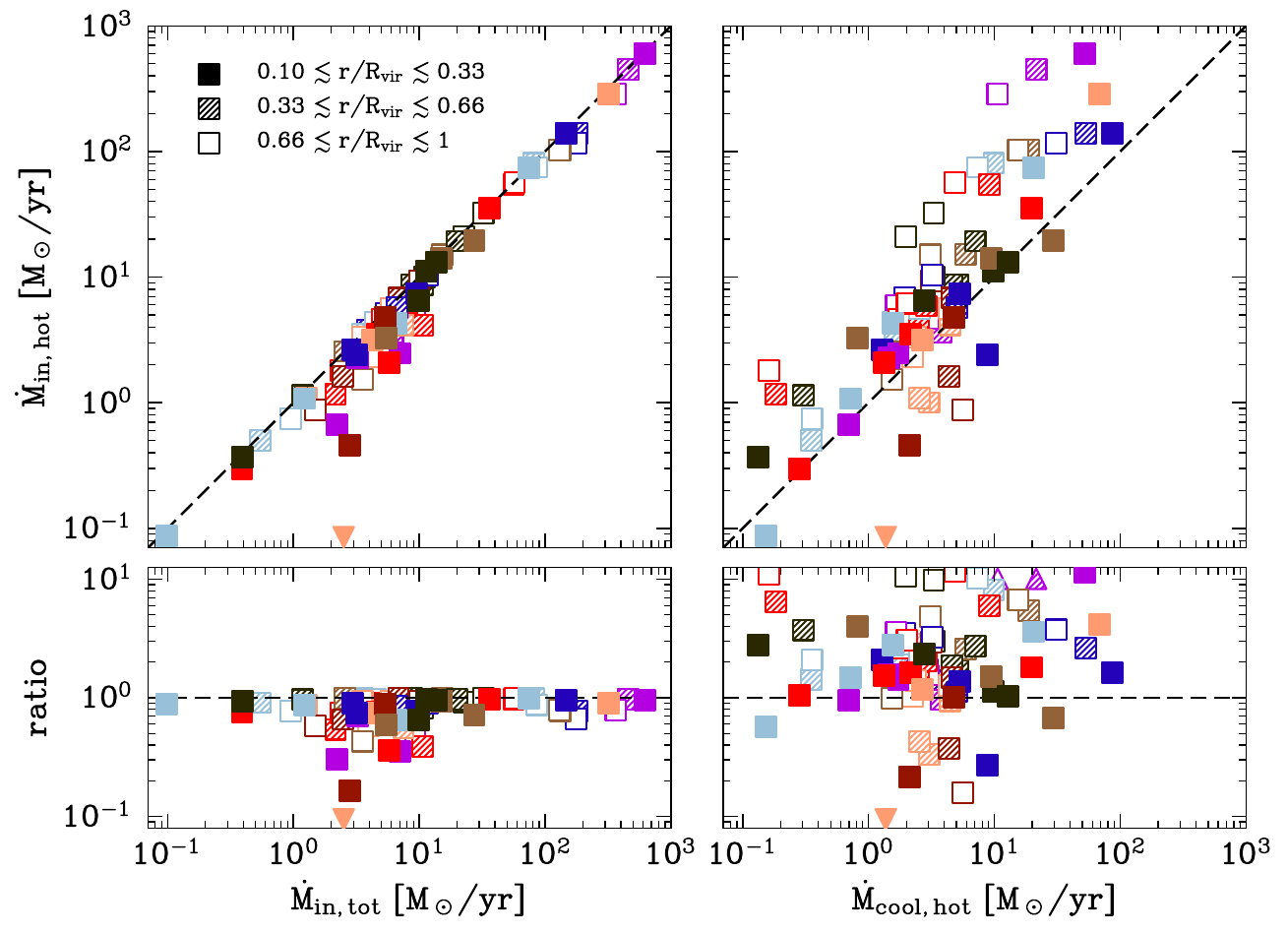}
    \caption{Comparison of hot CGM mass inflow rate $\dot{M}_{\mathrm{in,\ hot}}$ with the total CGM inflow rate $\dot{M}_{\mathrm{in,\ tot}}$ (left) and with the rate of cooling of the hot gas $\dot{M}_{\mathrm{cool, hot}}$ (right) in the FIRE simulations.
    The average inflow rates in three radial ranges are shown, representing the inner (solid points), middle (hatched points), and outer (empty points) halo.
    Ratios of $\dot{M}_{\mathrm{in,\ hot}}$ to $\dot{M}_{\mathrm{in,\ tot}}$ and $\dot{M}_{\mathrm{cool, hot}}$ are shown in the bottom panels.
    The dashed lines indicates a 1:1 relationship.
    Results are shown for FIRE-2 m12, FIRE-3 m12, and FIRE-3 m13 halos.
    Triangles indicate points that fall outside of the ranges plotted; points with $\dot{M}_i<0$ are omitted.
    Inflows are dominated by the hot phase in both the inner and outer halos.
    $\dot{M}_{\mathrm{in,\ hot}}$ is generally roughly consistent with the mass flow rate of cooling gas within the cooling radius, consistent with the expectation for cooling flows.
    For the m13 halos, which have the largest inflow rates, the middle and outer halo bins include radii beyond the cooling radius; the values $\dot{M}_{\mathrm{in,\ hot}} > \dot{M}_{\mathrm{cool, hot}}$ outside the cooling radius suggest the inflows are driven primarily by gravity rather than cooling.
    }
    \label{fig:Mdotratio}
\end{figure*}

\begin{figure}
	\includegraphics[width=3.12in]{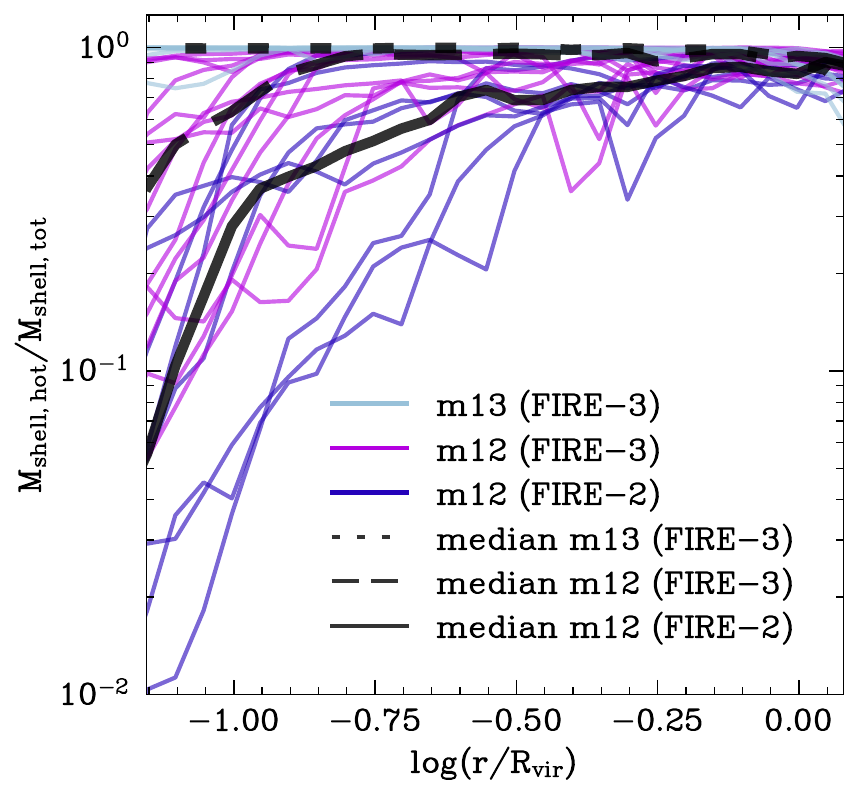}
    \caption{Hot-phase gas mass fractions of FIRE m12 and m13 simulations.
    In each radial shell centered at $r$, the fraction of all gas in the shell ($M_{\mathrm{shell, tot}}$) that belongs to the hot, virial phase ($M_{\mathrm{shell, hot}}$) is shown.
    The thick lines show the median fractions for the three halo subsets.
    Nearly all of the gas mass in the m13 halos is contained in the hot phase and the hot phase dominates the gas in the m12 simulations in most of the halo. The inner parts of the m12 halos contain larger median cold gas fractions (the FIRE-2 halos have a larger median cold gas fraction than the FIRE-3 halos), increasing with decreasing radius but only tens of percent for most halos outside $0.1 R_{\mathrm{vir}}$.
    }
    \label{fig:Mvir_ratio}
\end{figure}

\section{Analytic cooling flow models}\label{sec:AnalyticModels}
In this section we describe the cooling flow model we fit to the simulations.
The steady-state model describes gas in a gravitational potential that radiatively cools and flows towards the potential center.

\subsection{Cooling flow model with angular momentum}\label{sec:CFmodel_noturb}
\cite{sternCoolingFlowSolutions2019} derived solutions for an idealized hot CGM forming a steady-state inflow, assuming spherical symmetry and neglecting angular momentum, turbulence, and magnetic fields.
We include angular momentum in the cooling flow model using the 1D approximation  to the momentum conservation equation given by \cite{sternMaximumAccretionRate2020}, and also include the contribution of turbulence to the total pressure as described below. The full 2D effect of angular momentum on cooling flows has been derived by \cite{sternAccretionDiscGalaxies2024}, though this additional complication is not necessary for our analysis. 

The conservation equations for mass, momentum, and entropy are
\begin{align}
    &\dot{M} = 4 \pi r^2 \rho v\label{eq:conmass},\\
    &\frac{1}{2} \dv{v^2}{r} = -\frac{1}{\rho} \dv{P}{r} - \frac{v_c^2}{r}\left[1-\left( \frac{R_{\mathrm{circ}}}{r}\right)^2 \right]\label{eq:conp},\ \mathrm{and}\\
    &v \dv{\ln K}{r} = -\frac{1}{t_{\mathrm{cool}}}\label{eq:conK},
\end{align}
where $M$, $\rho$, $v$, and $P$ are the gas mass, density, radial velocity, and pressure, respectively.
The specific entropy $K = k_B T / n^{\gamma-1}$, where the total particle number density $n=\rho/(\mu m_p)$ and $\gamma$ is the adiabatic index.\footnote{This is the thermodynamic definition of entropy. For the results we present in this paper, we will show the quantity $K = k_B T / n^{2/3}$ by setting $\gamma=5/3$, the adiabatic index of a monatomic gas. This is a commonly used convention to define entropy profiles in studies of the hot CGM or intracluster medium.}
Note that $v<0$ for inflowing gas.
The circular velocity $v_c \equiv \sqrt{\frac{G M(<r)}{r}}$ is the velocity of a test particle orbiting in the gravitational potential at a radial distance $r$, where $M(<r)$ is the total mass contained within a distance $r$ from the halo center.

To include the angular momentum term in Equation \ref{eq:conp}, \cite{sternMaximumAccretionRate2020} assumed a specific angular momentum $v_c R_{\mathrm{circ}}$ in the halo; the circularization radius $R_{\mathrm{circ}}$ is the inner radius at which the radial inflow stalls due to angular momentum support in the halo preventing further collapse.
We set $R_{\mathrm{circ}}=0.05 R_{\mathrm{vir}}$, which is roughly the boundary of the CGM and ISM.\footnote{We have verified this is well outside the sonic radii of cooling flow solutions for the simulations we analyze, i.e., the flow at $r>R_{\mathrm{circ}}=0.05 R_{\mathrm{vir}}$ is in the subsonic regime for our halo mass range.}

Note the gas in cooling flows is nearly in hydrostatic equilibrium, since the deviation from the HSE condition (left hand side of Equation \ref{eq:conp}) is $\sim \mathcal{M}^2 \sim (t_{\mathrm{ff}}/t_{\mathrm{cool}})^2 \ll 1$, because $t_{\mathrm{ff}}/t_{\mathrm{cool}}<1$ in the virialized steady flow (see Equation 28 of \citealt{sternCoolingFlowSolutions2019}).
Here $\mathcal{M}=\frac{-v}{c_s}$ is the Mach number of the flow and $c_s=\sqrt{\frac{\gamma P}{\rho}}$ is the sound speed.

The cooling time,
\begin{equation}\label{eq:tcool}
    t_{\mathrm{cool}}=\frac{U}{n_{\mathrm{H}}^2 \Lambda} = \frac{P}{(\gamma-1)n_{\mathrm{H}}^2 \Lambda},
\end{equation}
is the time for the gas to lose all internal energy due to radiative cooling.
Here, $U$ is the energy density, $\gamma=5/3$ is the adiabatic index, and $\Lambda=\Lambda(T, n_{\mathrm{H}}, Z,z,)$ is the cooling function.
The hydrogen number density $n_{\mathrm{H}} = \frac{X \rho}{m_p}$, where $X$ is the hydrogen mass fraction and $m_p$ is the mass of a proton.
The cooling rate $n_{\mathrm{H}}^2 \Lambda$ is then energy lost to radiative cooling per unit time per unit volume.

As shown by \cite{sternCoolingFlowSolutions2019}, the conservation equations \ref{eq:conmass}--\ref{eq:conK} can be written in logarithmic form and numerically integrated to find solutions for temperature, density, and velocity as a function of radial distance.
We expect the solutions to be valid from $R_{\mathrm{circ}}$ (at which the radial inflow stalls due to angular momentum) out to the cooling radius $R_{\mathrm{cool}}$ (see Section 2.2 of \cite{sternCoolingFlowSolutions2019} for a detailed discussion).
The cooling radius, the radius at which the cooling time exceeds the Hubble time $t_H=H_0^{-1}$, marks the boundary outside which the hot CGM gas does not have time to radiatively cool within a cosmological timescale.

\subsection{Including turbulence in the cooling flow model}\label{sec:CFmodel_turb}
We show in Section \ref{sec:Results} that non-thermal turbulent pressure makes a significant contribution to the total pressure support of many halos we analyze, while magnetic pressure is negligible. 
We modify the cooling flow model to include turbulence as described below.

Given a halo with thermal pressure $P \equiv P_{\mathrm{th}}$ and turbulent pressure $P_{\mathrm{turb}}$, the total pressure is $P+P_{\mathrm{turb}}$.
We parameterize $P_{\mathrm{turb}}$ in terms of the turbulent pressure ratio $\alpha(r) \equiv P_{\mathrm{turb}}(r)/P(r)$, so that $P_{\mathrm{turb}} = \alpha P$.
The total pressure is then $P(1+\alpha)$, which we use to modify the momentum conservation equation (Equation \ref{eq:conp}).

The new momentum equation is
\begin{equation}\label{eq:conp_turb}
    \frac{1}{2} \dv{v^2}{r} = -\frac{1+\alpha}{\rho} \dv{P}{r} - \frac{v_c^2}{r}\left[1-\left( \frac{R_{\mathrm{circ}}}{r}\right)^2 \right].
\end{equation}
Note that we use a radially-dependent turbulent pressure ratio, but we assume $P \dv{\alpha}{r} \ll (1+\alpha)\dv{P}{r}$ in the halo for our cooling flow model.
In Section \ref{sec:Nonthermalpressure} we present $\alpha(r)$ measured in FIRE simulations.

\subsection{Fitting cooling flows to FIRE}\label{sec:CFmodel_fitting}
We fit the cooling flow model, defined by the free parameter $\dot{M}$, to FIRE simulations.
For each simulation in our analysis set, we begin by finding cooling flow solutions for a wide range of $\dot{M}$ values, given the gravitational potential $\Phi(r)$ (which sets $v_c$), hot-phase metallicity $Z(r)$ (which specifies $\Lambda = \Lambda(T, n_{\mathrm{H}}, Z,z)$), and turbulent pressure fraction $\alpha(r)$ of the simulated halo.
The details of our method for integrating a cooling flow for a FIRE simulation are given in Appendix \ref{app:CFintegration}.
We search for bound cooling flow solutions for 100 $\dot{M}$ parameter values logarithmically spaced in the range $10^{-1}\ \Msun/\mathrm{yr} \le \dot{M}/ \le 10^{3}\ \Msun/\mathrm{yr}$.

We fit the cooling flow model jointly to the density and temperature profiles measured in the simulation.
We define radial shells whose centers $r_i/R_{\mathrm{vir}}$ are placed equidistant in log space, with shell thickness $\Delta \log (r/R_{\mathrm{vir}})=0.05$.
We define the residual $\delta(r_i)$ as the sum of the relative errors,
\begin{equation}
    \delta(r_i) = \frac{|n_{\mathrm{H}, \mathrm{sim}} - n_{\mathrm{H}, \mathrm{CF}}|}{n_{\mathrm{H}, \mathrm{CF}}} + \frac{|T_{\mathrm{sim}} - T_{\mathrm{CF}}|}{T_{\mathrm{CF}}}
\end{equation}
$n_{\mathrm{H}, \mathrm{sim}}(r_i)$ and $T_{\mathrm{sim}}(r_i)$ are the volume-weighted average density and temperature in the shell, where we only consider gas particles belonging to the hot phase (see Appendix \ref{app:virialbranch}).
$n_{\mathrm{H}, \mathrm{CF}}(r_i)$ and $T_{\mathrm{CF}}(r_i)$ are the values predicted by the cooling flow model at $r_i$; we integrate the cooling flow solutions with resolution $\Delta (\ln r) \le 0.1$ and linearly interpolate the solutions to find the needed quantities at $r_i$.

We find the cooling flow solution that minimizes the total residual $\sum_{i} \delta(r_i)$, summing over radial shells that are in our `fitting region' (see Section \ref{sec:selectinghotphase}); we call this solution the best-fit cooling flow model, characterized by a mass inflow rate $\dot{M}_{\mathrm{fit}}$.

\section{Results}\label{sec:Results}

\begin{figure*}
	\includegraphics[width=6.5in]{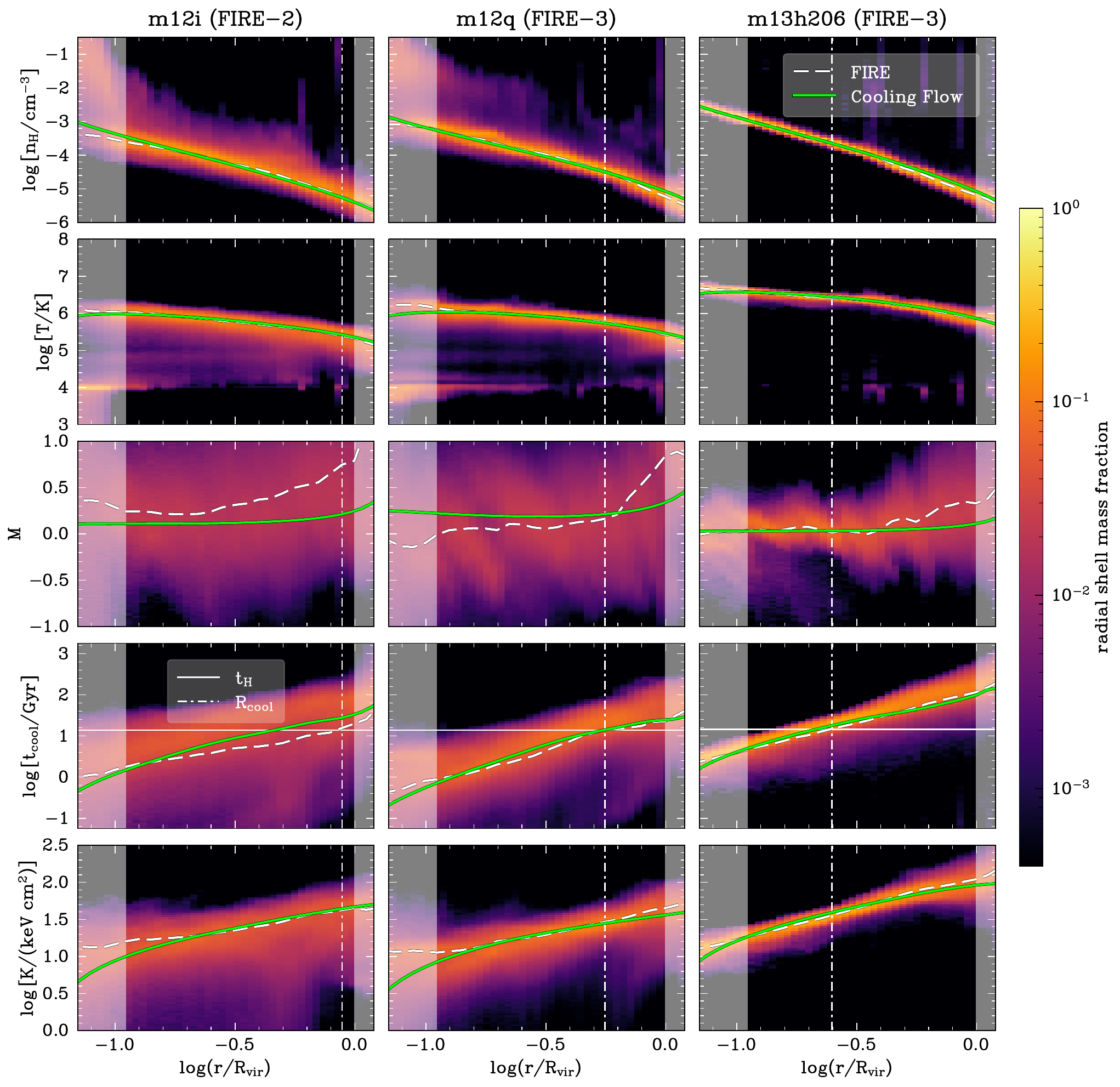}
    \caption{Radial profiles measured for three FIRE simulations plotted alongside the profiles of the best-fit cooling flow models.
    Results are shown for m12i (FIRE-2), m12q (FIRE-3), and m13h206 (FIRE-3).
    Hydrogen number density, temperature, Mach number, cooling time, and entropy profiles are shown in rows one to five, respectively; each column shows a different halo at $z=0$.
    Mass-weighted fractions of each quantity per radial bin are shown by the 2D histograms, which include \emph{all} gas mass.
    The white dashed lines show spherically averaged profiles calculated for gas particles belonging to the hot, virialized phase (see Appendix \ref{app:averaging} for details on our radial averaging method).
    The vertical dot-dashed line indicates the cooling radius, and the horizontal solid line shows the Hubble time.
    The profiles predicted by the best-fit cooling flow model are plotted as the solid curves; the model with angular momentum and turbulence was fit to the hot-phase density and temperature FIRE profiles.
    The shaded vertical bands indicate the radial shells that are outside the CGM ($0.1 R_{\mathrm{vir}} <r<R_{\mathrm{vir}}$).
    The 2D mass fractions show a multiphase CGM: in the m13 halo almost all of the gas within the cooling radius lies in the hot-phase branch, while the m12 halos contain a significant fraction of cold, dense gas in addition to the hot-phase gas, especially in the inner CGM.
    The cooling flows, despite being highly idealized and including a single free parameter, are able to model density and temperature in the hot CGM with good agreement.
    }
    \label{fig:CFprofiles_rep}
\end{figure*}

\subsection{Cooling flows in FIRE}
The first set of results we present test how well the FIRE simulations in our set are modeled by cooling flows.
As shown in Section \ref{sec:Nonthermalpressure}, non-thermal pressure support from magnetic fields is negligible in the hot FIRE halos we analyze, but turbulent pressure can be significant.

\subsubsection{Mass flows in simulations}\label{sec:Results_massflows}
Before we fit detailed models including turbulence and angular momentum, we begin by quantifying the hot-phase gas flowing towards the center of the halos. 
In the left column of Figure \ref{fig:Mdotratio}, we show the mass flow rate of gas belonging to the hot phase, $\dot{M}_{\mathrm{in,\ hot}}$, as a function of the mass flow rate of all gas, $\dot{M}_{\mathrm{in,\ tot}}$.
For all of the simulations in our sample, we show results for three radial ranges, representing the inner, middle, and outer parts of the halo.
Most points lie on or roughly on the 1:1 relationship, indicating that most of the inflows in the halos in our analysis are in the form of hot inflows.
Inflows are dominated by the hot phase in both the inner and outer halo.
Note for m12r (FIRE-3), m12r (FIRE-2), m12x (FIRE-3), and m13h223 (FIRE-3), we measure negative mass flow rates (i.e. outflows) in the hot phase and/or in all gas for at least one of the radial ranges; points with $\dot{M}_i<0$ are omitted from the figure.
The outflows in these halos may be powered by a recent burst of star formation.

In the right column of Figure \ref{fig:Mdotratio}, we plot $\dot{M}_{\mathrm{in,\ hot}}$ as a function of the rate of cooling of gas in the hot phase, $\dot{M}_{\mathrm{cool, hot}}$.
$\dot{M}_{\mathrm{cool, hot}}$ is the mass flow rate of hot-phase gas that cools in a cooling time.
Our procedure for calculating and averaging the mass inflow rates is described in Appendix \ref{app:averaging}.
There is rough agreement between the mass flow rate of hot-phase gas and the mass cooling rate (to an order of magnitude).
For many of the halos in our sample, points representing the inner radii have much better agreement with the 1:1 line than points measured in the middle and outer halo.
Although there is more scatter in this relationship compared to the relationship between $\dot{M}_{\mathrm{in,\ hot}}$ and $\dot{M}_{\mathrm{in,\ tot}}$ (left panels), note that for most of the $\sim 10^{13}\ \Msun$ halos, the points that systematically deviate from the 1:1 line correspond to radii outside the cooling radius.
The two outer radial ranges we plot lie outside the cooling radius of the m13 halos in our analysis, which we find ranges from $\sim 0.2R_{\mathrm{vir}} - 0.35R_{\mathrm{vir}}$.
There is not sufficient time for a steady state cooling flow to develop at these large radii, and gravitational collapse produces a larger hot inflow than one expected from cooling alone.
On the other hand, the innermost radial range shown, which is approximately within the cooling radius of the massive m13 halos, shows good agreement with the 1:1 line.

In short, Figure \ref{fig:Mdotratio} demonstrates that inflows in the $\sim 10^{12}-10^{13}\ \Msun$ halos we study are primarily hot. The results also indicate that most hot inflows are driven by cooling, especially within the cooling radius. 
The two results are consistent with expectations for cooling flows; building on these results, in the following sections we fit analytic cooling flow models to the simulations.

Figure \ref{fig:Mvir_ratio} shows the fraction of gas in radial shells that belongs to the hot phase.
The thick lines show median profiles for each of the three halo sets we analyze, calculated as the median hot-phase gas fraction per radial shell.
The m13 halos are dominated by the hot phase throughout the halo; nearly all of the gas is a part of the hot phase in these massive halos.
The m12 simulations contain more cold gas in the inner part of the halo.
In $0.1 R_{\mathrm{vir}} \lesssim r \lesssim 0.3R_{\mathrm{vir}}$, the median fraction of hot-phase gas in $\sim 10^{12}\ \Msun$ halos is $\sim 90\%$ for FIRE-3 and $\sim 60\%$ FIRE-2
However, the hot phase dominates the gas in the m12 simulations in the middle and outer halo.
As we show in the next sections, the higher fractions of cold gas in the inner parts of the m12 halos correspond to deviations in the inner halos from idealized cooling flows, 
while the hot phase-dominated outer halos are modeled well by cooling flows.

\subsubsection{Radial profiles of select halos}\label{sec:Results_radpro}
We next show CGM properties for three example halos in our set.
In Figure \ref{fig:CFprofiles_rep} we show results at $z=0$ for m12i (FIRE-2), m12q (FIRE-3), and m13h206 (FIRE-3).
These three halos are the same as those visualized in Figures \ref{fig:Sigmamap} and \ref{fig:Tmap}.
In the next section, we show results for the cooling flow fits over our entire simulation set.
We show radial profiles of hydrogen number density, temperature, Mach number, cooling time, and entropy measured in the simulations.
The 2D histograms show the mass fraction per radial bin of the quantities, calculated over \emph{all} gas particles within the bin.
The dashed lines are spherically averaged profiles computed for gas belonging to the hot, virialized phase; in Appendix \ref{app:averaging} we give details on our spherical averaging of each quantity.

The 2D mass fractions of all gas plotted in Figure \ref{fig:CFprofiles_rep} reveal a multiphase CGM.
The bright branches that are clearly visible in many panels, containing high fractions of the total gas mass, is gas we associate with the hot, virialized phase that is the focus of this paper; we describe our selection of this phase in Appendix \ref{app:virialbranch}.
For the two m12 halos shown there is also a nonnegligible fraction of cold, dense gas, especially in the inner CGM.
This is consistent with the low hot-phase gas mass fractions measured at small radii for many m12 halos (see Figure \ref{fig:Mvir_ratio}).
The cold gas, apparent at $T \sim 10^4$ K, can be seen in the surface density and temperature maps of the m12 halos (Figures \ref{fig:Sigmamap} and \ref{fig:Tmap}) in the form of the cold, dense structures, especially in the inner CGM.
These structures may represent cold gas flows into and out of the galaxy (though note that the hot gas typically dominates the mass flow, see Fig.~\ref{fig:Mdotratio}), and our process for selecting the hot phase excludes this gas.
Our spherically averaged profiles of the hot phase gas are in excellent agreement with the bright virialized branch, indicating that our method is effective in selecting the hot phase.

The solid curves in Figure \ref{fig:CFprofiles_rep} show our modeled cooling flows; we show the model with turbulence that best fits the FIRE simulation. 
We begin plotting at $0.07 R_{\mathrm{vir}}$; this is outside our choice of $R_{\mathrm{circ}}=0.05 R_{\mathrm{vir}}$ we use to integrate the model, where the modeled flow stalls since angular momentum supports the halo from collapsing further in. 
Note our regions over which we fit the cooling flows to the simulations ($r\gtrsim0.1 R_{\mathrm{vir}}$) are well outside this radius, and are only weakly affected by effects at this boundary.

The vertical dot-dashed line in each panel indicates the cooling radius, $R_{\mathrm{cool}}$.
The cooling times within $R_{\mathrm{cool}}$ are less than the Hubble time (plotted as the solid horizontal line), so radiative losses are significant and an inflow has time to develop.
The cooling times increase with increasing radius until $t_{\mathrm{cool}} > t_{\mathrm{H}}$ and radiative losses become small. 

Overall, the cooling flows, which only include a single free parameter ($\dot{M}$), have good agreement with the FIRE density, temperature, and entropy profiles in the CGM.
As we show in Section \ref{sec:Results_Mdot}, the mass flow rates $\dot{M}$ that best fit the cooling flow to the simulations have overall good agreement with the rates we measure in the hot-phase gas, with the best-fit $\dot{M}$ value falling within a factor of two of the rate measured in the inner halos of $>60$\% of simulations we analyze.
The cooling flow excellently models the profiles measured for the m13h206 (FIRE-3) halo, reproducing the profiles to within $\sim 10\%$ over $0.1 R_{\mathrm{vir}} <r<R_{\mathrm{cool}}$.
The agreement between the cooling flow and m13 halo is good even outside $R_{\mathrm{cool}}$, where the density profile is reproduced to $\lesssim 25\%$, and the temperature and entropy profiles are reproduced to $\lesssim 10\%$; as we discuss in Section \ref{sec:disc_outsideRcool}, this agreement of cooling flows with simulations outside the cooling radius may be due to approximately the same entropy profile being valid both within and outside $R_{\mathrm{cool}}$.
For the two m12 halos shown in Figure \ref{fig:CFprofiles_rep}, the profiles converge to the best-fit cooling flow in the middle and outer halo, with agreement generally within $\sim 10\%$.
The good agreement of the cooling flows with the measured profiles indicates that the halos contain a hot inflow in an approximate steady state.

The radial range over which we fit the cooling flow ($0.1 R_{\mathrm{vir}} <r<R_{\mathrm{vir}}$, i.e. the CGM) is indicated by the non-shaded area of the plots.
Within the fitting regions, the density and temperature of the two m12 shown in Figure \ref{fig:CFprofiles_rep} slightly differ from the cooling flows at small radii: the cooling flows over-predict density and under-predict temperature in the inner CGM (by $\lesssim 30\%$ at $r=0.1 R_{\mathrm{vir}}$).
This leads to an under-prediction of entropy by $\lesssim 60\%$ at $r=0.1 R_{\mathrm{vir}}$.
As we discuss in Section \ref{sec:innerCGMm12}, the deviation of the $\sim 10^{12}\ \Msun$ halo profiles in the inner halo from the cooling flow model may be due to the presence of rotating cooling flows (our cooling flow model includes only a 1D approximation of angular momentum) and significant cold gas.
We show results for how well the cooling flows fit our entire set of FIRE halos in the next section.

The spherically averaged profiles show a subsonic inflow of hot gas, i.e., $\mathcal{M} < 1$, over a wide radial range in the halos.
The best-fit cooling flows have Mach numbers $\mathcal{M} < 1$, which is expected since $\mathcal{M} \sim t_{\mathrm{ff}}/t_{\mathrm{cool}}$ in a cooling flow, and hot gas in the cooling flow slowly radiates at a timescale longer than the free fall time.

\begin{figure*}
	\includegraphics[width=6in]{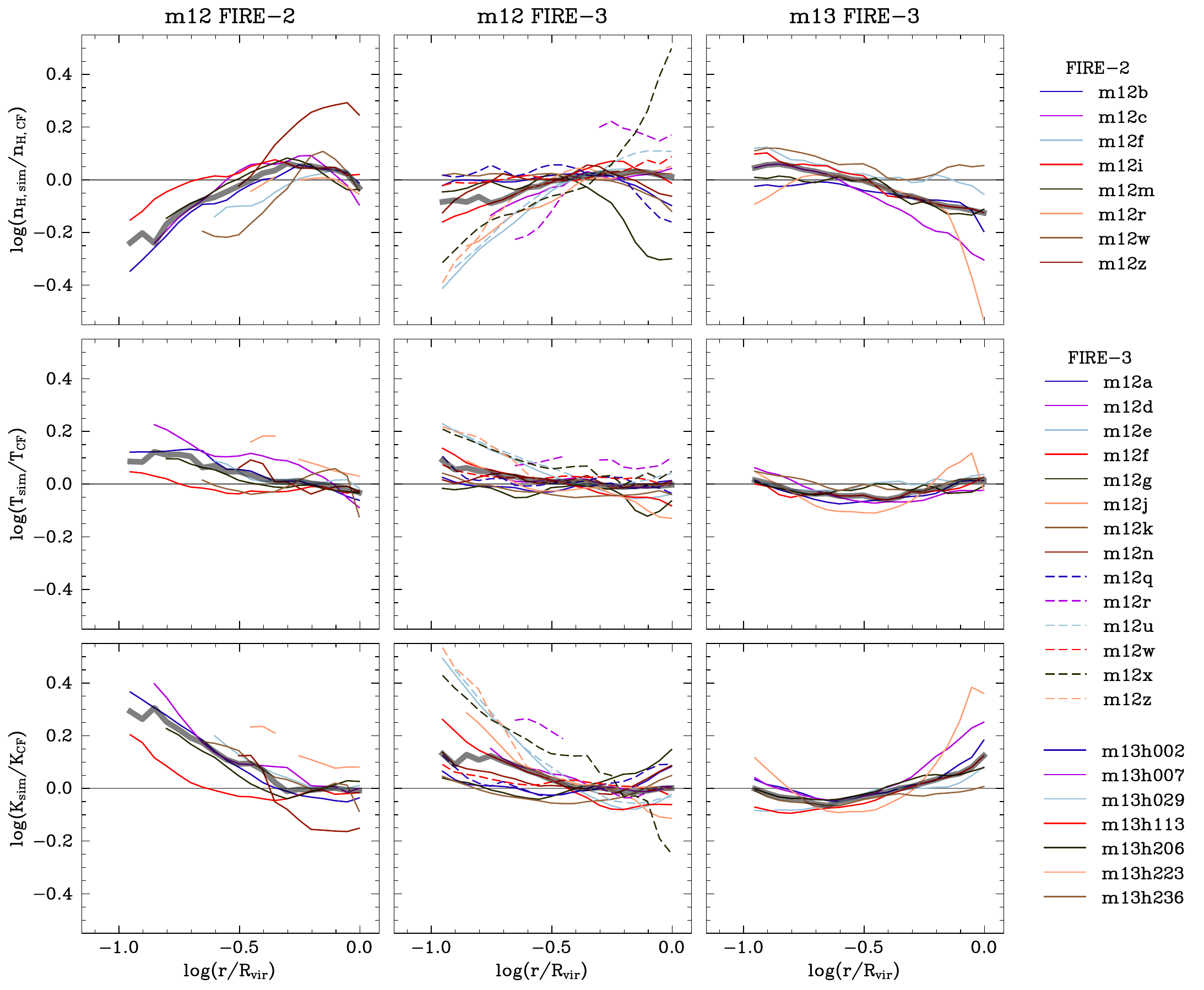}
    \caption{Ratios of spherically averaged FIRE profiles with respect to profiles of the best-fit cooling flow models with angular momentum and turbulence.
    Hydrogen number density, temperature, and entropy ratios are shown in rows one to three, respectively.
    Columns one to three show results for the FIRE-2 m12 halos, FIRE-3 m12 halos, and FIRE-3 m13 halos, respectively.
    Thin lines show results for individual halos, while the thick transparent lines show the median ratio measured in each panel.
    The cooling flow models were fit jointly to the hot-phase density and temperature profiles in each simulation.
    Ratios are plotted in the radial range over which we fit the cooling flow model; the fitting range covers radii where the majority of the gas belongs to the hot phase of the CGM (see Appendix \ref{app:virialbranch}).
    The m13 halos are excellently modeled by cooling flows, with the density and temperature profiles generally agreeing to within $\sim$20\% and $\sim$10\% in the median, respectively, over the full range shown. %
    For the m12 halos, the thermodynamic profiles converge to the cooling flow solutions in the outer CGM, with median errors in density, temperature, and entropy of $\lesssim$10\%.
    In the inner CGM of m12 halos, cooling flows systematically predict higher densities (by $\lesssim$25\%) and lower temperatures (by $\lesssim$20\%). 
    The inner halo deviations may be due to limitations of our 1D treatment of angular momentum and/or to the higher mass fractions in cool gas in these regions (see \S \ref{sec:innerCGMm12}). 
    }
    \label{fig:CFratio_turb}
\end{figure*}

\subsubsection{Cooling flow fits to simulation set}\label{sec:Results_fits}
Next we expand our analysis to all of the simulations listed in Table \ref{tab:sims}.
In Figure \ref{fig:CFratio_turb}, we present a summary of how well the cooling flows are able to model all of the simulations in our set.
We show ratios of the FIRE profiles (i.e., spherically averaged profiles measured for the hot phase) to the best-fit cooling flow model, where cooling flows have been fit jointly to the density and temperature profiles.

We show ratios of density, temperature, and entropy profiles of the simulations in our set to the predicted profiles of our best-fit cooling flow model with turbulence.
For each simulation, the ratios are plotted in the radial range over which we fit the cooling flow model.
This fitting range covers radial shells within the CGM ($0.1 R_{\mathrm{vir}} <r<R_{\mathrm{vir}}$) in which the hot phase gas dominates (i.e., where the majority of the gas belongs to the hot phase of the CGM; see Appendix \ref{app:virialbranch}).
The first two columns of Figure \ref{fig:CFratio_turb} show ratios for the m12 halos simulated in FIRE-2 (left) and FIRE-3 (middle), and the last column shows ratios for the FIRE-3 m13 halos. 
In each panel, the thick solid line shows the median ratio, calculated as the median value of all curves in a radial shell.

In the hot phase of the CGM of the $\sim 10^{13}\ \Msun$ halos, there is generally excellent agreement between the measured profiles and the modeled cooling flows, both in the inner CGM and extending to the outer regions of the halo.
Within the cooling radius, which we find ranges from $\sim 0.2R_{\mathrm{vir}} - 0.35R_{\mathrm{vir}}$ for the seven m13 halos shown, the density and temperature ratios are tightly distributed near the unity line, with a scatter of about 0.2-dex in density and 0.1-dex in temperature.
In comparison with the m12 halos, there is less scatter in the density and temperature ratios for the m13 halos.
Within the cooling radius, the median density and temperature ratios are $\lesssim 15\%$ and $\lesssim 10\%$, respectively.
The agreement between the cooling flows and m13 halos is good beyond the cooling radius as well, where the median temperature ratio remains $\lesssim 10\%$, and the median density ratio slightly increases to $\lesssim 20\%$.
The median entropy deviation of the m13 halos is generally $\lesssim 20\%$ throughout the halo.
The success of the cooling flows in modeling the halos even outside $R_{\mathrm{cool}}$, where the model is not expected to be applicable since the gas does not have enough time to cool, may be due to the halos having approximately the same entropy profile both within and outside the cooling radius (see the discussion in Section \ref{sec:disc_outsideRcool}).
Outside of the innermost-halo, at $r \gtrsim 0.2 R_{\mathrm{vir}}$ the cooling flows slightly over-predict temperature by $\lesssim$10\%.

In the outer CGM ($0.5 R_{\mathrm{vir}} \lesssim r \lesssim R_{\mathrm{vir}}$), the $\sim 10^{12}\ \Msun$ halo profiles generally converge to the cooling flow solutions.
At these radii, the profiles are scattered around the unity lines, and the cooling flows excellently fit the halos with median $n_\mathrm{H}$, $T$, and $K$ ratios of $\lesssim$10\% (for both the FIRE-2 and FIRE-3 m12 halos).
The m12 halos we analyze slightly deviate from the cooling flow model in the inner CGM, with the deviations increasing with decreasing radius as the inner boundary of the CGM (and roughly the outer boundary of the galaxies' gas in the ISM) is approached at $\sim 0.1 R_{\mathrm{vir}}$.
At small radii, the cooling flows systematically over-predict density (by $\lesssim 20\%$ for FIRE-3 and $\lesssim 40\%$ for FIRE-2) and under-predict temperature (by $\lesssim 15\%$ for FIRE-3 and $\lesssim 30\%$ for FIRE-2).
Since $K \propto T/n_{\mathrm{H}}^{2/3}$, we expect both of these deviations to lead to under-predicting entropy in the inner CGM by $\lesssim 30\%$ ($\lesssim 80\%$) for the FIRE-3 (FIRE-2) halos, which is indeed the result shown by the median entropy ratios.
The reason for the deviations in the inner part of the $\sim 10^{12}\ \Msun$ halos from our cooling flow model, despite hot gas dominating the inflows in these halos as we showed in Figure \ref{fig:Mdotratio}, may be due to the presence of rotation in the inner halos requiring a more accurate 3D treatment of angular momentum (our cooling flow model includes a 1D approximation of angular momentum) and/or the non-linear precipitation of cold gas out of the hot inner halo.
We discuss the effects further in Section \ref{sec:innerCGMm12}.

\begin{figure}
	\includegraphics[width=3.12in]{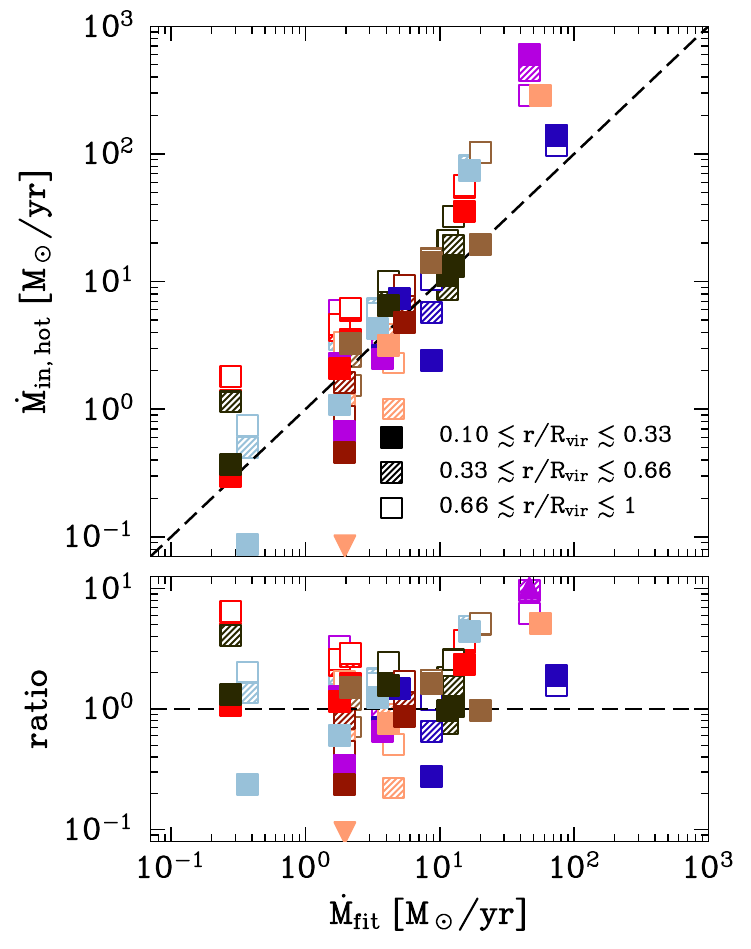}
    \caption{Comparison of mass inflow rates measured in the simulations, and mass inflow rates of the best-fit cooling flows models with turbulence, $\dot{M}_{\mathrm{fit}}$.
    $\dot{M}_{\mathrm{in,\ hot}}$ is the mass inflow rate of gas in the hot, virial phase, averaged in three radial ranges as in Figure \ref{fig:Mdotratio} (see Appendix \ref{app:averaging}).
    The cooling flow model was fit to the hot-phase density and temperature profiles in the simulations; the models shown here are the same models shown in Figure \ref{fig:CFratio_turb}.
    The dashed lines indicates a 1:1 relationship; the ratio of $\dot{M}_{\mathrm{in,\ hot}}$ to $\dot{M}_{\mathrm{fit}}$ is shown in the bottom panel.
    Results are shown for FIRE-2 m12, FIRE-3 m12, and FIRE-3 m13 halos.
    Triangles indicate points that fall outside of the ranges plotted; points with $\dot{M}_{\mathrm{in,\ hot}}<0$ are omitted. 
    Nearly 70\% of the points shown for the inner halo (solid points) are within a factor of two from the 1:1 lines, indicating overall good agreement between the measured and predicted mass flow rates within the cooling radius.
    }
    \label{fig:Mdot_scatter}
\end{figure}

\subsubsection{Mass flow rate comparison}\label{sec:Results_Mdot}
Finally, we compare the mass flow rate that best fits the cooling flow to a simulation versus the average mass flow rate of the hot phase measured directly in the simulation.
Figure \ref{fig:Mdot_scatter} shows $\dot{M}_{\mathrm{fit}}$, the mass flow rate of the best-fit cooling flow model with turbulence, plotted as a function of $\dot{M}_{\mathrm{in,\ hot}}$, the average mass flow rate of hot-phase gas measured from the simulation (see Appendix \ref{app:averaging} for our calculation of $\dot{M}_{\mathrm{in,\ hot}}$).
Note three simulations (FIRE-3 m12r, m12x, and m13h223) were measured to have negative mass flow rates in the hot phase in at least one of the radial ranges we show, and we omit them from the plot.
The net outflows in these halos may be powered by recent star formation bursts.

The dashed line in the figure marks $\dot{M}_{\mathrm{fit}} = \dot{M}_{\mathrm{in,\ hot}}$. 
The mass flow rates that best fit the cooling flow model to the simulations have overall good agreement with the actually inflow rates we measure in the hot-phase gas: nearly all points lie within 1-dex of the 1:1 line, and $\sim70$\% of the solid points shown that represent the inner halo radial range (which is within the cooling radius for most simulations) are within a factor of two from the 1:1 line. 
This consistency check confirms the cooling flow interpretation of the hot-gas thermodynamic profiles.

\subsection{Non-thermal pressure in FIRE halos}\label{sec:Nonthermalpressure}
We quantify two sources of non-thermal pressure support in FIRE halos, turbulence and magnetic fields.
\subsubsection{Turbulence}
We first analyze radial profiles of turbulent pressure $P_{\mathrm{turb}}$, which we calculate in radial shells as $P_{\mathrm{turb}}=\rho \sigma^2$. 
Here, $\sigma^2 = \frac{1}{3}(\sigma_r^2 + \sigma_\theta^2 + \sigma_\phi^2)$ is the square of the gas velocity dispersion, which we calculate in spherical coordinates using gas particles within the radial shell.
The square of the velocity dispersion (i.e. variance) $\sigma_j^2 = \sum_i \frac{(v_j-\bar{v}_j)^2}{N}$ for $j=\{r,\theta,\phi\}$, where the sum is over all particles $i$ within the radial shell and $N$ is the total number of particles in the shell; we only consider particles that belong to the hot phase.

\begin{figure}
	\includegraphics[width=3.12in]{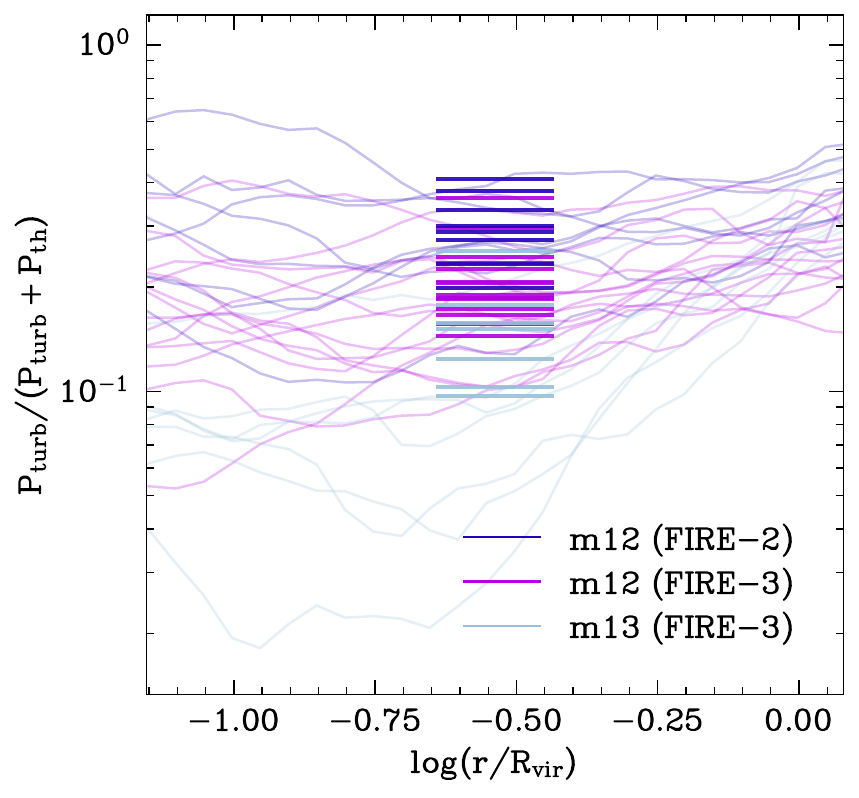}
    \caption{Turbulent pressure fractions of FIRE m12 and m13 simulations.
    Radial profiles of the contribution of the turbulent pressure to the summed thermal and turbulent pressure are shown as the faint solid lines. 
    The thick horizontal lines show the average turbulent pressure fraction for each simulation.
    Significant amounts of turbulence are measured for the halos in our set (turbulence contributes $\sim 10-40\%$ of the total pressure in the CGM); m12 halos are generally more turbulent than m13 halos. 
    }
    \label{fig:Pturb_ratio}
\end{figure}

\begin{figure}
	\includegraphics[width=3.12in]{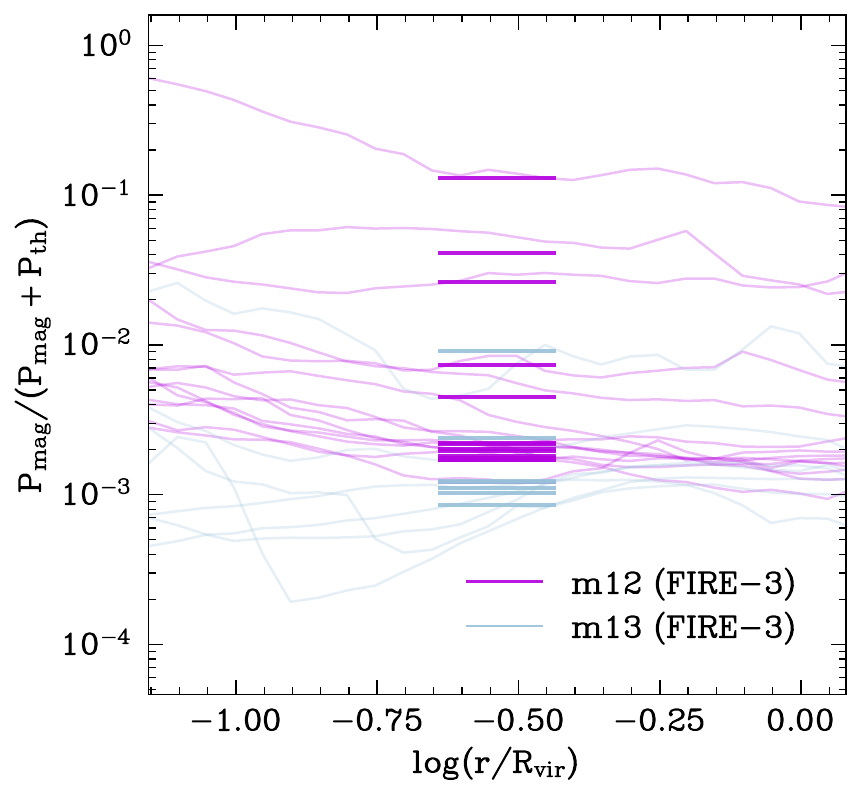}
    \caption{Magnetic pressure fractions in FIRE-3 m12 and m13 simulations.
    The magnetic pressure $P_{\mathrm{mag}}$ and thermal pressure $P_{\mathrm{th}}$ are spherically averaged profiles calculated for gas in the hot phase.
    The contribution of the magnetic pressure to the summed thermal and magnetic pressure are shown.
    The FIRE-3 simulations we analyze include magnetic fields.
    The thick horizontal lines show the average magnetic pressure fraction for each simulation.
    Magnetic pressure fractions are negligible for most of the halos.
    The average magnetic pressure fraction is slightly higher for the m12 halos, but typically less than one percent.
    }
    \label{fig:Pmag_ratio}
\end{figure}

\begin{figure*}
    \includegraphics[width=6in]{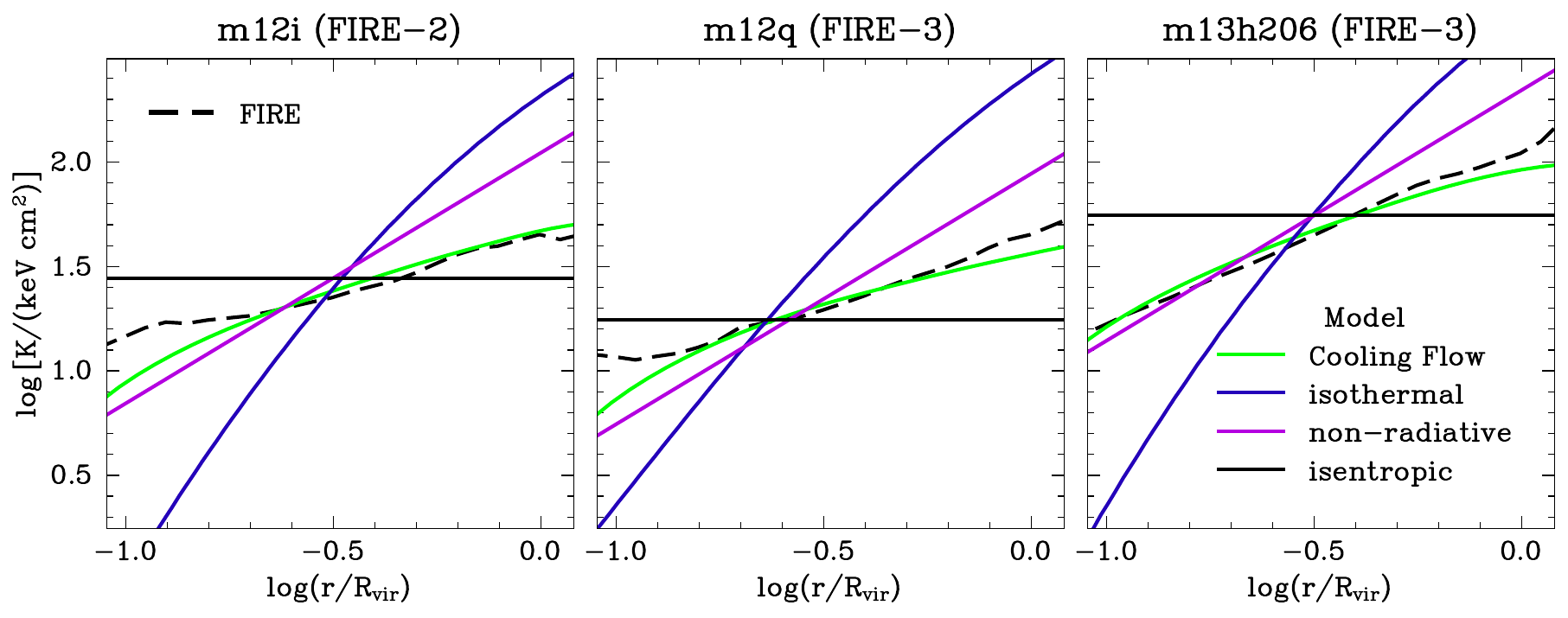}
    \caption{Entropy profiles for three FIRE simulations, overlaid with entropy slope predictions of analytic models of the hot CGM. 
    The spherically averaged entropy profile of hot-phase gas is indicated by the dashed line, and the profile of the best-fit cooling flow is plotted as the solid lime-green line (the same as the last row of Figure \ref{fig:CFprofiles_rep}). 
    The other solid lines show the slopes predicted by analytic models of the hot phase: an isothermal model with constant temperature equal to the virial temperature, the power law found by \protect\cite{voitBaselineIntraclusterEntropy2005} to best fit in non-radiative cosmological hydrodynamical simulations of clusters, and an isentropic model with constant entropy.
    The three slopes are normalized arbitrarily to facilitate comparison with the simulation profile within the CGM. The cooling flow has good agreement in the outer parts of the m12 halos, and throughout the m13 halo. The other models shown predict slopes that generally deviate more from the FIRE slopes.
    }
    \label{fig:entropyprofiles_NoBH}
\end{figure*}

Figure \ref{fig:Pturb_ratio} shows turbulent pressure results for our entire set of simulations.
We plot the turbulent pressure fraction $P_{\mathrm{turb}}/(P_{\mathrm{turb}}+P_{\mathrm{th}})$, which is the contribution of turbulence to the summed thermal and turbulent pressure in the hot phase.
Here the thermal pressure $P_{\mathrm{th}}$ is the spherically averaged profile calculated for gas in the hot phase.
Note the turbulent pressure fractions shown in Figure \ref{fig:Pturb_ratio} are equal to $\frac{\alpha}{1+\alpha}$, since $\alpha = P_{\mathrm{turb}}/P_{\mathrm{th}}$.
We show turbulent pressure fractions for m12 and m13 simulations.
There is significant scatter in the turbulence between the simulations.

The thick horizontal lines show the average turbulent pressure fraction for each simulation, calculated as the hot-phase mass-weighted average fraction of radial shells we identified dominated by the hot phase (within a maximum averaging window of $0.1 R_{\mathrm{vir}} <r<R_{\mathrm{vir}}$).
The average turbulent pressure fractions range from $P_{\mathrm{turb}}/(P_{\mathrm{turb}}+P_{\mathrm{th}}) \sim 0.1-0.4$.
m12 halos generally have higher average fractions then m13 halos (the mean average fraction is 0.25 for the m12 halos, and 0.15 for the m13 halos).
This is potentially due to the lower $t_{\mathrm{cool}}/t_{\mathrm{ff}}$ in the m12 halos (Goldner et al., in prep.).

The significant turbulent pressure fractions within the CGM that are the case for most simulations in our analysis set imply turbulence is a non-negligible contribution to the total pressure; as demonstrated in Figure \ref{fig:Pturb_ratio}, this effect is slightly bigger for m12-mass halos than m13-mass halos.
In Appendix \ref{app:CFturbcomparison}, we demonstrate the importance of including turbulence in the cooling flow model to reproduce the simulated temperature profiles. 

\subsubsection{Magnetic fields}
Next we consider pressure support from magnetic fields in the hot phase.
The FIRE-3 simulations in our analysis include magnetic fields; magnetohydrodynamics is part of the default physics implemented in FIRE-3 simulations (the FIRE-2 simulations in our set did not include magnetic fields).
For a magnetic field $\vb B$, the magnetic pressure is given by $|\vb B|^2/(2 \mu_0)$, where $\mu_0$ is the permeability in a vacuum.

In Figure \ref{fig:Pmag_ratio}, we show the magnetic pressure fraction $P_{\mathrm{mag}}/(P_{\mathrm{mag}}+P_{\mathrm{th}})$.
The magnetic pressure $P_{\mathrm{mag}}$ and thermal pressure $P_{\mathrm{th}}$ are spherically averaged profiles calculated for gas in the hot phase.
As shown in the figure, the magnetic pressure in the hot phase of the CGM is negligible for nearly all of the FIRE-3 m12 and m13 halos we analyze.
The contribution of magnetic pressure to the total pressure is $\lesssim 1$\% for most of the halos.
This result is consistent with \cite{hopkinsWhatCosmicRays2020}, who found a sub-dominant magnetic pressure relative to the thermal pressure ($P_{\mathrm{mag}}\ll P_{\mathrm{th}}$) in the CGM of a FIRE-2 `MHD+' m12 simulation (see also, \citealt{jiPropertiesCircumgalacticMedium2020}). As was the case for turbulent pressure fractions, the m13 halos generally have lower magnetic pressure fractions than their less-massive peers, with most m13 halos having magnetic pressure fractions of $< 10^{-3}$ in parts of the CGM.

The thick horizontal lines show the average magnetic pressure fraction for each simulation, calculated as the hot-phase mass-weighted average fraction of radial shells where we identified a significant hot phase (within a maximum averaging window of $0.1 R_{\mathrm{vir}} <r<R_{\mathrm{vir}}$; see Appendix \ref{app:virialbranch}).
m12 halos generally have higher average fractions then m13 halos; the mean average fraction is 0.02 for the m12 halos, and 0.002 for the m13 halos.
We measure a much higher average magnetic pressure fraction of 0.13 for one simulation in our set, m12z (FIRE-3).
For a FIRE-2 simulation run with the same initial condition, \cite{hopkinsFIRE2SimulationsPhysics2018} note that there is a halo merger at $z \approx 0$.
This merger may explain the significantly higher magnetic pressure we measure in this halo compared to the other FIRE-3 simulations.
The halo undergoing a merger may be in a temporarily disturbed state during which there is a much higher magnetic flux in the halo due to a high density of tangled field lines.

\subsection{Comparison of cooling flows with other analytic models}\label{sec:Results_othermodels}
We use the entropy profile to compare the agreement of various analytic models of the hot phase of the CGM with FIRE simulations and our modeled cooling flows.
We compare entropy profiles measured in FIRE simulations to the slopes that arise from various assumptions.
For our results we define the specific entropy as $K = k_B T/n^{2/3}$, where the total particle density $n=\rho/(\mu m_p)$ is related to the hydrogen number density we present elsewhere in this paper by $n_{\mathrm{H}} = n X \mu$.

In Figure \ref{fig:entropyprofiles_NoBH}, the dashed lines show spherically averaged entropy profiles we measured for m12i (FIRE-2), m12q (FIRE-3), and m13h206 (FIRE-3).
The solid lime-green lines show the result of the best-fit cooling flow (these results are identical to the last row of Figure \ref{fig:CFprofiles_rep}).

We overlay the figure with slopes of entropy predicted by three other analytic models, indicated by the additional solid lines.
The three slopes are normalized by eye to allow comparison with the simulation profile within the CGM.
The solid blue line shows the slope predicted by an isothermal model, where we assume a single-phase halo with constant temperature equal to the virial temperature.
Under hydrostatic equilibrium, $K \propto \exp \left[ \frac{2}{3}\xi \Phi(r) \right]\xi^{-5/3}$, where $\xi \equiv \frac{\mu m_p}{k_B T_{\mathrm{vir}}}$.
We measure the gravitational potential $\Phi(r)$ and virial temperature $T_{\mathrm{vir}}(M_{\mathrm{vir}})$ from the simulation, and assume a mean molecular weight of $\mu=0.62$.

The solid purple line is the power law $K \propto r^{1.2}$.
This slope was found by \cite{voitBaselineIntraclusterEntropy2005} to best fit the entropy profiles of non-radiative cosmological hydrodynamical simulations of galaxy clusters, over radial distances $0.2 \le r/R_{200c} \le 1.0$.
This slope is similar to the $K \propto r^{1.1}$ power law found in earlier studies by \cite{tozziEvolutionXRayClusters2001} and others. 

The solid black line represents an isentropic model, where entropy is assumed to be constant throughout the CGM.
Isentropic models have been used by, e.g., \cite{faermanMassiveWarmHot2020}, to model the CGM of Milky Way-mass galaxies. 

Both m12 halos shown are well fit by the cooling flow model in the outer region of the halo.
m13h206 is excellently fit by the cooling flow over nearly the entire CGM radial region plotted.
In contrast, the models which assume constant entropy or temperature in the halo produce slopes that are in clear disagreement with the slopes of the simulations.
Also, the slope of the power law measured in simulations of clusters that do not radiate by \cite{voitBaselineIntraclusterEntropy2005} has better, but still not as good, agreement with the FIRE slopes; their slope of $\dv{\ln K}{\ln r} = 1.2$ is steeper than the slopes of the three halos shown.

The FIRE-2 m12i halo has a shallow entropy profile in the inner CGM at $r \lesssim 0.3 R_{\mathrm{vir}}$.
The entropy profile of the FIRE-3 m12q halo gets slightly shallower at these smaller radii.
We find shallow entropy profiles in the inner CGM that are nearly isentropic for a significant fraction of the m12 halos in our set, in both FIRE-2 and FIRE-3.
Shallow inner entropy profiles were also found by \cite{esmerianThermalInstabilityCGM2021} in their study of FIRE-2 m12 halos. %
None of the m13 halos we analyze have flat entropy profiles; most of the m13 halos have entropy that approximately follows a power law in most of the halo. 
We discuss possible reasons for deviations from our idealized cooling flow models in the inner CGM of some m12 simulations in \S \ref{sec:innerCGMm12}.

As shown in Figure \ref{fig:Mdotratio}, the inflows we measure for the $\sim 10^{12}-10^{13}\ \Msun$ halos are generally dominated by hot phase gas.
This is in contrast to ‘precipitation’ models (which we do not show here) that predict inflows are cold and caused by gas that cools out of the hot phase and precipitates onto the central galaxy from larger radii (e.g., \citealt{voitAmbientColumnDensities2019}).

\section{Discussion}\label{sec:Discussion}

\subsection{Cooling flows as a benchmark model for the hot CGM without AGN feedback}\label{sec:disc_CF}
Previous studies have suggested the use of various analytic models as `benchmark' models that can serve as a reference point when analyzing different physical processes in the hot gaseous media surrounding galaxies.
For example, \cite{voitBaselineIntraclusterEntropy2005} showed that a power law was able to fit the entropy profiles of non-radiative cosmological hydrodynamical simulations of galaxy clusters.
They proposed the use of their entropy power law, which captures the effects of gravity and hydrodynamics, as a `baseline' model of the hot intracluster medium when considering additional physics in the halo such as radiative cooling and feedback.

In this work we focus on the massive galaxy mass range ($M_{\mathrm{halo}} \sim 10^{12}-10^{13}\ \Msun$), which is lower than the cluster masses studied by \cite{voitBaselineIntraclusterEntropy2005}.
At these lower masses, the cooling radius is a significant fraction of the virial radius, so we expect radiative cooling to play a role.
As shown in Section \ref{sec:Results_othermodels}, a comparison of the entropy slopes in three simulations reveals that the cooling flow model has a slope that is in significantly better agreement with the simulations than the steeper non-radiative cluster power law, and describes the hot gas in FIRE halos much better than the isothermal or isentropic assumptions.

We compared the realized density, temperature, and entropy profiles of the hot phase of the CGM in our simulations to the predictions of the best-fit cooling flow model, exploring CGM-radii $0.1 R_{\mathrm{vir}} <r<R_{\mathrm{vir}}$.
We found that $\sim 10^{13}\ \Msun$ FIRE halos are very well described by cooling flows within $R_{\mathrm{vir}}$. %
Simulated halos at the $\sim 10^{12}\ \Msun$ mass scale are also generally better described by cooling flows than by other analytic models in the literature, but with significantly flatter entropy profiles in the inner CGM in some halos (see Fig. \ref{fig:CFratio_turb}). 

These results imply that the hot CGM in the $\sim 10^{12}-10^{13}\ \Msun$ halos we analyze is primarily in the form of a hot inflow driven by radiative cooling. This is supported both by our analysis of the mass inflow rates of the gas in Figure \ref{fig:Mdotratio} and our fits to the full thermodynamic profiles. 
This indicates that in our simulations, feedback from stars does not significantly affect the hot CGM at the mass scales analyzed. 
This applies to most halos but there can be exceptions, e.g., following intense bursts of star formation, such as can be triggered by mergers.
This is consistent with the weak outflows measured for low-redshift $\gtrsim 10^{12}\ \Msun$ halos in previous FIRE studies \citep{muratovGustyGaseousFlows2015, angles-alcazarCosmicBaryonCycle2017, pandyaCharacterizingMassMomentum2021}.

The simulations we analyzed include multi-channel stellar feedback but neglect AGN feedback. 
The simulations also neglect some stellar processes, such as cosmic rays, which could potentially affect the CGM \citep[e.g.,][]{jiPropertiesCircumgalacticMedium2020}. 
Our results establish cooling flows as a useful reference solution to identify the effects of physical processes neglected here in future theoretical and observational studies. 

\subsection{Validity of cooling flow solutions outside of the cooling radius}\label{sec:disc_outsideRcool}
We find good agreement between the $n_\mathrm{H}$, $T$, and $K$ profiles predicted by the cooling flows and the profiles measured in the simulations out to $R_{\mathrm{vir}}$.
For many halos, especially at the $\sim 10^{13}\ \Msun$ mass scale, this agreement extends past $R_{\mathrm{cool}}$.
For example, for the $\sim 10^{13}\ \Msun$ halos we analyze (for which $R_{\mathrm{cool}}$ ranges from $\sim 0.2R_{\mathrm{vir}} - 0.35R_{\mathrm{vir}}$), the temperature profile is generally reproduced to $\lesssim 10\%$ throughout the halo, and the median density deviation slightly increases from $\lesssim 15\%$ within $R_{\mathrm{cool}}$ to $\lesssim 20\%$ beyond $R_{\mathrm{cool}}$ (see Figure \ref{fig:CFratio_turb}).
The $\sim 10^{12}\ \Msun$ halos also show good agreement (to within $\sim 10\%$ ) with the cooling flows in $n_\mathrm{H}$, $T$, and $K$ out to $R_{\mathrm{vir}}$, which is beyond the median $R_{\mathrm{cool}}$ of $\sim 0.52 R_{\mathrm{vir}}$ ($\sim 0.84 R_{\mathrm{vir}}$) measured for FIRE-3 (FIRE-2) halos.

One possible explanation for why the profiles predicted by the cooling flow models agree with the simulations at $r > R_{\mathrm{cool}}$ is that the entropy profiles predicted by cooling flows and by non-radiative simulations are not that different. 
For the three halos shown in Figure \ref{fig:entropyprofiles_NoBH}, note that the entropy power law found for cluster simulations without radiative cooling \citep{voitBaselineIntraclusterEntropy2005} is close to but slightly steeper than the slope of the cooling flows.
If the power law found for non-radiative galaxy clusters is also valid for halos at the $\sim 10^{12}-10^{13}\ \Msun$ halo mass scale, this implies that the entropy profile should follow roughly the same power law both inside and outside the cooling flow (i.e. the cooling radius).
For halos in approximate hydrostatic equilibrium, 
this would also imply similar $n_\mathrm{H}$ and $T$ profiles.

\subsection{Inner CGM of $\sim 10^{12}\ \Msun$ halos}\label{sec:innerCGMm12}
We found that the $\sim 10^{12}\ \Msun$ halos in FIRE deviate somewhat from pure cooling flows in the inner CGM, with the deviations increasing closer to the center of the halo.
In the inner CGM, the cooling flows over-predict density 
and under-predict temperature.
This leads to the cooling flows under-predicting entropy in the inner CGM,  
producing steeper entropy profiles in the inner CGM compared to the flatter entropy cores we measure in the simulations.
These deviations, which are stronger in FIRE-2 than in FIRE-3, may be due to a few factors that we discuss in this section.

Despite the deviations in the thermodynamic profiles from the cooling flow model in the inner CGM of some $\sim 10^{12}\ \Msun$ halos, we found that both the $\sim 10^{12}\ \Msun$ and $\sim 10^{13}\ \Msun$ inner halos are in most cases dominated by inflows of hot, virialized gas driven by radiative cooling (Figure \ref{fig:Mdotratio}).
These hot inflows are consistent with expectations from cooling flows.
\cite{hafenFatesCircumgalacticMedium2020} also previously found that most CGM gas that accreted onto the central galaxy in $\sim 10^{12}\ \Msun$ halos at $z=0$ was hot at $z\sim 0.25$.

The FIRE halos contain angular momentum and the hot gas inflows are therefore affected by rotation as the circularization radius $R_{\mathrm{circ}}$ is approached at the center. 
Indeed, \cite{hafenHotmodeAccretionPhysics2022} showed that hot inflows tend to form a rotating disk before accreting onto the central galaxy. 
The cooling flow model we adopt in this paper includes angular momentum using a 1D approximation 
\citep{sternMaximumAccretionRate2020}.
\cite{sternAccretionDiscGalaxies2024} show, for example, that including a 2D angular momentum treatment in the cooling flow model can decrease the predicted density by up to a factor of $\sim 2$ along the rotation axis rotation relative to a cooling flow without rotation (see their Figure 5). 
Neglecting angular variations in the density and temperature profiles expected in 2D or 3D in our 1D approximation could introduce significant discrepancies in average profiles. 
This effect is expected to be stronger for the m12 halos than for the m13 halos in our dataset because our m12 halos have larger average ratios of $R_{\mathrm{circ}} / R_{\rm vir}$. 
It would be interesting to investigate in future work whether the FIRE simulations can be modeled more accurately using multidimensional cooling flow models.

Alongside rotating cooling flows in the inner CGM, there may be a significant amount of cold gas that cooled out of the hot phase through nonlinear thermal instabilities, decreasing the hot-phase density of the inner FIRE halos.
In the surface density and temperature maps we show for two m12 halos (Figures \ref{fig:Sigmamap} and \ref{fig:Tmap}), there are cold, dense gas flows extending far into the CGM.
Additionally, in 2D mass fractions of density and temperature for these two halos (Figure \ref{fig:CFprofiles_rep}), we find there is a significant amount of cold, dense gas coexisting with the main, virialized branch of gas mass.
The hot-phase gas mass fractions plotted in Figure \ref{fig:Mvir_ratio} show that while nearly all of the gas in the m13 simulations belongs to the hot phase throughout halos, the m12 simulations (especially the FIRE-2 halos) contain more cold gas in the inner and middle parts of the halo.
In $0.1 R_{\mathrm{vir}} \lesssim r \lesssim 0.3R_{\mathrm{vir}}$, the median fraction of cold gas in $\sim 10^{12}\ \Msun$ halos is $\sim 10\%$ for FIRE-3 and $\sim 40\%$ FIRE-2.

In their analysis of FIRE-2 m12 halos, \cite{esmerianThermalInstabilityCGM2021} found that large density perturbations in the CGM created nonlinear thermal instabilities, causing cold gas to cool out of the hot phase.
They found that a majority ($\sim 70-90$\%) of low-entropy cold-phase gas at $z=0.25$ had cooled from the high-entropy hot phase at earlier times.

It is apparent in the 2D mass fractions shown in Figure \ref{fig:CFprofiles_rep} that the m12 halos contain gas with large variations in density throughout the inner halo, while nearly all of the gas in the m13 halo is contained in the hot phase and there are significantly less large variations in density.
\cite{esmerianThermalInstabilityCGM2021} found that the large density perturbations that cause a cold phase to form in the m12 halos can be triggered by winds from satellite galaxies, accretion from the IGM, winds from the central galaxy, and tidal interactions.
All of these processes are present in the $\sim 10^{12}\ \Msun$ halos we analyze, giving rise to an inner halo that deviates from an idealized, spherically symmetric state.
Further evidence for significant deviations from a smooth, spherically symmetry
CGM at this halo mass scale are the high turbulent pressure fractions we measure in the FIRE m12 halos (see Figure. \ref{fig:Pturb_ratio}).
\cite{oppenheimerDeviationsHydrostaticEquilibrium2018} also found significant differences in $\sim 10^{12}\ \Msun$ halos in EAGLE zoom-in simulations from hydrostatic equilibrium, due in large part to tangential motions in the halo that include uncorrelated motions.

Additionally, a theoretical analysis by \cite{sormaniEffectRotationThermal2019} found that rotation in the CGM could promote the condensation of cold gas from the hot phase by enhancing thermal instabilities in the medium.
Although these authors assumed that heating balances cooling in the hot CGM, which is not an assumption of our simulations, there may be related effects in the halos we analyze.

In the inner CGM of $\sim 10^{12}\ \Msun$ halos, we found the cooling flow model fit the FIRE-3 simulations with better agreement than the FIRE-2 halos in a median sense.
We carried out two tests to explore this systematic difference.
(1) We measured $R_{\mathrm{circ}}/R_{\mathrm{vir}}$ for the m12 halos, where $R_{\mathrm{circ}} = \left<j_{\mathrm{hot, z}}/v_c\right>_M$ is the circularization radius in the inner halo, and $j_{\mathrm{hot}, z}$ is the component of the specific angular momentum of hot-phase gas in the $z$ direction (which we average by mass).
For this test, we defined our coordinate system such that the $z$-axis is the axis of rotation of the galaxy (i.e., the total angular momentum vector of stars within $0.1 R_{\mathrm{vir}}$ of the center of the halo).
We found no systematic difference in $R_{\mathrm{circ}}/R_{\mathrm{vir}}$ between FIRE-2 and FIRE-3, suggesting that angular momentum may not play a larger role at the radii analyzed in FIRE-2 than in FIRE-3. 
(2) As shown above, we measure a $\sim 30\%$ higher median cold gas fraction in the inner CGM of FIRE-2 halos than FIRE-3 halos (Figure \ref{fig:Mvir_ratio}).
This suggests cold gas formation out of the hot phase (due to nonlinear instabilities, as found by \citealt{esmerianThermalInstabilityCGM2021}) may play a larger role in the FIRE-2 halos than in the FIRE-3 halos, and partly drive the deviations from cooling flows in their inner CGM. 
This is plausible since the CGM is non-trivially different at the m12 mass scale in FIRE-2 vs. FIRE-3, e.g. higher metallicities in FIRE-2 vs. FIRE-3 \citep[see][]{wijersNeVIIIWarmhot2024}. 

Another possibility for the flatter inner entropy profiles in the $\sim 10^{12}\ \Msun$ simulations vs. pure cooling flows is that stellar feedback, which is included in the simulations, could play some role. Even though this feedback does not drive significant CGM-scale outflows in the simulations, it can generate significant entropy in and around the central galaxy. 
Since an inverted entropy gradient is unstable to convection, convection would tend to create entropy cores.

In future work, it would be valuable to investigate the physics of the inner entropy profiles in this mass regime more thoroughly.

\subsection{Observations of the Hot Phase}\label{sec:observations}
As we discuss above, our results 
establish cooling flows as a reference solution for the hot CGM. %
In this section, we briefly discuss cooling flows in the context of existing observations of the hot CGM in $M_{\mathrm{halo}}\sim 10^{12}-10^{13}\ \Msun$ halos.

\subsubsection{O absorption and emission}
\cite{sternCoolingFlowSolutions2019} performed a detailed comparison of analytic cooling flow models to various observations, including O VII and O VIII absorption and emission in the Milky Way halo, and O VI absorption in the halos of the Milky Way and $z \sim 0.2$ star-forming galaxies. 
They found that cooling flows with a mass flow rate equal to the star formation rate of the Milky Way agreed with observations of X-ray absorption (O VII and O VIII column densities using \citealt{guptaHUGERESERVOIRIONIZED2012}), as well as observations of X-ray emission in the same lines; the slope of the density profile predicted by the cooling flow agreed with density slope derived from O VII and O VIII emission MW observations in \citealt{liPropertiesGalacticHot2017}). 
The rotation profile of the Milky Way's hot CGM has also been constrained using O VII absorption lines by \cite{2016ApJ...822...21H}. \cite{sternAccretionDiscGalaxies2024} modeled cooling flows with angular momentum in 3D (but with axisymmetry) and showed that \cite{2016ApJ...822...21H}'s observations are consistent with a rotating cooling flow.

However, \cite{sternCoolingFlowSolutions2019} found that cooling flows with a mass flow rate equal to the star formation rate of the galaxy predicted a factor of $\sim 5$ lower O VI column densities than those measured by UV observations of O VI absorption around the Milky way and $z \sim 0.2$ star-forming galaxies.
As they discuss, this may be explained by, e.g., processes that selectively heat the O VI-enriched outer halo, or by O VI tracing low-pressure photoionized gas outside the accretion shock of the halo in thermal equilibrium with the UV background rather than tracing the cooling flow within the accretion shock. 
One limitation of the comparison in \cite{sternCoolingFlowSolutions2019} to observations, however, is that their models did not include a fully realistic cosmological environment. 
To more firmly establish possible divergences from cooling flow models, it would be important in the future to not only compare the observations with idealized models but also to cosmological simulations like the FIRE simulations, which we show in this paper well approximated by cooling flows in some regimes.

Another important issue is accurately matching observed halos to modeled halos, especially around the $\sim 10^{12}\ \Msun$ mass scale where the properties of the CGM are predicted to change strongly owing to virialization \citep[e.g.,][]{sternMaximumAccretionRate2020}. 
Note that many of the halos around $z \sim 0.2$ star-forming galaxies observed by the COS-Halos survey \citep{werkCOSHALOSSURVEYEMPIRICAL2013} which \cite{sternCoolingFlowSolutions2019} included in their comparison 
have inferred halo masses of $\sim 10^{11.5}-10^{12}\ \Msun$ (e.g., see \citealt{mcquinnImplicationsLargeVi2018}).
Virialization of the inner CGM may not have completed yet in these lower-mass halos. %
Halos that have not completely virialized cannot sustain steady cooling flows to the center and may host larger-scale stellar feedback \citep[e.g.,][]{sternVirializationInnerCGM2021} that could heat the outer halo. %
Thus, the lower-mass halos in the COS-Halos sample may be in a fundamentally different physical state than the fully virialized Milky Way-mass ($\gtrsim 10^{12}\ \Msun$) halos in our m12 simulation sample that we find are well described by cooling flows. 
If this is the case, then discrepancies with COS-Halos would not invalidate cooling flow models for slightly more massive halos. 

\subsubsection{Ne VIII absorption}\label{sec:NeVIII}
Another probe of the hot CGM is Ne VIII absorption in the extreme UV (for gas in collisional ionization equilibrium, Ne VIII is most abundant at temperatures $\sim 10^{5.6}-10^{6.2}$ K, which is slightly higher than the peak CIE ionization fraction of OVI at $\sim 10^{5.5}$ K).
\cite{wijersNeVIIIWarmhot2024} analyzed Ne VIII in FIRE simulations of $\sim 10^{12}-10^{13}\ \Msun$ halos, including the simulations we analyze in this study, and compared Ne VIII column densities predicted by the simulations with observations made by the CASBaH \citep{burchettCOSAbsorptionSurvey2019} and CUBS \citep{quCosmicUltravioletBaryon2024} surveys. 
\cite{wijersNeVIIIWarmhot2024} showed that most of the Ne VIII-observed halos are comparable in mass to the $\sim 10^{12}\ \Msun$ FIRE-2 halos in their study. 

For the $\sim 10^{12}\ \Msun$ FIRE-2 halos, 
\cite{wijersNeVIIIWarmhot2024} found that the median Ne VIII column densities measured in the simulations at $z=0.5-1$ were consistent with CASBaH and CUBS data (which included many upper limits).
Taking into account our result that the FIRE-2 m12 halos are well-described by cooling flows in the outer CGM, this implies that the observations are consistent with the presence of cooling flows in $\sim 10^{12}\ \Msun$ halos.
\cite{wijersNeVIIIWarmhot2024} found evidence that the scatter in Ne VIII column densities in the FIRE-2 halos was smaller than that predicted by the observations, although this result had only modest statistical significance.

\subsubsection{X-ray surface brightness}\label{sec:xrays}
Recent studies have used eROSITA to measure stacked X-ray surface brightness profiles of $\sim 10^{12}-10^{13}\ \Msun$ halos (see e.g. \citealt{zhangHotCircumgalacticMedium2024a, popessoAverageXrayProperties2024, liRobustDetectionHot2024}). %

We discuss the predicted bolometric and X-ray surface brightness profiles of cooling flows in Appendix \ref{app:analytic_obs}. 
For $\sim 10^{12} \ \Msun$ halos, the eROSITA observations imply much shallower soft X-ray profiles $S_{X} \propto b^{-1}$ (where $b$ is the impact parameter) than predicted for cooling flows. 
As we show in the appendix, cooling flows are expected to have bolometric surface brightness profiles with slope on the order of $S_{\rm bol} \propto b^{-2.2}$, but much steeper than this in X-rays for $\sim 10^{12} \ \Msun$ halos because of the drop in gas temperature with increasing radius.
\cite{hopkinsCosmicRaysMasquerading2025} recently investigated this strong apparent discrepancy and suggested it could be resolved if the X-rays observed by eROSITA are dominated by inverse Compton (IC) scattering of cosmic microwave background (CMB) photons by $\sim$GeV cosmic ray (CR) electrons.
Their model, with CR electron injection rates consistent with those inferred from standard sources  in galaxies (i.e., supernovae and AGN), predicts a normalization and shallow slope 
that are both consistent with X-ray surface brightness observations by \cite{zhangHotCircumgalacticMedium2024a}. 
In this interpretation, the eROSITA surface brightness profiles in $\sim 10^{12} \ \Msun$ halos are not diagnostic of thermal emission from the hot gas but rather of emission powered by CRs. 
If correct, this means the thermal gas in the halos could be well-described by cooling flows despite the apparent discrepancy in the X-ray surface brightness profiles.

For more massive $\sim 10^{13} \ \Msun$ halos, there appears to be a real tension between the X-ray surface brightness profiles measured by eROSITA and the predictions of cooling flow models. 
At these higher masses, \cite{hopkinsCosmicRaysMasquerading2025} predict more thermal emission and CR-powered IC emission is not expected to dominate, so apparent X-ray discrepancies cannot be explained by CRs. 
At this mass scale, eROSITA observations also find shallow slopes (e.g., $ \dd{(\ln S_X)} / \dd{(\ln b)} \sim -1.2$ in \citealt{popessoAverageXrayProperties2024}) vs. the steeper slopes predicted by cooling flows. For example,  \cite{vandevoortImpactStellarFeedback2016} found slopes of $\dd{(\ln S_X)} / \dd{(\ln b)} \lesssim -3.5$ in earlier FIRE m13 simulations, also neglecting AGN feedback. 
The results of this paper imply that the discrepancy between the observed X-ray surface brightness profiles and cooling flow models tell us about missing physics in the simulations we analyzed at this halo mass scale. 
This is not surprising as there is a large body of previous work indicating that AGN feedback is likely required to explain the properties of galaxies and their halos at the massive end \citep[for reviews, see e.g.][]{somervillePhysicalModelsGalaxy2015, 2017ARA&A..55...59N}.

We also briefly consider X-ray data beyond surface brightness profiles. 
In particular, the slope of the entropy profile, $\dd(\ln K)/\dd(\ln r)$, is a useful test when comparing theoretical models to observations of the hot CGM.
\cite{babykUniversalEntropyProfile2018} used X-ray observations by Chandra to calculate entropy profiles for a sample of 40 massive galaxy halos (which consisted of elliptical and massive spiral galaxies, and faint groups), in addition to 110 cluster-scale halos. 
The halo masses for the massive galaxies in their sample range from $\sim 2\times10^{12} \ \Msun$ to $\sim 10^{14} \ \Msun$. 
When combining all the systems in their sample, including clusters, they found that a broken power law was able to fit the entropy profile, with $K \propto r^{2/3}$ in the innermost halo ($r \lesssim 0.1 R_{2500}$, which is at smaller radii than the CGM scales we study in this work), and $K \propto r^{1}$ for $r \gtrsim 0.1 R_{2500}$ (note the data extend only to $\lesssim 0.3 R_{2500}$ for the galaxy-scale halos in their sample). 
Here, $R_{2500}$ is the radius at which the mean enclosed density is 2500 times the critical density of the universe.
In comparison, non-radiative cosmological simulations predict $K \propto r^{1.2}$ (see \S \ref{sec:Results_othermodels}) and we measure a shallower slope $\dd(\ln K)/\dd(\ln r) \sim 0.8$ for the cooling flow model that successfully fits the m13h206 halo (see Figure \ref{fig:entropyprofiles_NoBH}). 
This suggests that while processes such as cooling and feedback must flatten entropy profiles relative to non-radiative simulations, cooling alone can possibly explain the observed flattering outside $0.1R_{2500}$, with feedback playing a more crucial role inside this radius. 
We caution, however, that \cite{babykUniversalEntropyProfile2018} fitted their ``universal'' entropy profile to the full set of systems in their study, and the entropy profile for galaxy-scale halos specifically is not well constrained outside $0.1R_{2500}$ due to the limited dynamic range probed by current X-ray observations, which are not sensitive to much larger radii at these masses.

\subsubsection{Thermal SZ effect}\label{sec:tSZdisc}

The thermal Sunyaev–Zeldovich (tSZ) signal, which is proportional to the integrated thermal pressure of electrons along a line of sight, is another probe of the hot CGM.
\cite{bregmanHotExtendedGalaxy2022} combined observations by Planck and WMAP, and presented stacked results of the Compton $y$ parameter for a sample of 11 $\sim L^*$ galaxies, all at distances of $< 10$ Mpc (at a distance of 10 Mpc, their Compton $y$-maps have a spatial resolution of 29 kpc, based on the 10' angular resolution of Planck).
For their sample of nearby galaxies, they were able to spatially resolve the halos and separate the contribution to the tSZ signal from contamination by the optical galaxy.
They measure the stacked $y$ profile for impact parameters up to $\sim 500$ kpc.

In Appendix \ref{app:analytic_obs}, we derive a predicted slope of the tSZ signal, $y \propto b^{-0.74}$, for an approximate, power-law cooling flow model.
This is roughly consistent with the slope of $d\ln{y}/d\ln b\approx -0.8$ found by \cite{bregmanHotExtendedGalaxy2022}, which follows from their best-fit $\beta$ profile 
at CGM scales (see their Equation 2).
Thus, the stacked tSZ results appear consistent with the presence of cooling flows in local $L^*$ galaxies.

While the tSZ signal is in principle stronger for more massive halos, more massive halos are on average more distant and existing measurements are not well resolved within $R_{\rm vir}$ at the $\sim 10^{13}$ M$_{\odot}$ mass scale \citep[e.g.,][]{2025arXiv250208850L}.

\subsubsection{Hidden cooling flows}
Finally, we comment on a recent development in observational studies of cooling flows. 
Namely, the cooling flow problem posed by the lack of observations of gas cooling below $\sim 1$ keV in galaxy clusters (as well as groups and elliptical galaxies) may not be as severe as initially suspected.
\cite{fabianHiddenCoolingFlows2022, fabianHiddenCoolingFlows2023a, fabianHiddenCoolingFlows2023} detected spectroscopic signatures of `hidden cooling flows' in clusters, in groups, and around elliptical galaxies.
In their interpretation, the observed X-ray spectra can be modeled as emission from cooling flows absorbed by embedded dust.

\cite{iveyHiddenCoolingFlows2024} recently applied the intrinsic multilayer dust absorption model to seven $z \sim 0$ elliptical galaxies, and inferred the presence of hidden cooling flows in all the galaxies in their sample.
The mass cooling rates of the hidden cooling flows are $\sim 20$ times the galaxies' star formation rates. 
This still poses a problem in terms of the fate of the cooling gas, but it suggests cooling flows could be present in the CGM. 
To account for the discrepancies with star formation rates, the authors favor low-mass star formation as the fate of the gas in the cooling flow. 
While the spectroscopic signatures suggest cooling flow models may be more broadly applicable than previously assumed, we note that these observations do not spatially resolve the X-ray emission. 
Therefore, the challenges associated with the surface brightness profiles discussed in \S \ref{sec:xrays} remain.

In future work, it would be interesting to create mock observables of simulated cooling flows, as well as simulations including AGN feedback, to study in more detail which scenarios are consistent with the growing set of observational constraints.

\subsection{Other Observational Diagnostics and Physics}\label{sec:other_diagnostics}
In this paper, we focus on modeling the hot phase, so we compared the cooling flow predictions with observations directly sensitive to the hot gas in the previous section. Here, we comment on other observational diagnostics and physical processes. We focus on halos around the $M_{\mathrm{halo}}\sim 10^{12}\ \Msun$ mass scale, since observations of the hot phase already imply the CGM of $M_{\mathrm{halo}}\sim 10^{13}\ \Msun$ halos is not well described by a pure cooling flow model.

Many CGM measurements are sensitive to cool gas ($T \sim 10^{4}$ K), and quasar absorption lines have shown that halos at the mass scales we analyze contain significant amounts of cool gas \citep[for a review, see][]{tumlinsonCircumgalacticMedium2017}. Some cool gas is also evident in our simulations (e.g., in the temperature maps in Fig. \ref{fig:Tmap} and in the quantitative diagnostics in Figs. \ref{fig:Mvir_ratio} and \ref{fig:CFprofiles_rep}), especially for the $\sim 10^{12}\ \Msun$ halos. 
We note that the presence of some cool CGM gas is compatible with cooling flows being a useful description for the bulk of the hot phase in realistic situations where halos are perturbed, for example, by satellite galaxies and their outflows \citep[e.g.,][]{esmerianThermalInstabilityCGM2021}. In realistic halos, non-linear perturbations can stimulate the condensation of cool gas. When embedded in a hot phase in a virialized CGM, this cool gas is generally subdominant in mass and by volume by large factors relative to hot gas. In this regime, the small-scale, cool clouds are sensitive to numerical resolution \citep[e.g.,][]{2019ApJ...882..156H, 2019MNRAS.482L..85V, 2019ApJ...873..129P} and are likely not robust tracers of the volume-filling hot phase. 

\cite{2025arXiv250417001K} analyze a subset of the FIRE-2 m12 simulations included in the present paper with a focus on the observational properties of cool gas. They show that the cool gas in the simulations can explain observed strong (equivalent width $\sim 1$ \AA) low- and mid-ionization absorbers from a variety of quasar absorption line surveys. Interestingly, though, \cite{2025arXiv250417001K} find that these strong absorbers generally arise in the inner CGM of halos that are not yet fully virialized (since the CGM is predicted to complete viralization around $M_{\mathrm{halo}}\sim 10^{12}\ \Msun$, many sight lines through halos at that mass scale are expected to probe a pre-virialization CGM). In the inner CGM of such halos, the cool gas dominates by mass, has a volume-filling fraction of order 50\%, and its properties are relatively well converged \citep[see][for related results on damped Ly$\alpha$ absorbers arising in the CGM at high redshift]{2021MNRAS.507.2869S}.\footnote{The analysis in \cite{2025arXiv250417001K} includes a range of redshifts. The m12 halos that are fully virialized by $z=0$ in our analysis have a cool inner CGM at earlier times.} Overall, we conclude that the available cool gas observations are consistent with the presence of cooling flows in fully virialized halos at the $M_{\mathrm{halo}}\sim 10^{12}\ \Msun$ mass scale.

One of our results is that turbulence is a significant component to the total pressure support in our cooling flow models for $\sim 10^{12}\ \Msun$ halos, with $P_{\rm turb}/(P_{\rm turb}+P_{\rm th}) \sim 15-40\%$ for the m12 simulations at $z=0$ (Fig. \ref{fig:Pturb_ratio}). As for column densities and equivalent widths, it is important to compare our cooling flow models with observations of halos sufficiently massive to have a fully virialized CGM. The existing measurements sensitive to turbulence in the relevant halo mass range are quite limited, but we performed the following check. Turbulence in the hot phase can be probed by the kinematics of high-ionization absorbers, such as Ne VIII and O VI, which probe temperatures comparable to the virial temperature at $M_{\mathrm{halo}}\sim 10^{12}\ \Msun$. \cite{quCosmicUltravioletBaryon2024} report such measurements. For the more massive galaxies in their samples ($M_{\star}\sim 10^{10-11}\ \Msun$),
they find that about half of their Ne VIII and O VI absorbers have line-of-sight velocity dispersions in the range $\sim50-100$ km s$^{-1}$. The turbulent pressure fraction can be estimated from observations as
\begin{equation}
\frac{P_{\rm turb}}{P_{\rm turb}+P_{\rm th}} \approx \frac{0.9 m_{\rm p} \sigma_{\rm los}^{2}}{1.5 k_{\rm B} T + 0.9 m_{\rm p} \sigma_{\rm los}^{2}},
\end{equation}
where $\sigma_{\rm los}$ is the line-of-sight (1D) velocity dispersion and the formula assumes a mean molecular weight $\mu \approx 0.6$ appropriate for an ionized cosmic plasma \citep[see eq. 6 in][]{2022MNRAS.516.4882Q}.

For $T\approx10^{6}$ K, the velocity dispersion range $\sim50-100$ km s$^{-1}$ implies $P_{\rm turb}/(P_{\rm turb}+P_{\rm th}) \sim 15-40\%$, in excellent agreement with what we found for our m12 simulations. For O VI especially, there are several measured velocity dispersions smaller than 50 km s$^{-1}$, however. At face value, this suggests lower turbulence energy fractions in some halos, but we note that the O VI collisional ionization fraction peaks at temperatures $T \sim 10^{5.5}$ K, which is lower than most of the hot gas in the fully virialized m12 halos modeled in this work (e.g., Fig. \ref{fig:CFprofiles_rep}). Therefore, some of the O VI measurements may probe localized structures in halos rather than full sight lines through the hot gas. If so, then the measured velocity dispersions could substantially underestimate the true velocity dispersion of the hot gas. 
Based on the existing observations, we conclude that the turbulence fractions predicted by our models appear broadly consistent with observations. In the future, X-ray observations could also constrain turbulence in $\sim 10^{12}\ \Msun$ halos, but current instruments lack either the sensitivity or the spectral resolution to do this.

Next, we comment briefly on other non-thermal processes in the CGM. Some of the simulations analyzed in this paper include magnetic fields (the FIRE-3 runs), while others do not (the FIRE-2 runs). We find no significant differences in the properties of the hot phase directly attributable to magnetic fields. This is consistent with our result that the magnetic pressure in the hot phase of the CGM is $\lesssim 1\%$ of the thermal pressure for most of the halos we analyze with magnetic fields included.

\cite{2022MNRAS.516.4417P} previously compared synthetic Faraday rotation measure (RM) profiles for FIRE m12 galaxies to observational constraints in the CGM. \cite{2022MNRAS.516.4417P} found the m12 simulations are consistent with the RM upper limit from a fast radio burst sight line through the halo of a single $M_{\star} \approx 10^{10.7}$ M$_{\odot}$ galaxy from \cite{2019Sci...366..231P}, as well as with upper limits from a statistical analysis of RMs in high-redshift radio sources with foreground galaxies from the DESI Legacy Imaging Surveys from \cite{2020MNRAS.496.3142L}. \cite{2023A&A...670L..23H} reported a tentative detection of magnetic field strengths of a few microgauss using LOFAR data in the CGM of foreground galaxies, but they only find evidence of a detection when restricting their analysis to the subset of sight lines from their full dataset that pass through the halos of galaxies with higher inclination angles, at azimuthal angles close to the minor axis of the galaxies and for impact parameters less than 100 kpc. In their conservative estimate, the significance of this detection is $2.8\sigma$. If confirmed, this could indicate significant magnetization of regions of the CGM affected by outflows along the minor axis (either stellar or driven by AGN). While this is an interesting possibility, we note that the median stellar mass of the galaxies for the detection in the \cite{2023A&A...670L..23H} sample is $M_{\star} = 10^{9.1 \pm 0.9}$ M$_{\odot}$. Given the empirically inferred relation between stellar mass and halo mass, this implies halo masses $\sim 2 \times 10^{11}$ M$_{\odot}$ \citep[see e.g. Appendix A in][]{wijersNeVIIIWarmhot2024}. This is below the halo masses for the m12 simulations analyzed in this paper (see Table \ref{tab:sims}). At these low masses, the CGM is not expected to be fully virialized in most cases, so we do not expect most of the galaxies in the \cite{2023A&A...670L..23H} subset to be well described by the cooling flow model. In the more massive halos we focus on, star formation-driven outflows do not reach far into the CGM in FIRE, but outflows can be much more important in lower-mass halos \citep[e.g.,][]{muratovGustyGaseousFlows2015, angles-alcazarCosmicBaryonCycle2017, pandyaCharacterizingMassMomentum2021}. Overall, we conclude that the existing observational constraints on magnetic fields in the CGM are consistent with our results regarding cooling flows at the $\sim 10^{12}\ \Msun$ mass scale, when restricting attention to the halos sufficiently massive for the CGM to be fully virialized.

Cosmic rays are neglected in the simulations analyzed in this paper but, depending on assumptions, could have significant effects on the CGM \citep[see][for a recent review of cosmic rays in the context of galaxy formation]{2023A&ARv..31....4R}. Several studies have shown that non-thermal CR pressure gradients can accelerate cooler galactic winds and/or support a cooler CGM \citep[e.g.][]{2013ApJ...777L..16B, 2014MNRAS.437.3312S, jiPropertiesCircumgalacticMedium2020}. However, the effects of CRs remain highly uncertain because they are highly dependent on the assumed CR transport model \citep[e.g.,][]{2021MNRAS.501.3663H, 2022MNRAS.517.5413H}. In this paper, we find that cooling flow models that neglect CRs appear broadly consistent with a variety of observations of low-redshift $\sim 10^{12}\ \Msun$ halos. This suggests that cosmic rays do not strongly modify the CGM in this regime. This is consistent with the study of \cite{2023MNRAS.521.2477B}, who argued that COS-Halos observations imply that the CR transport rate increases rapidly away from the interstellar medium as the cosmic rays stream out into the diffuse CGM. In \S \ref{sec:xrays}, we noted that inverse Compton emission powered by cosmic ray electrons could potentially explain the shallow soft X-ray surface brightness profiles in $\sim 10^{12}\ \Msun$ halos \citep{hopkinsCosmicRaysMasquerading2025}. This is not necessarily inconsistent with cosmic rays playing a small dynamical or thermodynamical role in the CGM, since the CR protons (which generally carry most of the energy at acceleration) need not follow the CR electrons. The role of CRs in shaping the CGM in different regimes remains uncertain, so future work on refining the constraints is certainly warranted. For example, \cite{2025arXiv250201753Q} recently suggested that cosmic rays from massive black holes could have important effects in group-mass halos, so CRs may play a role in explaining why $\sim 10^{13}\ \Msun$ halos are not well modeled by cooling flows.

\section{Conclusions}\label{sec:Conclusions}
There is currently a lack of consensus in modeling the hot phase of the CGM of massive galaxies ($M_{\mathrm{halo}}\sim 10^{12}-10^{13}\ \Msun$) since the common assumption of hydrostatic equilibrium does not uniquely constrain the thermodynamic ($n$, $T$, $K$) profiles of hot gas.
In this paper, we use cosmological zoom-in simulations of galaxy formation from the FIRE project to test the predictions of idealized cooling flow models against the hot CGM self-consistently realized in the simulations. 
We focus on 
Milky Way-mass halos ($M_{\mathrm{halo}} \sim 10^{12}\ \Msun$; our `m12' halos) and more massive halos ($M_{\mathrm{halo}} \sim 10^{13}\ \Msun$; our `m13' halos). 
We analyze simulations that include multi-channel stellar feedback but neglect AGN feedback. 
Our dataset consists of 29 different simulations. Some simulations include magnetic fields. 

We compare the simulation predictions for the hot CGM to 
the predictions of the cooling flow model presented in \cite{sternCoolingFlowSolutions2019, sternMaximumAccretionRate2020}. 
Cooling flows are an idealized steady-state model that describe gas in a gravitational potential that radiatively cools and flows towards the center of the halo.
We modified the cooling flow model %
to include non-thermal pressure support from turbulence in the halo.

Our main results are summarized below:
\begin{enumerate}
    \item Inflows in the simulated $\sim 10^{12}-10^{13}\ \Msun$ halos are in most cases dominated by hot gas, and the hot inflows are consistent with being driven by radiative cooling, especially within the cooling radius (Figure \ref{fig:Mdotratio}).

    \item The thermodynamic profiles of the hot phase of $\sim 10^{13}\ \Msun$ halos in FIRE are excellently modeled by cooling flows, with agreement in their density and temperature profiles to within $\sim 20\%$ and $\sim 10\%$ in the median, respectively (Figure \ref{fig:CFratio_turb}).
    This indicates that the CGM in these halos is approximately in steady state, with radiative cooling losses in the hot gas compensated by compressive heating in the inflow. 
    Moreover, stellar feedback does not have a significant effect at the halo scale.
    
    \item For the $\sim 10^{12}\ \Msun$ halos, the thermodynamic profiles converge to the cooling flow solutions in the outer CGM, 
    with median errors in density and temperature of $\lesssim$10\%. 
    In the inner CGM, cooling flows systematically predict higher densities (by $\lesssim 40\%$) and lower temperatures (by $\sim$30\%) than what we measure in the hot phase of some simulations (Figure \ref{fig:CFratio_turb}).
    The deviations from the cooling flow model in the inner CGM may be due to limitations of our 1D treatment of angular momentum and/or to the larger cold gas fractions due to non-linear thermal instabilities in these regions (\S \ref{sec:innerCGMm12}).

    \item The entropy profiles predicted by the cooling flow model are in better agreement with FIRE than several other models from the literature. 
    The cooling flow model is in qualitatively better agreement with the FIRE profiles than the isothermal and isentropic assumptions, which predict much steeper and shallower entropy profiles, respectively, than measured in FIRE (Figure \ref{fig:entropyprofiles_NoBH}). 
    While the power-law entropy profile predicted from non-radiative galaxy cluster simulations by \cite{voitBaselineIntraclusterEntropy2005} matches the FIRE halos better than the isothermal and isentropic assumptions, the cooling flow models agree better with the FIRE simulations especially at large radii.

    \item Non-thermal pressure from turbulence is a significant contribution to the pressure support in the CGM of FIRE halos.
    We measure average turbulent pressure fractions in the hot phase ranging from $\sim 10-40\%$.
    The $\sim 10^{12}\ \Msun$ halos have higher average turbulent pressure fractions than the $\sim 10^{13}\ \Msun$ halos; the overall mean fraction is 25\% for the $\sim 10^{12}\ \Msun$ halos and 15\% for the $\sim 10^{13}\ \Msun$ halos (Figure \ref{fig:Pturb_ratio}). 
    Turbulence must be included in the cooling flow model to accurately predict the temperature profiles.
    
    \item Non-thermal pressure from magnetic fields is not a significant contribution to the total pressure in FIRE halos.
    The magnetic pressure in the hot phase of the CGM is negligible ($\lesssim 1\%$) for most of the 
    halos we analyze with magnetic fields included (Figure \ref{fig:Pmag_ratio}).

    \item Predictions by cooling flows agree with a variety of observations made for $\sim 10^{12}\ \Msun$ halos, including O VII and O VIII absorption and emission in the Milky Way halo, Ne VIII absorption in $z \sim 0.5-1$ halos, and stacked tSZ observations of nearby $\sim L^*$ galaxies.
    These observations suggest the possibility of CGM-scale cooling flows at this halo mass scale.

     \item For $\sim 10^{13}\ \Msun$ halos, the predicted X-ray surface brightness profiles from thermal emission in cooling flows are much steeper than recent stacked eROSITA observations. %
     This discrepancy may be due to the effects of AGN feedback in massive halos neglected in pure cooling flow models.
\end{enumerate}

Our results imply that cooling flows can be used as a useful baseline model for the hot CGM of halos of mass $M_{\mathrm{halo}} \sim 10^{12}-10^{13}\ \Msun$.  
For example, since the simulations we analyzed neglect AGN feedback, deviations from cooling flow predictions can be used to quantify the effects of AGN feedback, either in simulations or in observations. 
Of course, there can be other physical processes we neglected that impact the hot CGM (e.g., cosmic rays), so deviations from cooling flows more generally probe effects not included in the FIRE simulations we used to validate the cooling flow model.

The observational landscape over the coming decade and beyond is promising.
X-ray missions, such as the recently launched XRISM \citep{tashiroXRISMXrayImaging2022}, planned Athena \citep{barconsAthenaESAsXray2017}, and proposed HUBS \citep{cuiHUBSHotUniverse2020}
and AXIS \citep{reynoldsOverviewAdvancedXray2023} missions will enable absorption \citep{wijersWarmhotCircumgalacticMedium2020} and emission \citep{wijersWarmhotCircumgalacticMedium2022} studies of the hot CGM of galaxy-scale halos.
In addition to X-ray observations, upcoming cosmic microwave background experiments, e.g., the Simons Observatory \citep{adeSimonsObservatoryScience2019} and CMB-S4 \citep{abazajianCMBS4ScienceBook2016}, will have the sensitivity and resolution needed to enable studies of the Sunyaev–Zeldovich effect in the CGM of such halos.
Fast radio burst experiments such as CHIME/FRB will also soon greatly increase the number of dispersion measure measurements that can be used to study the CGM of galaxies \citep{chime/frbcollaborationCHIMEFastRadio2018}.

\section*{Acknowledgements}
We thank Sarah Wellons for help using FIRE Studio to create visualizations, and Drummond Fielding, Elizabeth Moné, and Phil Hopkins for useful discussions and collaboration on related projects.
I.S.\ was supported by the NSF Graduate Research Fellowship under Grant No.\ DGE-2234667.
C.-A.F.-G.\ was supported by NSF through grants AST-1715216, AST-2108230, AST-2307327, and CAREER award AST-1652522; by NASA through grants 17-ATP17-0067, 21-ATP21-0036, and 23-ATP23-0008; by STScI through grant JWST-AR-03252.001-A; and by the Research Corporation for Science Advancement through a Cottrell Scholar Award.
JS was supported by the Israel Science Foundation (grant No. 2584/21). 
L.B.\ was supported by the DOE Computer Science
Graduate Fellowship through grant DE-SC0020347.
N.A.W.\ was supported by a CIERA Postdoctoral Fellowship.
This work was performed in part at the Aspen Center for Physics, which is supported by National Science Foundation grant PHY-2210452.
Numerical calculations were run on the Northwestern computer cluster Quest, the Caltech computer cluster Wheeler, Frontera allocation FTA-Hopkins/AST20016 supported by the NSF and TACC, XSEDE/ACCESS allocations ACI-1548562, TGAST140023, and TG-AST140064 also supported by the NSF and TACC, and NASA HEC allocations SMD-16-7561, SMD-17-1204, and SMD-16-7592. 
Some of the calculations in this study utilize publicly available code developed by Alex Gurvich, which can be accessed at \url{github.com/agurvich/abg_python}.

\section*{Data Availability}
A public version of the GIZMO code is available at \url{http://www.tapir.caltech.edu/~phopkins/Site/GIZMO.html}. FIRE data products, including FIRE-2 simulation snapshots, initial conditions, and derived data products are available at \url{http://fire.northwestern.edu/data/}.

\input{refsoutput.bbl}

\appendix
\section{Virial branch selection procedure}\label{app:virialbranch}
\begin{figure*}
	\includegraphics[width=6in]{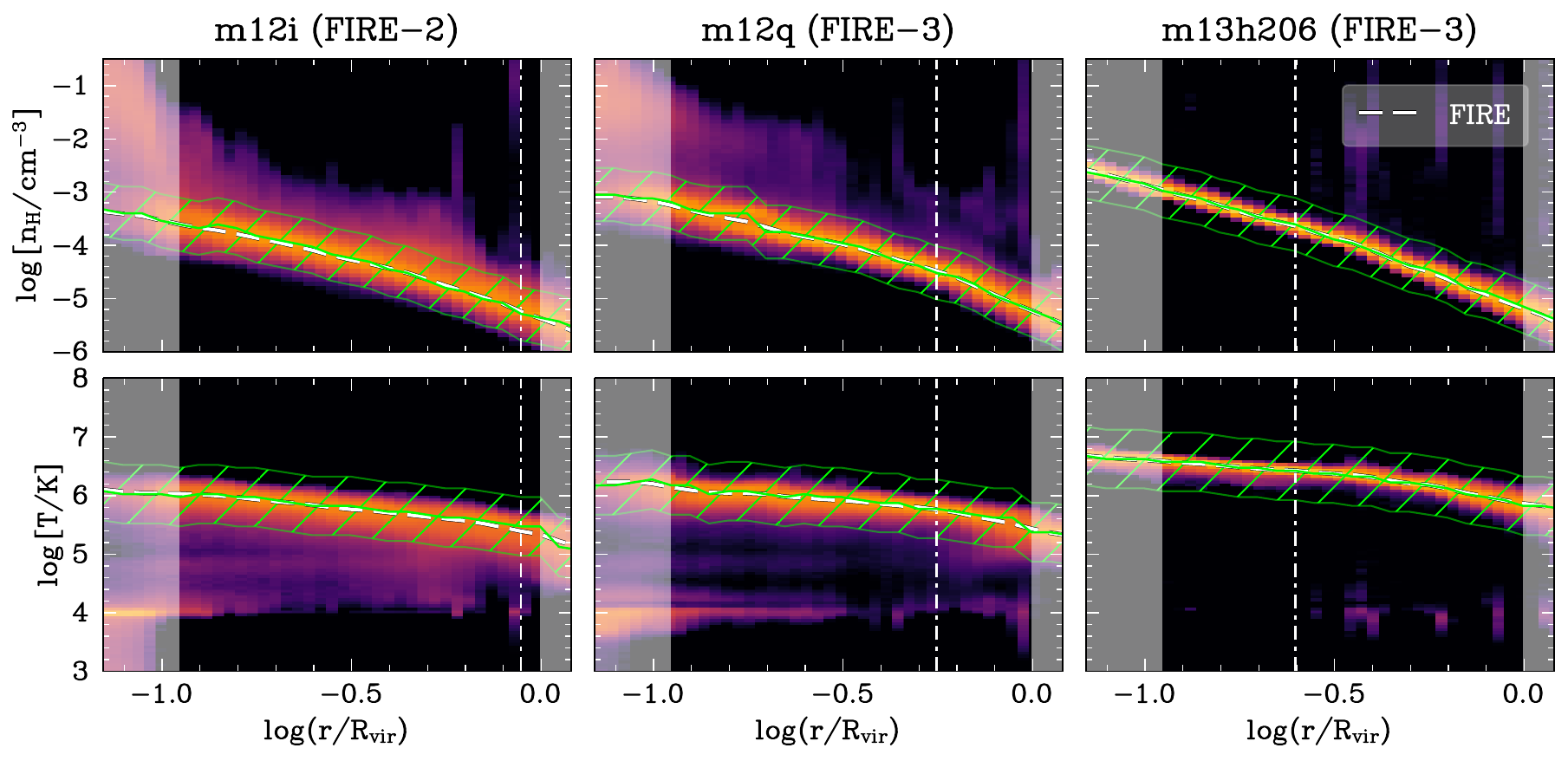}
    \caption{Virial branch selection in density and temperature for three FIRE simulations.
    FIRE results are identical to the first two rows of Figure \ref{fig:CFprofiles_rep}.
    The green hatched region in each panel is our identification of the virial branch, which we use to select the hot phase of the CGM (see Appendix \ref{app:virialbranch}).
    The shaded vertical bands indicate the radial shells that are either outside the CGM ($0.1 R_{\mathrm{vir}} <r<R_{\mathrm{vir}}$), or shells we do not identify as being dominated by the hot phase (in these examples, there are no such shells in the CGM region).
    Our fitting region is the radial range outside of the shaded bands.
    }
    \label{fig:mblabel}
\end{figure*}

We give details of our procedure for selecting the hot, virialized phase of the CGM in our simulations here.
In 2D temperature histograms of the particle mass distribution (the second rows of Figures \ref{fig:CFprofiles_rep} and \ref{fig:mblabel}), this phase is generally discernible as the bright `branch' at $\sim T_{\mathrm{vir}}$ containing most of the total gas mass (at most radii).
There is a corresponding virial branch visible in the 2D density histograms.

To select the virial branch, we begin by dividing the volume into radial shells whose centers $r/R_{\mathrm{vir}}$ are placed equidistant in log space, with $\Delta \log (r/R_{\mathrm{vir}})=0.05$.
In each shell, our procedure is essentially to find the modes of the temperature and density distributions, and select all gas particles with $T$ and $n_{\mathrm{H}}$ close to the modes. 

When computing the modes, we only consider particles with $T > 10^5$ K; this temperature is below the virial temperature for our halo mass range (see Equation \ref{eq:Tvir}).
The temperature cut ensures we select the mode of the correct branch; visually in the 2D temperature histograms, $T \sim 10^5$ K separates the virial branch from cool gas at $T \sim 10^4$ K.

Due to the high dynamic range in temperature and density, we calculate the temperature and density modes in log-space.
To find the temperature mode $(\log T)_{\mathrm{mode}}$, we plot a histogram of $\log T$ using particles in the radial shell with $T > 10^5$ K. 
$(\log T)_{\mathrm{mode}}$ is then the midpoint of the histogram bin containing the most particles. 
We follow the same steps to find $(\log n_{\mathrm{H}})_{\mathrm{mode}}$, again considering particles in the radial shell with $T > 10^5$ K. 

We then consider particles with $|\log T - (\log T)_{\mathrm{mode}}| \le 0.5$ and
$|\log n_{\mathrm{H}} - (\log n_{\mathrm{H}})_{\mathrm{mode}}| \le 0.5$ to belong to the hot phase.
Particles that fail to meet either of these two criteria are considered to not be part of the hot phase.
The results of our virial branch selection are shown in Figure \ref{fig:mblabel} for three FIRE simulations.
The green hatched regions shows the virial branch found in temperature and density.
The center line in the hatched regions is the mode, and the boundaries show $\pm$0.5-dex.
For the example simulations shown, the green hatched regions do an excellent job selecting the visibly-bright main branch, while excluding particles not belonging to the hot phase.
We find a similar result with all of the simulations in our analysis set.

Additionally, we remove radii where the hot phase of the CGM is not dominant from our analysis.
We consider the CGM to be found within $0.1 R_{\mathrm{vir}} <r<R_{\mathrm{vir}}$.
For each radial shell in the CGM, if the majority of the gas does not belong to the hot phase (i.e., $<50$\% of the total gas in the shell is found in the hot phase), we exclude that shell from our analysis of the hot phase.

\begin{figure}
	\includegraphics[width=3.12in]{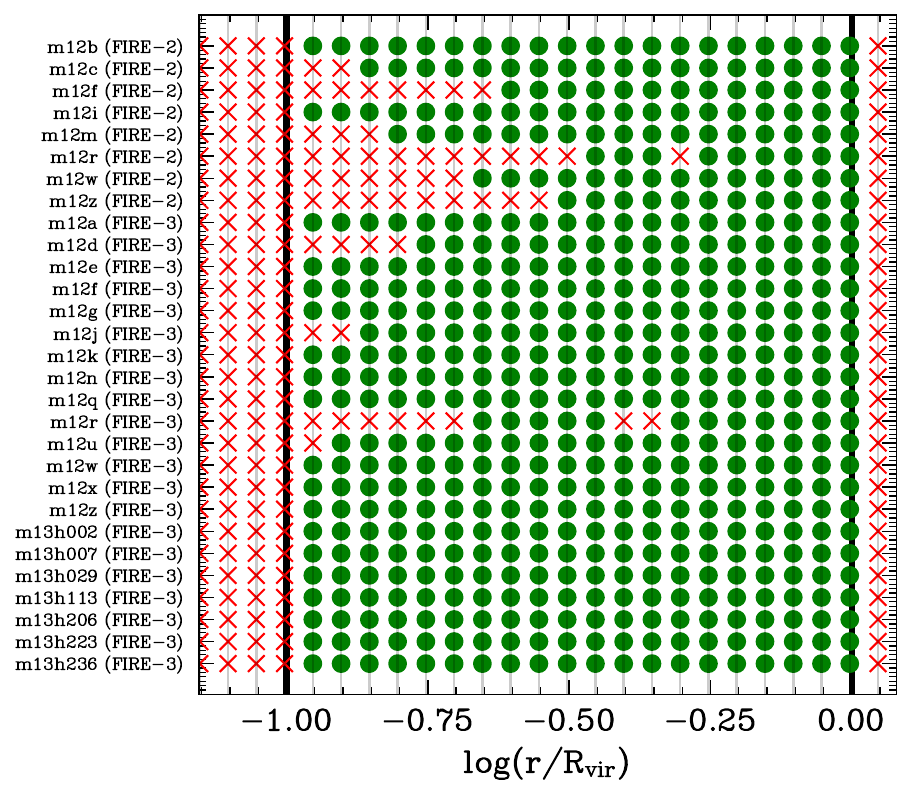}
    \caption{Fitting regions for all FIRE simulations in our analysis set.
    Each point marks the center of a radial shell centered on the halo.
    The green points indicate shells in the CGM ($0.1 R_{\mathrm{vir}} <r<R_{\mathrm{vir}}$); the thick vertical lines in the plot) that have $\ge 50$\% of their total gas mass in the hot phase.
    Shells that do not meet these criteria are indicated by the red crosses.
    The green points define our fitting region in which we fit the cooling flow model to FIRE density and temperature profiles, and compute various average quantities.
    }
    \label{fig:fittingregions}
\end{figure}

Figure \ref{fig:fittingregions} shows the radial shells in the CGM dominated by the hot phase (green points) for all simulations in our analysis set.
The green points indicate the radial bins over which we perform our analyses in this work--- this is our `fitting region' in which we fit the cooling flow model, and we compute average quantities (i.e. metallicity, mass inflow rate, and turbulent pressure fraction) using these radial bins.
The red crosses indicate shells that are either outside $0.1 R_{\mathrm{vir}} <r<R_{\mathrm{vir}}$ or do not contain a significant hot phase, and we exclude these shells from our analysis.
For nearly all of the simulations in our set, our shell removal procedure has the effect of either keeping the fitting region unchanged ($0.1 R_{\mathrm{vir}} <r<R_{\mathrm{vir}}$), or extending the inner boundary farther out.
For two simulations (FIRE-2 m12r and FIRE-3 m12r), there are intermediate radii for which the majority of the gas is not in the hot phase.

\section{Finding cooling flow solutions in FIRE}\label{app:CFintegration}
Given a FIRE simulation of a halo in a gravitational potential $\Phi(r)$, hot-phase gas metallicity $Z(r)$, and turbulent pressure fraction $\alpha(r)$, we find the corresponding solution for a cooling flow in a potential $\Phi(r)$, with a mass inflow rate $\dot{M}$ and radiative cooling set by the cooling function $\Lambda(T, n_{\mathrm{H}}, Z, z)$.
Note $\dot{M}$ is the only free parameter of the cooling flow model, with all other parameters set by the simulation.

We numerically integrate logarithmic forms of the fluid equations for a cooling flow with angular momentum (Equations \ref{eq:conmass}-\ref{eq:conK}) using the cooling flow package of \cite{sternCoolingFlowSolutions2019, sternMaximumAccretionRate2020}\footnote{\url{https://sites.northwestern.edu/jonathanstern/the-cooling_flow-package/}}.
We integrate subsonic flows from the circularization radius $R_{\mathrm{circ}}$ (at which the radial inflow stalls due to angular momentum support) out to a maximum radius $R_{\mathrm{max}}$, where we choose $R_{\mathrm{circ}}=0.05 R_{\mathrm{vir}}$ and $R_{\mathrm{max}}=1.5 R_{\mathrm{vir}}$.
To find the solution for a cooling flow with $\dot{M}$, the code uses a shooting method: the boundary condition $T(R_{\mathrm{circ}})$ is iteratively varied and a flow is integrated out to $R_{\mathrm{max}}$ until the integrated solutions converge to a single bound solution (see Appendix A of \citealt{sternCoolingFlowSolutions2019}).
We allow the boundary condition to vary within $10^4 < T(R_{\mathrm{circ}})/\mathrm{K} < 10^7$, which includes the inner CGM temperatures of the halo mass range we study.

We calculate the halo potential and circular velocity for our simulations as described below.
The two quantities are related by $\dv{\Phi}{r} = -\frac{v_c^2}{r}$ and are determined by the total enclosed mass $M(<r)$ in the halo; $M(<r)$ includes dark matter, gas, and stars. 
We compute $M(<r)$ in logarithmically spaced radial shells $r_j$, and calculate $v_c(r_j)$ and $\Phi(r_j)$ in each bin.
We apply a Savitzky-Golay filter to smooth our numerical result for $v_c(r_j)$.
We then use linear interpolation to find $v_c(r)$ and $\Phi(r)$.

To find $Z(r)$ and $\alpha(r) \equiv P_{\mathrm{turb}}(r)/P(r)$ in the simulation, we start by dividing the volume $0.04 R_{\mathrm{vir}} <r<1.6 R_{\mathrm{vir}}$ into logarithmically spaced radial shells $r_j$.
In each shell we calculate $Z(r_j)$ and $\alpha(r_j) = P_{\mathrm{turb}}(r_j)/P(r_j)$, where $Z(r_j)$ is the mass-weighted average metallicity, $P(r_j)$ is the mass-weighted average thermal pressure, and $P_{\mathrm{turb}}(r_j)$ is the turbulent pressure ($P_{\mathrm{turb}}=\rho \sigma^2$; see Section \ref{sec:Nonthermalpressure}).
Note we only consider particles that belong to the hot gas phase when calculating the three quantities.

We then interpolate $\log Z(r_j)$ and $\alpha(r_j)$ using a cubic smoothing spline, which smooths small-scale noise in the radial profiles (using \texttt{scipy.interpolate.UnivariateSpline} with the default cubic smoothing condition).
For our interpolation, we only consider points $r_j$ that are within the fitting region identified for each simulation (represented by the green points in Figure \ref{fig:fittingregions}; see Appendix \ref{app:virialbranch}).
For points in our interpolation range $0.04 R_{\mathrm{vir}} <r<1.6 R_{\mathrm{vir}}$ that fall outside of the two outermost endpoints of the fitting region, we fix $Z(r_j)$ and $\alpha(r_j)$ to their value at the nearest endpoint.

\section{Effects of turbulence and assumed angular momentum in cooling flows}\label{app:CFturbcomparison}
In this paper we present the results of cooling flows that include angular momentum and turbulence.
This is because in the FIRE halos we analyze, we find a significant contribution to the pressure support in the hot halos is in the form of turbulence (see Section \ref{sec:Nonthermalpressure}).

\begin{figure*}
    \includegraphics[width=6in]{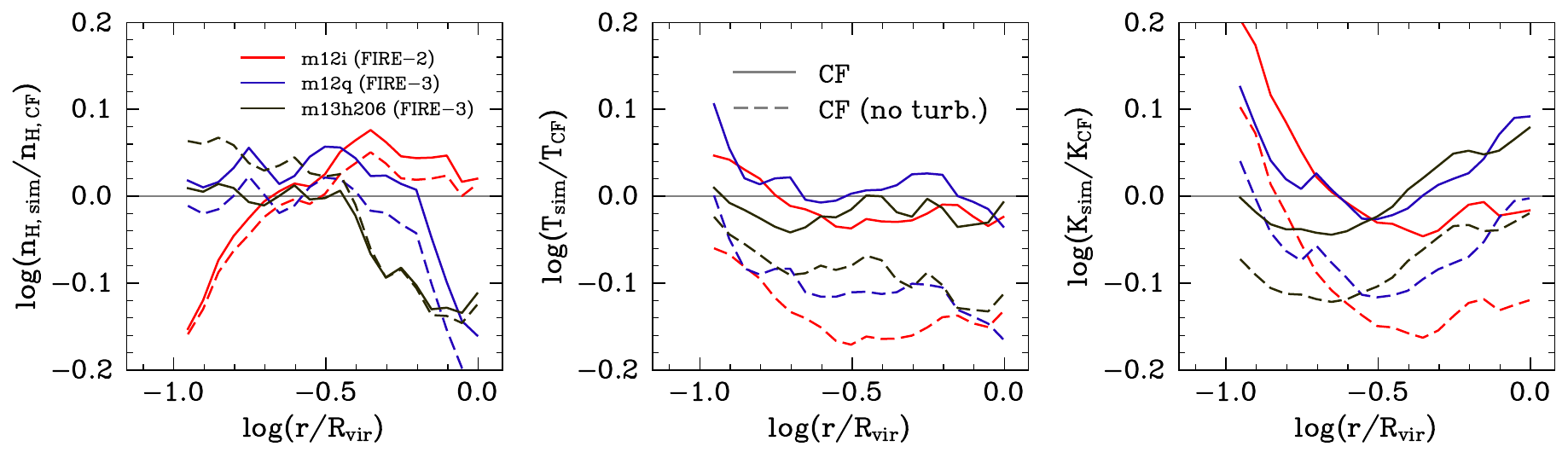}
    \caption{Comparison of how well cooling flows with and without turbulence fit three FIRE simulations.
    Ratios of spherically averaged FIRE profiles with respect to profiles of the best-fit cooling flow models are shown.
    Hydrogen number density, temperature, and entropy ratios are shown in rows one to three, respectively.
    As in Figure \ref{fig:CFratio_turb}, the cooling flow models were fit to the hot-phase density and temperature profiles in the simulations.
    Ratios are plotted in the radial range over which we fit the cooling flow model.
    Adding non-thermal pressure to the cooling flow model increases the predicted temperature by $\sim 10-30$\%, improving the fit of the model to the simulation.
    }
    \label{fig:CFratio_turbcomparison}
\end{figure*}

In Figure \ref{fig:CFratio_turbcomparison}, we examine the effect of turbulence in the cooling flow model.
Ratios of spherically averaged FIRE profiles are shown with respect to the predictions of the cooling flow models with and without turbulence that best fit the simulations.
The best-fit cooling flow model without turbulence, represented by the dashed lines, is given by Equations \ref{eq:conmass} to \ref{eq:conK}.
The best-fit model with turbulence is shown as the solid lines, which are identical to the ratios shown in Figure \ref{fig:CFratio_turb}.
We focus on three FIRE halos.
Adding non-thermal pressure to the cooling flow model increases the predicted temperature by $\sim 10-30$\%, improving the fit of the model to the simulation.
Adding turbulence has a bigger impact on the predicted temperature for the two m12 halos than the m13 halo.
This is consistent with the higher average turbulent pressure fractions we measured for FIRE-2 m12i and FIRE-3 m12q compared to FIRE-3 m13h206, of $\alpha=$35\%, 30\%, and 14\%, respectively (see Figure \ref{fig:Pturb_ratio}).

In our simulation set containing m12 and m13 halos, adding turbulence to the cooling flows has the general effect of increasing the predicted temperature by $\sim 20$\%, improving the agreement of the cooling flow for most halos.
There is not a similar clear effect for density, which does not consistency increase or decrease as a result of including non-thermal pressure in the cooling flow.

\begin{figure*}
    \includegraphics[width=6in]{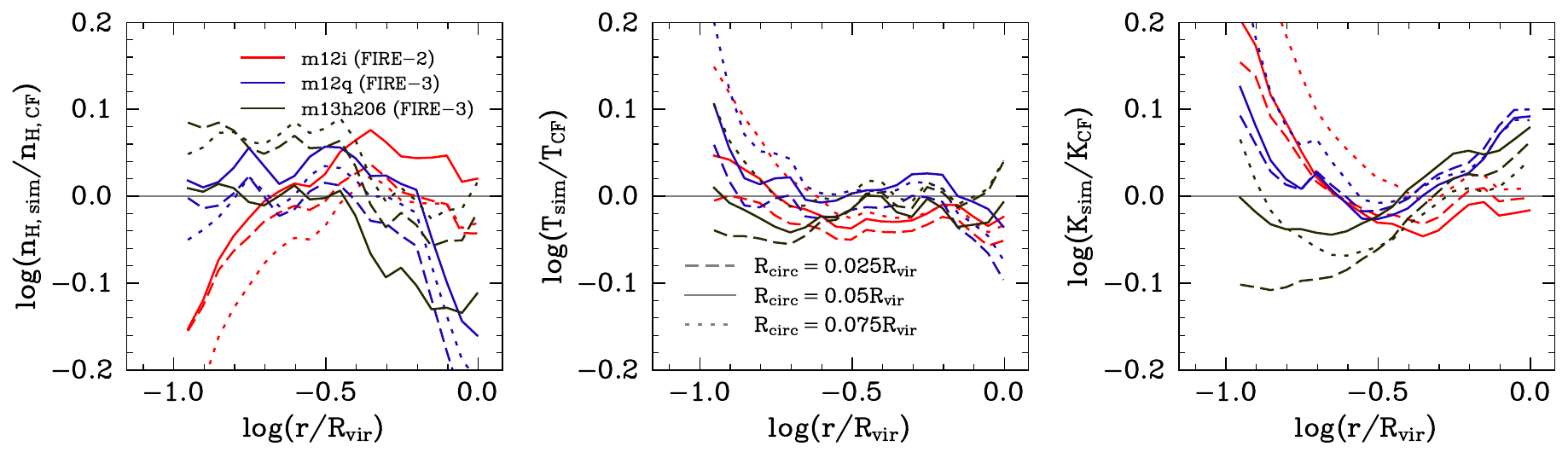}
    \caption{
    Comparison of how well cooling flows with different angular momentum assumptions fit three FIRE simulations.
    This figure is similar to Figure \ref{fig:CFratio_turbcomparison}.
    Results for cooling flows with turbulence that assume three different $R_{\mathrm{circ}}$ values are shown.
    Adding angular momentum in the cooling flow, i.e. increasing the circularization radius, decreases the predicted temperature for the three simulations.
    }
    \label{fig:CFratio_rotcomparison}
\end{figure*}

We also consider the effect of the angular momentum we assume in the cooling flows.
For the results shown in this paper, we assumed the radial inflow stalls due to angular momentum support at a circularization radius of $R_{\mathrm{circ}}=0.05 R_{\mathrm{vir}}$.
In Figure \ref{fig:CFratio_rotcomparison}, we test how sensitive our results are to this assumption by varying $R_{\mathrm{circ}}$ from $0.025 R_{\mathrm{vir}}$ to $0.075 R_{\mathrm{vir}}$.

We show results for three FIRE simulations.
Adding angular momentum in the cooling flow, i.e. increasing the circularization radius, has the effect of decreasing the predicted temperature for the three simulations.
There is not a systematic effect in predicted density when we vary $R_{\mathrm{circ}}$.

Although varying the circularization radius affects the profiles predicted by the cooling flows (especially in the inner halo), there is not a single choice of the radius which consistently produces better fits to the FIRE simulations in our analysis.
This indicates that the amount of angular momentum may vary from halo to halo.

\section{Calculating spherically averaged quantities}\label{app:averaging}
We calculate radial profiles of various thermodynamic quantities in this work.
As above, we construct radial bins whose centers $r/R_{\mathrm{vir}}$ are placed equidistant in log space, with $\Delta \log (r/R_{\mathrm{vir}})=0.05$.
For a radial shell with gas particles $i$ belonging to the hot virialized phase, and volume $V_{\mathrm{shell}}=\sum_i V_i$, we define the volume-weighted average of quantity $f$ as
\begin{equation}
    \langle f(r) \rangle_V = \frac{\sum_i f_i V_i}{V_{\mathrm{shell}}}.
\end{equation}
Similarly, the mass-weighted average of $f$ in each radial bin is
\begin{equation}
    \langle f(r) \rangle_M = \frac{\sum_i f_i m_i}{M_{\mathrm{shell}}},
\end{equation}
where $M_{\mathrm{shell}}=\sum_i m_i$ is the sum of all hot-phase gas particles in the shell.
The volume of particle $i$ is defined as $V_i=m_i/\rho_i$, where $m_i$ and $\rho_i$ are the mass and density of the particle, respectively.

Throughout this paper, we show volume-weighted spherically averaged profiles for hydrogen number density $n_{\mathrm{H}}$, temperature $T$, Mach number $\mathcal{M}$, entropy $K$, thermal pressure $P_{\mathrm{th}}$, and magnetic pressure $P_{\mathrm{mag}}$. For gas particle $i$, we calculate
\begin{align}
    n_{\mathrm{H},i} &= \frac{X_i \rho_i}{m_p}\\
    T_i &= \mu_i m_p (\gamma_{\mathrm{th}}-1) u_i/k_B\\
    \mathcal{M}_i &= \frac{-v_{\mathrm{rad},i}}{c_{s,i}}\\
    c_{s,i} &= \sqrt{\gamma_{\mathrm{th}}(\gamma_{\mathrm{th}}-1)u_i}\\
    P_{\mathrm{th},i} &= (\gamma_{\mathrm{th}}-1) u_i \rho_i\\
    K_i &= \frac{k_B T_i}{n^{2/3}}\\
    n &= \frac{\rho_i}{\mu_i m_p}
\end{align}
where $X$ is the hydrogen mass fraction, $\mu$ is the mean molecular weight in units of $m_p$, $\rho$ is the gas density, $m_p$ is the mass of a proton, $\gamma_{\mathrm{th}}=5/3$ is the adiabatic index, and $u=u_{\mathrm{th}}$ is the specific internal energy.

We calculate the cooling time of a gas particle in our simulation as $t_{\mathrm{cool}}=\frac{U}{n_{\mathrm{H}}^2 \Lambda}$ (Equation \ref{eq:tcool}), where $U$ is the energy density.
For the cooling function $\Lambda(T, n_{\mathrm{H}}, Z,z)$, we interpolate the tables of \cite{wiersmaEffectPhotoionizationCooling2009}, consistent with our cooling function used to integrate the cooling flow model.
We calculate the cooling function for 1000 metallicities spanning the range of metallicities of all gas particles in the simulation for computational efficiency.

We calculate the spherically averaged cooling time in a radial shell as
\begin{equation}
    \left< t_{\mathrm{cool}} \right> =  \frac{\sum_i (u_i m_i)}{ \sum_i \left[ n_{\mathrm{H},i}^2 \Lambda(T_i, n_{\mathrm{H},i}, Z_j,z) V_i \right]},
\end{equation}
where the numerator is the total internal energy of the shell, and the denominator is the luminosity (i.e., the total cooling rate) of the shell.
The sum is over all particles in the shell that belong to the hot phase.

Finally, we calculate the mass flow rate in each logarithmically spaced radial shell $r_j$, $\dot{M}(r_j)$. 
In each radial shell of width $\Delta r$, $\dot{M}(r_j)=\frac{1}{\Delta r} \sum_i v_i m_i$, where the sum is over all hot-phase gas particles in the bin; $v_i$ is the radial velocity of a particle, with $v_i>0$ for an inflow.
We find the mass inflow rate of all gas, $\dot{M}_{\mathrm{in,\ tot}}(r_j)$, by summing over all gas particles in the shell; we also find the mass inflow rate of hot-phase gas, $\dot{M}_{\mathrm{in,\ hot}}(r_j)$, by summing over only particles that belong to the hot phase in the shell.
Additionally, we define the rate of cooling of the hot-phase gas by $\dot{M}_{\mathrm{cool, hot}}(r_j)= \frac{M_{\mathrm{shell, hot}}}{ \left< t_{\mathrm{cool}} \right>}\frac{r_j}{\Delta r}$, where $M_{\mathrm{shell, hot}}=\sum_{i \in \mathrm{hot\ phase}} m_i$ is the mass of hot-phase gas in the shell.

For the results shown in Figures \ref{fig:Mdotratio} and \ref{fig:Mdot_scatter}, we calculated average values for $\dot{M}_{\mathrm{in,\ tot}}(r_j)$, $\dot{M}_{\mathrm{in,\ hot}}(r_j)$, and $\dot{M}_{\mathrm{cool, hot}}(r_j)$ by calculating the mass-weighted average rate of radial shells in the CGM with a significant hot phase ($\ge 50$\% of the total gas in the shell is found in the hot phase).
For a given radial range $[r_0,r_1]$, the average mass inflow rate $\dot{M}_i = \frac{\sum_{r_0 < r_j < r_1} \dot{M}_i(r_j) M_{\mathrm{shell, hot}}\
}{\sum_{r_0 < r_j < r_1} M_{\mathrm{shell, hot}}}$, where the sum is over shells with $M_{\mathrm{shell, hot}} \ge 0.5 M_{\mathrm{shell, tot}}$.
Unless specified, the radial range of our averaging is $[0.1 R_{\mathrm{vir}}, R_{\mathrm{vir}}]$.

\section{Analytic derivation of bolometric surface brightness and thermal SZ profiles}\label{app:analytic_obs}
In this appendix, we show how the cooling flow model can be used to analytically to predict bolometric surface brightness and thermal SZ profiles. 

\subsection{Surface brightness profiles}
In a cooling flow, we can exploit the fact that the total energy radiated away is provided by the loss of gravitational potential energy during the inflow to predict the surface brightness profile. 
In steady-state, energy conservation implies that the  mass inflow rate 
\begin{equation}
    \dot{M} \approx \frac{4 \pi r^2 n_{\mathrm{H}}^2 \Lambda}{\dd{\Phi}/\dd{r}}
\end{equation}
\citep[e.g.,][]{sternMaximumAccretionRate2020}. 
The power radiated per unit volume, $n_{\mathrm{H}}^2 \Lambda$, is then 
\begin{equation}
    n_{\mathrm{H}}^2 \Lambda \approx \dv{\Phi}{r} \frac{\dot{M}}{4 \pi r^2} = \frac{\dot{M} v_c(r)^2}{4 \pi r^3},
\end{equation}
where we used $\dd{\Phi}/\dd{r} = v_c(r)^2/r$.

The bolometric surface brightness $S_{\mathrm{bol}}$ at an impact parameter $b$ can be calculated by integrating $n_{\mathrm{H}}^2 \Lambda$ along a sightline going through the halo:
\begin{equation}
    S_{\mathrm{bol}}(b) = \int_{-\infty}^{\infty} n_{\mathrm{H}}^2 \Lambda \dd{s} \approx \int_{-\infty}^{\infty} \frac{\dot{M} v_c(r)^2}{4 \pi r^3} \dd{s},
\end{equation}
where $S_{\mathrm{bol}}$ has units of power radiated per unit area. 
Here, $s$ is the distance along the sightline and is related to the impact parameter and the distance from the halo center $r$ via $r = \sqrt{s^2 + b^2}$. 
Thus,
\begin{equation}
    S_{\mathrm{bol}}(b) \approx \frac{\dot{M}}{4 \pi} \int_{-\infty}^{\infty} \frac{v_c(\sqrt{s^2 + b^2})^2}{(s^2 + b^2)^{\frac{3}{2}}} \dd{s},
\end{equation}
where we have moved the constant mass inflow rate outside the integral.

As in \cite{sternCoolingFlowSolutions2019}, 
we write the circular velocity as a power law: 
\begin{equation}\label{eq:vc}
    v_c(r) = v_c(R_{\mathrm{vir}}) \qty(\frac{r}{R_{\mathrm{vir}}})^m.
\end{equation}
The surface brightness profile is then
\begin{align}
    S_{\mathrm{bol}}(b) &\approx \frac{\dot{M} v_c(R_{\mathrm{vir}})^2}{4 \pi R_{\mathrm{vir}}^{2m}} \int_{-\infty}^{\infty} (s^2+b^2)^{m-3/2} \dd{s}\\
        &= \frac{\dot{M} v_c(R_{\mathrm{vir}})^2}{4 \pi R_{\mathrm{vir}}^{2m}} \sqrt{\pi} \frac{\Gamma(1-m)}{\Gamma(\frac{3}{2}-m)} b^{2m-2},\label{eq:Sb}
\end{align}
where $\Gamma$ is the gamma function defined by $\Gamma(z) = \int_0^\infty t^{z-1} e^{-t} \dd{t}$ for all $\Re(z)>0$, and Equation \ref{eq:Sb} assumes $m<1$.
For the case of a constant circular velocity profile (i.e. $m=0$), Equation \ref{eq:Sb} simplifies to
\begin{align}
        S_{\mathrm{bol}}(b) &\approx \frac{\dot{M} v_c(R_{\mathrm{vir}})^2}{2 \pi} b^{-2}\\
            &= \SI{2e37}{\mathrm{\frac{erg}{\ s\ kpc^2}}} \qty(\frac{\dot{M}}{20\ \Msun/\mathrm{yr}}) \qty(\frac{v_c(R_{\mathrm{vir}})}{300\ \mathrm{km/s}})^2 \qty(\frac{b}{100\ \mathrm{kpc}})^{-2},
\end{align}
where, in the second equality, we use representative values of $\dot{M}$ and $v_c(R_{\rm vir})$ for $\sim 10^{13}\ \Msun$ halos.

For the FIRE simulations, we calculate approximate values for the logarithmic slope using $m \equiv \dd{(\ln v_c)} / \dd{(\ln r)} \approx \frac{\ln [v_c(1R_{\mathrm{vir}})] - \ln [0.1v_c(R_{\mathrm{vir}})]}{\ln (R_{\mathrm{vir}}) - \ln (0.1R_{\mathrm{vir}})}$.
We measure a median value of $m \approx -0.1$ for both our set of m12 and our set of m13 simulations.
We therefore find the cooling flow prediction of $S_{\mathrm{bol}} \propto b^{-2.2}$ for both halo mass scales.

Since $S_{\mathrm{bol}} \propto \dot{M}$, we emphasize that the normalization of the surface brightness can be somewhat arbitrarily varied depending on the mass flow rate of the cooling flow.\footnote{However, the mass flow rate is constrained by $\dot{M} \lesssim f_b M_{\mathrm{halo}}/t_{\mathrm{H}}$, where $f_b = \Omega_b / \Omega_m$ is the cosmic baryon fraction.} 
The main result here is that the slope of the bolometric surface brightness profile is related to the slope of the gravitational potential.

Although this result is beautifully simple, care must be taken when comparing to observations in specific bands, especially for $\sim 10^{12}\ \Msun$ halos. 
This is because cooling flows in $\sim 10^{12}\ \Msun$ halos radiate their energy in both UV and X-ray bands, which are generally not captured by the same instruments (see the temperature profiles in Figure \ref{fig:CFprofiles_rep}, which show that $10^5\ \mathrm{K} \lesssim T \lesssim 10^6$ K in the outer parts).
\cite{hopkinsCosmicRaysMasquerading2025} roughly predict analytically the X-ray emission in the 0.5-2 keV band detected by eROSITA, which is predominantly metal-line emission, to have a profile $S_X \propto b^{-4.5}$.
Even steeper 
soft X-ray surface brightness profiles were found in an earlier analysis of FIRE simulations by \cite{vandevoortImpactStellarFeedback2016} (see their Appendix A).
\cite{janaXraySignaturesGalactic2024} also predicted X-ray surface brightness profiles at 0.5-2 keV for rotating cooling flows in halos around Milky Way-type galaxies, but using more idealized simulations that neglected small-scale galactic processes such as stellar feedback.
They found a shallow slope at small impact parameters $b \lesssim 50$ kpc that is consistent with stacked results from eROSITA observations.
However, in their idealized simulations, the steeper slopes predicted at $b \gtrsim 50$ kpc are also strongly discrepant with the X-ray observations discussed in Section \ref{sec:xrays}.

\subsection{Thermal SZ profiles}
We can also predict the thermal Sunyaev–Zeldovich (tSZ) signal for a cooling flow.
The SZ effect is the shift in the temperature of the CMB due to the inverse Compton scattering of CMB photons by free electrons in the halo gas \citep{sunyaevSmallScaleFluctuationsRelic1970, sunyaevObservationsRelicRadiation1972}.
The tSZ signal is given by the Compton $y$ parameter, found by integrating the thermal pressure of electrons $P_e = n_e k_B T$ along a sightline through the halo at an impact parameter $b$,
\begin{equation}\label{eq:y}
    y(b) = \frac{\sigma_T}{m_e c^2} \int_{-\infty}^{\infty} n_e k_B T \dd{s} \approx
    \frac{1.2 \sigma_T}{m_e c^2} \int_{-\infty}^{\infty} n_{\mathrm{H}} k_B T \dd{s}
\end{equation}
where $\sigma_T$ is the Thomson cross section, $m_e$ is the electron mass, $c$ is the speed of light in a vacuum, $n_e$ is the number density of free electrons, and in the second equality, we assumed $n_e \approx 1.2 n_{\mathrm{H}}$ for a fully ionized cosmic plasma. 

We can express $n_{\mathrm{H}}$ and $T$ as power laws by again assuming the circular velocity is a power law given by $\dd{(\ln v_c)} / \dd{(\ln r)} = m$ (Equation \ref{eq:vc}).
As shown by \cite{fabianCoolingFlowsClusters1984}, in a subsonic cooling flow, $v_c^2/c_s^2$ and $t_{\mathrm{cool}}/t_{\mathrm{flow}}$ are approximately constant ratios of order unity, where $t_{\mathrm{flow}}=r/v$ is the crossing time.
The temperature of the cooling flow is given by a critical solution for which $T$ is finite as $r \rightarrow \infty$ and the ratio $v_c^2/c_s^2$ is a constant (also note the flow is approximately under hydrostatic equilibrium, so $v_c^2 \sim c_s^2$; see Section \ref{sec:CFmodel_noturb}).
This implies that 
\begin{equation}
    T \propto v_c^2 \propto r^{2m}\label{eq:powerT}.
\end{equation}

Additionally, to maintain a steady inflow, the cooling time ($\propto T/\Lambda n_{\rm H}$) must approximately be equal to the crossing time, which results in
\begin{equation}
    \dv{\ln n_{\mathrm{H}}}{\ln r} = \dv{\ln T}{\ln r} + \dv{\ln v}{\ln r} - \dv{\ln \Lambda}{\ln r} - 1.
\end{equation}
Substituting $\dd{(\ln T)} / \dd{(\ln r)}=2m$ from Equation \ref{eq:powerT}, and $\dd{(\ln v)} / \dd{(\ln r)} = -2 - \dd{(\ln n_{\mathrm{H}})} / \dd{(\ln r)}$, which follows from the conservation of mass (Equation \ref{eq:conmass}), yields
\begin{equation}
    n_{\mathrm{H}} \propto r^{m - \frac{3}{2} - \frac{1}{2}\dv{\ln \Lambda}{\ln r}}\label{eq:powern}.
\end{equation}
Equations \ref{eq:powerT} and \ref{eq:powern} are similar to the self-similar cooling flow solution derived in \S 2.3 of \cite{sternCoolingFlowSolutions2019}; however, while \cite{sternCoolingFlowSolutions2019} assumed a constant cooling function, here we allow $\Lambda$ to depend on $r$.

Integrating Equation \ref{eq:y} using Equations \ref{eq:powerT} and \ref{eq:powern} and assuming $\dv{\ln \Lambda}{\ln r}={\rm const.}$, we find
\begin{equation}
    y(b) \propto b^{3m - \frac{1}{2} - \frac{1}{2}\dv{\ln \Lambda}{\ln r}},
\end{equation}
which is valid for $3m - \frac{1}{2} - \frac{1}{2}\dv{\ln \Lambda}{\ln r} < 0$.
We use the approximation for $\Lambda$ in the metal-dominated regime given by Equation 11 of \cite{sternMaximumAccretionRate2020}\footnote{\cite{sternMaximumAccretionRate2020} note that this approximation is valid for $T \sim 10^5 - 10^7$ K and $Z \gtrsim 0.3\  \mathrm{Z_{\odot}}$. However, the FIRE metallicity profiles of the hot-phase CGM gas we measure are typically lower, in the range $Z \sim 0.05 - 0.3\ \mathrm{Z_\odot}$. While this approximation remains useful, applying it in our context requires some caution.},
\begin{equation}
    \Lambda = \SI{0.5e-22}{erg\ cm^3\ s^{-1}} \left( \frac{T}{10^6\ \mathrm{K}} \right)^{-0.7} \left( \frac{Z}{0.3 \mathrm{Z}_\odot} \right)^{0.9}.
\end{equation}
We can then write
\begin{equation}
    \dv{\ln \Lambda}{\ln r} = -0.7 \dv{\ln T}{\ln r} + 0.9 \dv{\ln Z}{\ln r} = -1.4m + 0.9 \dv{\ln Z}{\ln r},
\end{equation}
where, to obtain the second equality, we used Equation \ref{eq:powerT}.

We measure the slope of the metallicity profile in the hot CGM, and find a median result of $\dd{(\ln Z)}/\dd{(\ln r)} \approx -0.3$ for FIRE-2 $\sim 10^{12}\ \Msun$ halos (see Appendix \ref{app:CFintegration} for a description of our method for finding $Z(r)$ for a given halo).
We calculate the median metallicity slope for FIRE-2 $\sim 10^{12}\ \Msun$ halos because we compare our prediction of $y$ to tSZ observations of halos of similar mass (see Section \ref{sec:tSZdisc}), and because FIRE-2 better reproduces Ne VIII absorption observations compared to FIRE-3 for this halo mass scale (see Section \ref{sec:NeVIII} and \citealt{wijersNeVIIIWarmhot2024}).

Finally, using $m = -0.1$, approximately the median value of $\dd{(\ln v_c)} / \dd{(\ln r)}$ that we measure in the CGM over our set of FIRE simulations, we find $y \propto b^{-0.74}$.

\bsp	%
\label{lastpage}

\begin{thebibliography}{}
\makeatletter
\relax
\def\mn@urlcharsother{\let\do\@makeother \do\$\do\&\do\#\do\^\do\_\do\%\do\~}
\def\mn@doi{\begingroup\mn@urlcharsother \@ifnextchar [ {\mn@doi@} {\mn@doi@[]}}
\def\mn@doi@[#1]#2{\def\@tempa{#1}\ifx\@tempa\@empty \href {http://dx.doi.org/#2} {doi:#2}\else \href {http://dx.doi.org/#2} {#1}\fi \endgroup}
\def\mn@eprint#1#2{\mn@eprint@#1:#2::\@nil}
\def\mn@eprint@arXiv#1{\href {http://arxiv.org/abs/#1} {{\tt arXiv:#1}}}
\def\mn@eprint@dblp#1{\href {http://dblp.uni-trier.de/rec/bibtex/#1.xml} {dblp:#1}}
\def\mn@eprint@#1:#2:#3:#4\@nil{\def\@tempa {#1}\def\@tempb {#2}\def\@tempc {#3}\ifx \@tempc \@empty \let \@tempc \@tempb \let \@tempb \@tempa \fi \ifx \@tempb \@empty \def\@tempb {arXiv}\fi \@ifundefined {mn@eprint@\@tempb}{\@tempb:\@tempc}{\expandafter \expandafter \csname mn@eprint@\@tempb\endcsname \expandafter{\@tempc}}}

\bibitem[\protect\citeauthoryear{Abazajian et~al.,}{Abazajian et~al.}{2016}]{abazajianCMBS4ScienceBook2016}
Abazajian K.~N.,  et~al., 2016, {{CMB-S4 Science Book}}, {{First Edition}}, \mn@doi{10.48550/arXiv.1610.02743}

\bibitem[\protect\citeauthoryear{Ade et~al.,}{Ade et~al.}{2019}]{adeSimonsObservatoryScience2019}
Ade P.,  et~al., 2019, \mn@doi [Journal of Cosmology and Astroparticle Physics] {10.1088/1475-7516/2019/02/056}, 2019, 056

\bibitem[\protect\citeauthoryear{Aghanim et~al.,}{Aghanim et~al.}{2020}]{aghanimPlanck2018Results2020}
Aghanim N.,  et~al., 2020, \mn@doi [Astronomy \& Astrophysics] {10.1051/0004-6361/201833910}, 641, A6

\bibitem[\protect\citeauthoryear{{Angl{\'e}s-Alc{\'a}zar}, {Faucher-Gigu{\`e}re}, Kere{\v s}, Hopkins, Quataert  \& Murray}{{Angl{\'e}s-Alc{\'a}zar} et~al.}{2017}]{angles-alcazarCosmicBaryonCycle2017}
{Angl{\'e}s-Alc{\'a}zar} D.,  {Faucher-Gigu{\`e}re} C.-A.,  Kere{\v s} D.,  Hopkins P.~F.,  Quataert E.,   Murray N.,  2017, \mn@doi [Monthly Notices of the Royal Astronomical Society] {10.1093/mnras/stx1517}, 470, 4698

\bibitem[\protect\citeauthoryear{Asplund, Grevesse, Sauval  \& Scott}{Asplund et~al.}{2009}]{asplundChemicalCompositionSun2009}
Asplund M.,  Grevesse N.,  Sauval A.~J.,   Scott P.,  2009, \mn@doi [Annual Review of Astronomy and Astrophysics] {10.1146/annurev.astro.46.060407.145222}, 47, 481

\bibitem[\protect\citeauthoryear{Babyk, McNamara, Nulsen, Russell, Vantyghem, Hogan  \& Pulido}{Babyk et~al.}{2018}]{babykUniversalEntropyProfile2018}
Babyk {\relax Iu}.~V.,  McNamara B.~R.,  Nulsen P. E.~J.,  Russell H.~R.,  Vantyghem A.~N.,  Hogan M.~T.,   Pulido F.~A.,  2018, \mn@doi [The Astrophysical Journal] {10.3847/1538-4357/aacce5}, 862, 39

\bibitem[\protect\citeauthoryear{Balbus \& Soker}{Balbus \& Soker}{1989}]{balbusTheoryLocalThermal1989}
Balbus S.~A.,  Soker N.,  1989, \mn@doi [The Astrophysical Journal] {10.1086/167521}, 341, 611

\bibitem[\protect\citeauthoryear{Barcons et~al.,}{Barcons et~al.}{2017}]{barconsAthenaESAsXray2017}
Barcons X.,  et~al., 2017, \mn@doi [Astronomische Nachrichten] {10.1002/asna.201713323}, 338, 153

\bibitem[\protect\citeauthoryear{Bassini, Feldmann, Gensior, {Faucher-Gigu{\`e}re}, Cenci, Moreno, Bernardini  \& Liang}{Bassini et~al.}{2024}]{bassiniInflowOutflowProperties2024}
Bassini L.,  Feldmann R.,  Gensior J.,  {Faucher-Gigu{\`e}re} C.-A.,  Cenci E.,  Moreno J.,  Bernardini M.,   Liang L.,  2024, \mn@doi [Monthly Notices of the Royal Astronomical Society] {10.1093/mnrasl/slae036}, 532, L14

\bibitem[\protect\citeauthoryear{Birnboim \& Dekel}{Birnboim \& Dekel}{2003}]{birnboimVirialShocksGalactic2003}
Birnboim Y.,  Dekel A.,  2003, \mn@doi [Monthly Notices of the Royal Astronomical Society] {10.1046/j.1365-8711.2003.06955.x}, 345, 349

\bibitem[\protect\citeauthoryear{{Booth}, {Agertz}, {Kravtsov}  \& {Gnedin}}{{Booth} et~al.}{2013}]{2013ApJ...777L..16B}
{Booth} C.~M.,  {Agertz} O.,  {Kravtsov} A.~V.,   {Gnedin} N.~Y.,  2013, \mn@doi [\apjl] {10.1088/2041-8205/777/1/L16}, \href {https://ui.adsabs.harvard.edu/abs/2013ApJ...777L..16B} {777, L16}

\bibitem[\protect\citeauthoryear{Bregman, Anderson, Miller, {Hodges-Kluck}, Dai, Li, Li  \& Qu}{Bregman et~al.}{2018}]{bregmanExtendedDistributionBaryons2018}
Bregman J.~N.,  Anderson M.~E.,  Miller M.~J.,  {Hodges-Kluck} E.,  Dai X.,  Li J.-T.,  Li Y.,   Qu Z.,  2018, \mn@doi [The Astrophysical Journal] {10.3847/1538-4357/aacafe}, 862, 3

\bibitem[\protect\citeauthoryear{Bregman, {Hodges-Kluck}, Qu, Pratt, Li  \& Yun}{Bregman et~al.}{2022}]{bregmanHotExtendedGalaxy2022}
Bregman J.~N.,  {Hodges-Kluck} E.,  Qu Z.,  Pratt C.,  Li J.-T.,   Yun Y.,  2022, \mn@doi [The Astrophysical Journal] {10.3847/1538-4357/ac51de}, 928, 14

\bibitem[\protect\citeauthoryear{Bryan \& Norman}{Bryan \& Norman}{1998}]{bryanStatisticalPropertiesXRay1998}
Bryan G.~L.,  Norman M.~L.,  1998, \mn@doi [The Astrophysical Journal] {10.1086/305262}, 495, 80

\bibitem[\protect\citeauthoryear{Burchett et~al.,}{Burchett et~al.}{2019}]{burchettCOSAbsorptionSurvey2019}
Burchett J.~N.,  et~al., 2019, \mn@doi [The Astrophysical Journal Letters] {10.3847/2041-8213/ab1f7f}, 877, L20

\bibitem[\protect\citeauthoryear{{Butsky}, {Nakum}, {Ponnada}, {Hummels}, {Ji}  \& {Hopkins}}{{Butsky} et~al.}{2023}]{2023MNRAS.521.2477B}
{Butsky} I.~S.,  {Nakum} S.,  {Ponnada} S.~B.,  {Hummels} C.~B.,  {Ji} S.,   {Hopkins} P.~F.,  2023, \mn@doi [\mnras] {10.1093/mnras/stad671}, \href {https://ui.adsabs.harvard.edu/abs/2023MNRAS.521.2477B} {521, 2477}

\bibitem[\protect\citeauthoryear{Byrne, {Faucher-Gigu{\`e}re}, Stern, {Angl{\'e}s-Alc{\'a}zar}, Wellons, Gurvich  \& Hopkins}{Byrne et~al.}{2023}]{byrneStellarFeedbackregulatedBlack2023}
Byrne L.,  {Faucher-Gigu{\`e}re} C.-A.,  Stern J.,  {Angl{\'e}s-Alc{\'a}zar} D.,  Wellons S.,  Gurvich A.~B.,   Hopkins P.~F.,  2023, \mn@doi [Monthly Notices of the Royal Astronomical Society] {10.1093/mnras/stad171}, 520, 722

\bibitem[\protect\citeauthoryear{Byrne et~al.,}{Byrne et~al.}{2024}]{byrneEffectsMultichannelActive2024}
Byrne L.,  et~al., 2024, \mn@doi [The Astrophysical Journal] {10.3847/1538-4357/ad67ca}, 973, 149

\bibitem[\protect\citeauthoryear{{CHIME/FRB Collaboration} et~al.,}{{CHIME/FRB Collaboration} et~al.}{2018}]{chime/frbcollaborationCHIMEFastRadio2018}
{CHIME/FRB Collaboration} et~al., 2018, \mn@doi [The Astrophysical Journal] {10.3847/1538-4357/aad188}, 863, 48

\bibitem[\protect\citeauthoryear{Colbrook, Ma, Hopkins  \& Squire}{Colbrook et~al.}{2017}]{colbrookScalingLawsPassivescalar2017}
Colbrook M.~J.,  Ma X.,  Hopkins P.~F.,   Squire J.,  2017, \mn@doi [Monthly Notices of the Royal Astronomical Society] {10.1093/mnras/stx261}, 467, 2421

\bibitem[\protect\citeauthoryear{Cui et~al.,}{Cui et~al.}{2020}]{cuiHUBSHotUniverse2020}
Cui W.,  et~al., 2020, \mn@doi [Journal of Low Temperature Physics] {10.1007/s10909-019-02279-3}, 199, 502

\bibitem[\protect\citeauthoryear{Donahue \& Voit}{Donahue \& Voit}{2022}]{donahueBaryonCyclesBiggest2022}
Donahue M.,  Voit G.~M.,  2022, \mn@doi [Physics Reports] {10.1016/j.physrep.2022.04.005}, 973, 1

\bibitem[\protect\citeauthoryear{{El-Badry} et~al.,}{{El-Badry} et~al.}{2018}]{el-badryGasKinematicsMorphology2018}
{El-Badry} K.,  et~al., 2018, \mn@doi [Monthly Notices of the Royal Astronomical Society] {10.1093/mnras/stx2482}, 473, 1930

\bibitem[\protect\citeauthoryear{Escala et~al.,}{Escala et~al.}{2018}]{escalaModellingChemicalAbundance2018}
Escala I.,  et~al., 2018, \mn@doi [Monthly Notices of the Royal Astronomical Society] {10.1093/mnras/stx2858}, 474, 2194

\bibitem[\protect\citeauthoryear{Esmerian, Kravtsov, Hafen, {Faucher-Gigu{\`e}re}, Quataert, Stern, Kere{\v s}  \& Wetzel}{Esmerian et~al.}{2021}]{esmerianThermalInstabilityCGM2021}
Esmerian C.~J.,  Kravtsov A.~V.,  Hafen Z.,  {Faucher-Gigu{\`e}re} C.-A.,  Quataert E.,  Stern J.,  Kere{\v s} D.,   Wetzel A.,  2021, \mn@doi [Monthly Notices of the Royal Astronomical Society] {10.1093/mnras/stab1281}, 505, 1841

\bibitem[\protect\citeauthoryear{Fabian, Nulsen  \& Canizares}{Fabian et~al.}{1984}]{fabianCoolingFlowsClusters1984}
Fabian A.~C.,  Nulsen P. E.~J.,   Canizares C.~R.,  1984, \mn@doi [Nature] {10.1038/310733a0}, 310, 733

\bibitem[\protect\citeauthoryear{Fabian, Ferland, Sanders, McNamara, Pinto  \& Walker}{Fabian et~al.}{2022}]{fabianHiddenCoolingFlows2022}
Fabian A.~C.,  Ferland G.~J.,  Sanders J.~S.,  McNamara B.~R.,  Pinto C.,   Walker S.~A.,  2022, \mn@doi [Monthly Notices of the Royal Astronomical Society] {10.1093/mnras/stac2003}, 515, 3336

\bibitem[\protect\citeauthoryear{Fabian, Sanders, Ferland, McNamara, Pinto  \& Walker}{Fabian et~al.}{2023a}]{fabianHiddenCoolingFlows2023a}
Fabian A.~C.,  Sanders J.~S.,  Ferland G.~J.,  McNamara B.~R.,  Pinto C.,   Walker S.~A.,  2023a, \mn@doi [Monthly Notices of the Royal Astronomical Society] {10.1093/mnras/stad507}, 521, 1794

\bibitem[\protect\citeauthoryear{Fabian, Sanders, Ferland, McNamara, Pinto  \& Walker}{Fabian et~al.}{2023b}]{fabianHiddenCoolingFlows2023}
Fabian A.~C.,  Sanders J.~S.,  Ferland G.~J.,  McNamara B.~R.,  Pinto C.,   Walker S.~A.,  2023b, \mn@doi [Monthly Notices of the Royal Astronomical Society] {10.1093/mnras/stad1870}, 524, 716

\bibitem[\protect\citeauthoryear{Faerman, Sternberg  \& McKee}{Faerman et~al.}{2017}]{faermanMASSIVEWARMHOT2017}
Faerman Y.,  Sternberg A.,   McKee C.~F.,  2017, \mn@doi [The Astrophysical Journal] {10.3847/1538-4357/835/1/52}, 835, 52

\bibitem[\protect\citeauthoryear{Faerman, Sternberg  \& McKee}{Faerman et~al.}{2020}]{faermanMassiveWarmHot2020}
Faerman Y.,  Sternberg A.,   McKee C.~F.,  2020, \mn@doi [The Astrophysical Journal] {10.3847/1538-4357/ab7ffc}, 893, 82

\bibitem[\protect\citeauthoryear{Fang, Buote, Bullock  \& Ma}{Fang et~al.}{2015}]{fangXMMNEWTONSURVEYLOCAL2015}
Fang T.,  Buote D.,  Bullock J.,   Ma R.,  2015, \mn@doi [The Astrophysical Journal Supplement Series] {10.1088/0067-0049/217/2/21}, 217, 21

\bibitem[\protect\citeauthoryear{{Faucher-Gigu{\`e}re}}{{Faucher-Gigu{\`e}re}}{2020}]{faucher-giguereCosmicUVXray2020}
{Faucher-Gigu{\`e}re} C.-A.,  2020, \mn@doi [Monthly Notices of the Royal Astronomical Society] {10.1093/mnras/staa302}, 493, 1614

\bibitem[\protect\citeauthoryear{{Faucher-Gigu{\`e}re} \& Kere{\v s}}{{Faucher-Gigu{\`e}re} \& Kere{\v s}}{2011}]{faucher-giguereSmallCoveringFactor2011}
{Faucher-Gigu{\`e}re} C.-A.,  Kere{\v s} D.,  2011, \mn@doi [Monthly Notices of the Royal Astronomical Society] {10.1111/j.1745-3933.2011.01018.x}, 412, L118

\bibitem[\protect\citeauthoryear{{Faucher-Gigu{\`e}re} \& Oh}{{Faucher-Gigu{\`e}re} \& Oh}{2023}]{faucher-giguereKeyPhysicalProcesses2023}
{Faucher-Gigu{\`e}re} C.-A.,  Oh S.~P.,  2023, \mn@doi [Annual Review of Astronomy and Astrophysics] {10.1146/annurev-astro-052920-125203}, 61, 131

\bibitem[\protect\citeauthoryear{{Faucher-Gigu{\`e}re}, Lidz, Zaldarriaga  \& Hernquist}{{Faucher-Gigu{\`e}re} et~al.}{2009}]{faucher-giguereNewCalculationIonizing2009}
{Faucher-Gigu{\`e}re} C.-A.,  Lidz A.,  Zaldarriaga M.,   Hernquist L.,  2009, \mn@doi [The Astrophysical Journal] {10.1088/0004-637X/703/2/1416}, 703, 1416

\bibitem[\protect\citeauthoryear{Feldmann et~al.,}{Feldmann et~al.}{2023}]{feldmannFIREboxSimulatingGalaxies2023}
Feldmann R.,  et~al., 2023, \mn@doi [Monthly Notices of the Royal Astronomical Society] {10.1093/mnras/stad1205}, 522, 3831

\bibitem[\protect\citeauthoryear{{Garrison-Kimmel} et~al.,}{{Garrison-Kimmel} et~al.}{2019}]{garrison-kimmelLocalGroupFIRE2019}
{Garrison-Kimmel} S.,  et~al., 2019, \mn@doi [Monthly Notices of the Royal Astronomical Society] {10.1093/mnras/stz1317}, 487, 1380

\bibitem[\protect\citeauthoryear{Gupta, Mathur, Krongold, Nicastro  \& Galeazzi}{Gupta et~al.}{2012}]{guptaHUGERESERVOIRIONIZED2012}
Gupta A.,  Mathur S.,  Krongold Y.,  Nicastro F.,   Galeazzi M.,  2012, \mn@doi [The Astrophysical Journal Letters] {10.1088/2041-8205/756/1/L8}, 756, L8

\bibitem[\protect\citeauthoryear{Gurvich}{Gurvich}{2022}]{gurvichFIREStudioMovie2022}
Gurvich A.~B.,  2022, Astrophysics Source Code Library, p. ascl:2202.006

\bibitem[\protect\citeauthoryear{Hafen et~al.,}{Hafen et~al.}{2020}]{hafenFatesCircumgalacticMedium2020}
Hafen Z.,  et~al., 2020, \mn@doi [Monthly Notices of the Royal Astronomical Society] {10.1093/mnras/staa902}, 494, 3581

\bibitem[\protect\citeauthoryear{Hafen et~al.,}{Hafen et~al.}{2022}]{hafenHotmodeAccretionPhysics2022}
Hafen Z.,  et~al., 2022, \mn@doi [Monthly Notices of the Royal Astronomical Society] {10.1093/mnras/stac1603}, 514, 5056

\bibitem[\protect\citeauthoryear{{Heesen} et~al.,}{{Heesen} et~al.}{2023}]{2023A&A...670L..23H}
{Heesen} V.,  et~al., 2023, \mn@doi [\aap] {10.1051/0004-6361/202346008}, \href {https://ui.adsabs.harvard.edu/abs/2023A&A...670L..23H} {670, L23}

\bibitem[\protect\citeauthoryear{Henley, Shelton, Kwak, Joung  \& Low}{Henley et~al.}{2010}]{henleyORIGINHOTGAS2010}
Henley D.~B.,  Shelton R.~L.,  Kwak K.,  Joung M.~R.,   Low M.-M.~M.,  2010, \mn@doi [The Astrophysical Journal] {10.1088/0004-637X/723/1/935}, 723, 935

\bibitem[\protect\citeauthoryear{{Hodges-Kluck}, {Miller}  \& {Bregman}}{{Hodges-Kluck} et~al.}{2016}]{2016ApJ...822...21H}
{Hodges-Kluck} E.~J.,  {Miller} M.~J.,   {Bregman} J.~N.,  2016, \mn@doi [\apj] {10.3847/0004-637X/822/1/21}, \href {https://ui.adsabs.harvard.edu/abs/2016ApJ...822...21H} {822, 21}

\bibitem[\protect\citeauthoryear{Hopkins}{Hopkins}{2015}]{hopkinsNewClassAccurate2015}
Hopkins P.~F.,  2015, \mn@doi [Monthly Notices of the Royal Astronomical Society] {10.1093/mnras/stv195}, 450, 53

\bibitem[\protect\citeauthoryear{Hopkins}{Hopkins}{2024}]{hopkinsImportanceSubtletiesScaling2024}
Hopkins P.~F.,  2024, The {{Importance}} of {{Subtleties}} in the {{Scaling}} of the '{{Terminal Momentum}}' {{For Galaxy Formation Simulations}}, \mn@doi{10.48550/arXiv.2404.16987}

\bibitem[\protect\citeauthoryear{Hopkins, Kere{\v s}, O{\~n}orbe, {Faucher-Gigu{\`e}re}, Quataert, Murray  \& Bullock}{Hopkins et~al.}{2014}]{hopkinsGalaxiesFIREFeedback2014}
Hopkins P.~F.,  Kere{\v s} D.,  O{\~n}orbe J.,  {Faucher-Gigu{\`e}re} C.-A.,  Quataert E.,  Murray N.,   Bullock J.~S.,  2014, \mn@doi [Monthly Notices of the Royal Astronomical Society] {10.1093/mnras/stu1738}, 445, 581

\bibitem[\protect\citeauthoryear{Hopkins et~al.,}{Hopkins et~al.}{2018}]{hopkinsFIRE2SimulationsPhysics2018}
Hopkins P.~F.,  et~al., 2018, \mn@doi [Monthly Notices of the Royal Astronomical Society] {10.1093/mnras/sty1690}, 480, 800

\bibitem[\protect\citeauthoryear{Hopkins et~al.,}{Hopkins et~al.}{2020}]{hopkinsWhatCosmicRays2020}
Hopkins P.~F.,  et~al., 2020, \mn@doi [Monthly Notices of the Royal Astronomical Society] {10.1093/mnras/stz3321}, 492, 3465

\bibitem[\protect\citeauthoryear{{Hopkins}, {Chan}, {Squire}, {Quataert}, {Ji}, {Kere{\v{s}}}  \& {Faucher-Gigu{\`e}re}}{{Hopkins} et~al.}{2021}]{2021MNRAS.501.3663H}
{Hopkins} P.~F.,  {Chan} T.~K.,  {Squire} J.,  {Quataert} E.,  {Ji} S.,  {Kere{\v{s}}} D.,   {Faucher-Gigu{\`e}re} C.-A.,  2021, \mn@doi [\mnras] {10.1093/mnras/staa3692}, \href {https://ui.adsabs.harvard.edu/abs/2021MNRAS.501.3663H} {501, 3663}

\bibitem[\protect\citeauthoryear{{Hopkins}, {Squire}, {Butsky}  \& {Ji}}{{Hopkins} et~al.}{2022}]{2022MNRAS.517.5413H}
{Hopkins} P.~F.,  {Squire} J.,  {Butsky} I.~S.,   {Ji} S.,  2022, \mn@doi [\mnras] {10.1093/mnras/stac2909}, \href {https://ui.adsabs.harvard.edu/abs/2022MNRAS.517.5413H} {517, 5413}

\bibitem[\protect\citeauthoryear{Hopkins et~al.,}{Hopkins et~al.}{2023}]{hopkinsFIRE3UpdatedStellar2023}
Hopkins P.~F.,  et~al., 2023, \mn@doi [Monthly Notices of the Royal Astronomical Society] {10.1093/mnras/stac3489}, 519, 3154

\bibitem[\protect\citeauthoryear{Hopkins, Quataert, Ponnada  \& Silich}{Hopkins et~al.}{2025}]{hopkinsCosmicRaysMasquerading2025}
Hopkins P.~F.,  Quataert E.,  Ponnada S.~B.,   Silich E.,  2025, Cosmic {{Rays Masquerading}} as {{Hot CGM Gas}}: {{An Inverse-Compton Origin}} for {{Diffuse X-ray Emission}} in the {{Circumgalactic Medium}} (\mn@eprint {arXiv} {2501.18696}), \mn@doi{10.48550/arXiv.2501.18696}

\bibitem[\protect\citeauthoryear{{Hummels} et~al.,}{{Hummels} et~al.}{2019}]{2019ApJ...882..156H}
{Hummels} C.~B.,  et~al., 2019, \mn@doi [\apj] {10.3847/1538-4357/ab378f}, \href {https://ui.adsabs.harvard.edu/abs/2019ApJ...882..156H} {882, 156}

\bibitem[\protect\citeauthoryear{Ivey, Fabian, Sanders, Pinto, Ferland, Walker  \& Jiang}{Ivey et~al.}{2024}]{iveyHiddenCoolingFlows2024}
Ivey L.~R.,  Fabian A.~C.,  Sanders J.~S.,  Pinto C.,  Ferland G.~J.,  Walker S.,   Jiang J.,  2024, Hidden {{Cooling Flows}} in {{Elliptical Galaxies}} (\mn@eprint {arXiv} {2411.03864}), \mn@doi{10.48550/arXiv.2411.03864}

\bibitem[\protect\citeauthoryear{Jana, Sarkar, Stern  \& Sternberg}{Jana et~al.}{2024}]{janaXraySignaturesGalactic2024}
Jana R.,  Sarkar K.~C.,  Stern J.,   Sternberg A.,  2024, \mn@doi [Monthly Notices of the Royal Astronomical Society] {10.1093/mnras/stae1248}, 531, 2757

\bibitem[\protect\citeauthoryear{Ji et~al.,}{Ji et~al.}{2020}]{jiPropertiesCircumgalacticMedium2020}
Ji S.,  et~al., 2020, \mn@doi [Monthly Notices of the Royal Astronomical Society] {10.1093/mnras/staa1849}, 496, 4221

\bibitem[\protect\citeauthoryear{{Kakoly}, {Stern}, {Faucher-Gigu{\`e}re}, {Fielding}, {Goldner}, {Sun}  \& {Hummels}}{{Kakoly} et~al.}{2025}]{2025arXiv250417001K}
{Kakoly} A.,  {Stern} J.,  {Faucher-Gigu{\`e}re} C.-A.,  {Fielding} D.~B.,  {Goldner} R.,  {Sun} G.,   {Hummels} C.~B.,  2025, arXiv e-prints, \href {https://ui.adsabs.harvard.edu/abs/2025arXiv250417001K} {p. arXiv:2504.17001}

\bibitem[\protect\citeauthoryear{Kere{\v s}, Katz, Weinberg  \& Dav{\'e}}{Kere{\v s} et~al.}{2005}]{keresHowGalaxiesGet2005}
Kere{\v s} D.,  Katz N.,  Weinberg D.~H.,   Dav{\'e} R.,  2005, \mn@doi [Monthly Notices of the Royal Astronomical Society] {10.1111/j.1365-2966.2005.09451.x}, 363, 2

\bibitem[\protect\citeauthoryear{Kere{\v s}, Katz, Fardal, Dav{\'e}  \& Weinberg}{Kere{\v s} et~al.}{2009}]{keresGalaxiesSimulatedLCDM2009}
Kere{\v s} D.,  Katz N.,  Fardal M.,  Dav{\'e} R.,   Weinberg D.~H.,  2009, \mn@doi [Monthly Notices of the Royal Astronomical Society] {10.1111/j.1365-2966.2009.14541.x}, 395, 160

\bibitem[\protect\citeauthoryear{{Lan} \& {Prochaska}}{{Lan} \& {Prochaska}}{2020}]{2020MNRAS.496.3142L}
{Lan} T.-W.,  {Prochaska} J.~X.,  2020, \mn@doi [\mnras] {10.1093/mnras/staa1750}, \href {https://ui.adsabs.harvard.edu/abs/2020MNRAS.496.3142L} {496, 3142}

\bibitem[\protect\citeauthoryear{Li \& Bregman}{Li \& Bregman}{2017}]{liPropertiesGalacticHot2017}
Li Y.,  Bregman J.,  2017, \mn@doi [The Astrophysical Journal] {10.3847/1538-4357/aa92c6}, 849, 105

\bibitem[\protect\citeauthoryear{Li et~al.,}{Li et~al.}{2024}]{liRobustDetectionHot2024}
Li D.,  et~al., 2024, Robust Detection of Hot Intragroup Medium in Optically Selected, Poor Galaxy Groups by {{eROSITA}} (\mn@eprint {arXiv} {2412.01261}), \mn@doi{10.48550/arXiv.2412.01261}

\bibitem[\protect\citeauthoryear{{Liu} et~al.,}{{Liu} et~al.}{2025}]{2025arXiv250208850L}
{Liu} R.~H.,  et~al., 2025, \mn@doi [arXiv e-prints] {10.48550/arXiv.2502.08850}, \href {https://ui.adsabs.harvard.edu/abs/2025arXiv250208850L} {p. arXiv:2502.08850}

\bibitem[\protect\citeauthoryear{Ma, Hopkins, {Faucher-Gigu{\`e}re}, Zolman, Muratov, Kere{\v s}  \& Quataert}{Ma et~al.}{2016}]{maOriginEvolutionGalaxy2016}
Ma X.,  Hopkins P.~F.,  {Faucher-Gigu{\`e}re} C.-A.,  Zolman N.,  Muratov A.~L.,  Kere{\v s} D.,   Quataert E.,  2016, \mn@doi [Monthly Notices of the Royal Astronomical Society] {10.1093/mnras/stv2659}, 456, 2140

\bibitem[\protect\citeauthoryear{Marszewski, Sun, {Faucher-Gigu{\`e}re}, Hayward  \& Feldmann}{Marszewski et~al.}{2024}]{marszewskiHighRedshiftGasPhaseMass2024}
Marszewski A.,  Sun G.,  {Faucher-Gigu{\`e}re} C.-A.,  Hayward C.~C.,   Feldmann R.,  2024, \mn@doi [The Astrophysical Journal] {10.3847/2041-8213/ad4cee}, 967, L41

\bibitem[\protect\citeauthoryear{McDonald, Gaspari, McNamara  \& Tremblay}{McDonald et~al.}{2018}]{mcdonaldRevisitingCoolingFlow2018}
McDonald M.,  Gaspari M.,  McNamara B.~R.,   Tremblay G.~R.,  2018, \mn@doi [The Astrophysical Journal] {10.3847/1538-4357/aabace}, 858, 45

\bibitem[\protect\citeauthoryear{McNamara \& Nulsen}{McNamara \& Nulsen}{2007}]{mcnamaraHeatingHotAtmospheres2007}
McNamara B.~R.,  Nulsen P. E.~J.,  2007, \mn@doi [Annual Review of Astronomy and Astrophysics] {10.1146/annurev.astro.45.051806.110625}, 45, 117

\bibitem[\protect\citeauthoryear{McQuinn \& Werk}{McQuinn \& Werk}{2018}]{mcquinnImplicationsLargeVi2018}
McQuinn M.,  Werk J.~K.,  2018, \mn@doi [The Astrophysical Journal] {10.3847/1538-4357/aa9d3f}, 852, 33

\bibitem[\protect\citeauthoryear{Muratov, Kere{\v s}, {Faucher-Gigu{\`e}re}, Hopkins, Quataert  \& Murray}{Muratov et~al.}{2015}]{muratovGustyGaseousFlows2015}
Muratov A.~L.,  Kere{\v s} D.,  {Faucher-Gigu{\`e}re} C.-A.,  Hopkins P.~F.,  Quataert E.,   Murray N.,  2015, \mn@doi [Monthly Notices of the Royal Astronomical Society] {10.1093/mnras/stv2126}, 454, 2691

\bibitem[\protect\citeauthoryear{{Naab} \& {Ostriker}}{{Naab} \& {Ostriker}}{2017}]{2017ARA&A..55...59N}
{Naab} T.,  {Ostriker} J.~P.,  2017, \mn@doi [\araa] {10.1146/annurev-astro-081913-040019}, \href {https://ui.adsabs.harvard.edu/abs/2017ARA&A..55...59N} {55, 59}

\bibitem[\protect\citeauthoryear{Oppenheimer}{Oppenheimer}{2018}]{oppenheimerDeviationsHydrostaticEquilibrium2018}
Oppenheimer B.~D.,  2018, \mn@doi [Monthly Notices of the Royal Astronomical Society] {10.1093/mnras/sty1918}, 480, 2963

\bibitem[\protect\citeauthoryear{Oren, Sternberg, McKee, Faerman  \& Genel}{Oren et~al.}{2024}]{orenSunyaevZeldovichSignals$L^$2024}
Oren Y.,  Sternberg A.,  McKee C.~F.,  Faerman Y.,   Genel S.,  2024, Sunyaev-{{Zeldovich Signals}} from \${{L}}{\textasciicircum}*\$ {{Galaxies}}: {{Observations}}, {{Analytics}}, and {{Simulations}}, \mn@doi{10.48550/arXiv.2403.09476}

\bibitem[\protect\citeauthoryear{Pandya et~al.,}{Pandya et~al.}{2021}]{pandyaCharacterizingMassMomentum2021}
Pandya V.,  et~al., 2021, \mn@doi [Monthly Notices of the Royal Astronomical Society] {10.1093/mnras/stab2714}, 508, 2979

\bibitem[\protect\citeauthoryear{{Peeples} et~al.,}{{Peeples} et~al.}{2019}]{2019ApJ...873..129P}
{Peeples} M.~S.,  et~al., 2019, \mn@doi [\apj] {10.3847/1538-4357/ab0654}, \href {https://ui.adsabs.harvard.edu/abs/2019ApJ...873..129P} {873, 129}

\bibitem[\protect\citeauthoryear{{Ponnada} et~al.,}{{Ponnada} et~al.}{2022}]{2022MNRAS.516.4417P}
{Ponnada} S.~B.,  et~al., 2022, \mn@doi [\mnras] {10.1093/mnras/stac2448}, \href {https://ui.adsabs.harvard.edu/abs/2022MNRAS.516.4417P} {516, 4417}

\bibitem[\protect\citeauthoryear{Popesso et~al.,}{Popesso et~al.}{2024}]{popessoAverageXrayProperties2024}
Popesso P.,  et~al., 2024, Average {{X-ray}} Properties of Galaxy Groups. {{From Milky Way-like}} Halos to Massive Clusters (\mn@eprint {arXiv} {2411.17120}), \mn@doi{10.48550/arXiv.2411.17120}

\bibitem[\protect\citeauthoryear{Power, Navarro, Jenkins, Frenk, White, Springel, Stadel  \& Quinn}{Power et~al.}{2003}]{powerInnerStructureLCDM2003}
Power C.,  Navarro J.~F.,  Jenkins A.,  Frenk C.~S.,  White S. D.~M.,  Springel V.,  Stadel J.,   Quinn T.,  2003, \mn@doi [Monthly Notices of the Royal Astronomical Society] {10.1046/j.1365-8711.2003.05925.x}, 338, 14

\bibitem[\protect\citeauthoryear{{Prochaska} et~al.,}{{Prochaska} et~al.}{2019}]{2019Sci...366..231P}
{Prochaska} J.~X.,  et~al., 2019, \mn@doi [Science] {10.1126/science.aay0073}, \href {https://ui.adsabs.harvard.edu/abs/2019Sci...366..231P} {366, 231}

\bibitem[\protect\citeauthoryear{{Qu} et~al.,}{{Qu} et~al.}{2022}]{2022MNRAS.516.4882Q}
{Qu} Z.,  et~al., 2022, \mn@doi [\mnras] {10.1093/mnras/stac2528}, \href {https://ui.adsabs.harvard.edu/abs/2022MNRAS.516.4882Q} {516, 4882}

\bibitem[\protect\citeauthoryear{Qu et~al.,}{Qu et~al.}{2024}]{quCosmicUltravioletBaryon2024}
Qu Z.,  et~al., 2024, \mn@doi [The Astrophysical Journal] {10.3847/1538-4357/ad410b}, 968, 8

\bibitem[\protect\citeauthoryear{{Quataert} \& {Hopkins}}{{Quataert} \& {Hopkins}}{2025}]{2025arXiv250201753Q}
{Quataert} E.,  {Hopkins} P.~F.,  2025, \mn@doi [arXiv e-prints] {10.48550/arXiv.2502.01753}, \href {https://ui.adsabs.harvard.edu/abs/2025arXiv250201753Q} {p. arXiv:2502.01753}

\bibitem[\protect\citeauthoryear{Rees \& Ostriker}{Rees \& Ostriker}{1977}]{reesCoolingDynamicsFragmentation1977}
Rees M.~J.,  Ostriker J.~P.,  1977, \mn@doi [Monthly Notices of the Royal Astronomical Society] {10.1093/mnras/179.4.541}, 179, 541

\bibitem[\protect\citeauthoryear{Reynolds et~al.,}{Reynolds et~al.}{2023}]{reynoldsOverviewAdvancedXray2023}
Reynolds C.~S.,  et~al., 2023, in {{UV}}, {{X-Ray}}, and {{Gamma-Ray Space Instrumentation}} for {{Astronomy XXIII}}. SPIE, pp 421--442, \mn@doi{10.1117/12.2677468}

\bibitem[\protect\citeauthoryear{{Ruszkowski} \& {Pfrommer}}{{Ruszkowski} \& {Pfrommer}}{2023}]{2023A&ARv..31....4R}
{Ruszkowski} M.,  {Pfrommer} C.,  2023, \mn@doi [\aapr] {10.1007/s00159-023-00149-2}, \href {https://ui.adsabs.harvard.edu/abs/2023A&ARv..31....4R} {31, 4}

\bibitem[\protect\citeauthoryear{{Salem} \& {Bryan}}{{Salem} \& {Bryan}}{2014}]{2014MNRAS.437.3312S}
{Salem} M.,  {Bryan} G.~L.,  2014, \mn@doi [\mnras] {10.1093/mnras/stt2121}, \href {https://ui.adsabs.harvard.edu/abs/2014MNRAS.437.3312S} {437, 3312}

\bibitem[\protect\citeauthoryear{Samuel et~al.,}{Samuel et~al.}{2020}]{samuelProfileFIREResolving2020}
Samuel J.,  et~al., 2020, \mn@doi [Monthly Notices of the Royal Astronomical Society] {10.1093/mnras/stz3054}, 491, 1471

\bibitem[\protect\citeauthoryear{Sharma, McCourt, Parrish  \& Quataert}{Sharma et~al.}{2012}]{sharmaStructureHotGas2012}
Sharma P.,  McCourt M.,  Parrish I.~J.,   Quataert E.,  2012, \mn@doi [Monthly Notices of the Royal Astronomical Society] {10.1111/j.1365-2966.2012.22050.x}, 427, 1219

\bibitem[\protect\citeauthoryear{Singh, Lau, Faerman, Stern  \& Nagai}{Singh et~al.}{2024}]{singhComparisonModelsWarmHot2024}
Singh P.,  Lau E.~T.,  Faerman Y.,  Stern J.,   Nagai D.,  2024, \mn@doi [Monthly Notices of the Royal Astronomical Society] {10.1093/mnras/stae1695}, 532, 3222

\bibitem[\protect\citeauthoryear{Somerville \& Dav{\'e}}{Somerville \& Dav{\'e}}{2015}]{somervillePhysicalModelsGalaxy2015}
Somerville R.~S.,  Dav{\'e} R.,  2015, \mn@doi [Annual Review of Astronomy and Astrophysics] {10.1146/annurev-astro-082812-140951}, 53, 51

\bibitem[\protect\citeauthoryear{Sormani \& Sobacchi}{Sormani \& Sobacchi}{2019}]{sormaniEffectRotationThermal2019}
Sormani M.~C.,  Sobacchi E.,  2019, \mn@doi [Monthly Notices of the Royal Astronomical Society] {10.1093/mnras/stz793}, 486, 215

\bibitem[\protect\citeauthoryear{Stern, Fielding, {Faucher-Gigu{\`e}re}  \& Quataert}{Stern et~al.}{2019}]{sternCoolingFlowSolutions2019}
Stern J.,  Fielding D.,  {Faucher-Gigu{\`e}re} C.-A.,   Quataert E.,  2019, \mn@doi [Monthly Notices of the Royal Astronomical Society] {10.1093/mnras/stz1859}, 488, 2549

\bibitem[\protect\citeauthoryear{Stern, Fielding, {Faucher-Gigu{\`e}re}  \& Quataert}{Stern et~al.}{2020}]{sternMaximumAccretionRate2020}
Stern J.,  Fielding D.,  {Faucher-Gigu{\`e}re} C.-A.,   Quataert E.,  2020, \mn@doi [Monthly Notices of the Royal Astronomical Society] {10.1093/mnras/staa198}, 492, 6042

\bibitem[\protect\citeauthoryear{{Stern} et~al.,}{{Stern} et~al.}{2021a}]{2021MNRAS.507.2869S}
{Stern} J.,  et~al., 2021a, \mn@doi [\mnras] {10.1093/mnras/stab2240}, \href {https://ui.adsabs.harvard.edu/abs/2021MNRAS.507.2869S} {507, 2869}

\bibitem[\protect\citeauthoryear{Stern et~al.,}{Stern et~al.}{2021b}]{sternVirializationInnerCGM2021}
Stern J.,  et~al., 2021b, \mn@doi [The Astrophysical Journal] {10.3847/1538-4357/abd776}, 911, 88

\bibitem[\protect\citeauthoryear{Stern, Fielding, Hafen, Su, Naor, {Faucher-Gigu{\`e}re}, Quataert  \& Bullock}{Stern et~al.}{2024}]{sternAccretionDiscGalaxies2024}
Stern J.,  Fielding D.,  Hafen Z.,  Su K.-Y.,  Naor N.,  {Faucher-Gigu{\`e}re} C.-A.,  Quataert E.,   Bullock J.,  2024, \mn@doi [Monthly Notices of the Royal Astronomical Society] {10.1093/mnras/stae824}, 530, 1711

\bibitem[\protect\citeauthoryear{Sunyaev \& Zeldovich}{Sunyaev \& Zeldovich}{1970}]{sunyaevSmallScaleFluctuationsRelic1970}
Sunyaev R.~A.,  Zeldovich {\relax Ya}.~B.,  1970, \mn@doi [Astrophysics and Space Science] {10.1007/BF00653471}, 7, 3

\bibitem[\protect\citeauthoryear{Sunyaev \& Zeldovich}{Sunyaev \& Zeldovich}{1972}]{sunyaevObservationsRelicRadiation1972}
Sunyaev R.~A.,  Zeldovich {\relax Ya}.~B.,  1972, Comments on Astrophysics and Space Physics, 4, 173

\bibitem[\protect\citeauthoryear{Tashiro}{Tashiro}{2022}]{tashiroXRISMXrayImaging2022}
Tashiro M.~S.,  2022, \mn@doi [International Journal of Modern Physics D] {10.1142/S0218271822300014}, 31, 2230001

\bibitem[\protect\citeauthoryear{Tozzi \& Norman}{Tozzi \& Norman}{2001}]{tozziEvolutionXRayClusters2001}
Tozzi P.,  Norman C.,  2001, \mn@doi [The Astrophysical Journal] {10.1086/318237}, 546, 63

\bibitem[\protect\citeauthoryear{Tumlinson et~al.,}{Tumlinson et~al.}{2011}]{tumlinsonLargeOxygenRichHalos2011}
Tumlinson J.,  et~al., 2011, \mn@doi [Science] {10.1126/science.1209840}, 334, 948

\bibitem[\protect\citeauthoryear{Tumlinson, Peeples  \& Werk}{Tumlinson et~al.}{2017}]{tumlinsonCircumgalacticMedium2017}
Tumlinson J.,  Peeples M.~S.,   Werk J.~K.,  2017, \mn@doi [Annual Review of Astronomy and Astrophysics] {10.1146/annurev-astro-091916-055240}, 55, 389

\bibitem[\protect\citeauthoryear{Voit}{Voit}{2019}]{voitAmbientColumnDensities2019}
Voit G.~M.,  2019, \mn@doi [The Astrophysical Journal] {10.3847/1538-4357/ab2bfd}, 880, 139

\bibitem[\protect\citeauthoryear{Voit, Kay  \& Bryan}{Voit et~al.}{2005}]{voitBaselineIntraclusterEntropy2005}
Voit G.~M.,  Kay S.~T.,   Bryan G.~L.,  2005, \mn@doi [Monthly Notices of the Royal Astronomical Society] {10.1111/j.1365-2966.2005.09621.x}, 364, 909

\bibitem[\protect\citeauthoryear{Werk, Prochaska, Thom, Tumlinson, Tripp, O'Meara  \& Peeples}{Werk et~al.}{2013}]{werkCOSHALOSSURVEYEMPIRICAL2013}
Werk J.~K.,  Prochaska J.~X.,  Thom C.,  Tumlinson J.,  Tripp T.~M.,  O'Meara J.~M.,   Peeples M.~S.,  2013, \mn@doi [The Astrophysical Journal Supplement Series] {10.1088/0067-0049/204/2/17}, 204, 17

\bibitem[\protect\citeauthoryear{Wetzel et~al.,}{Wetzel et~al.}{2023}]{wetzelPublicDataRelease2023}
Wetzel A.,  et~al., 2023, \mn@doi [The Astrophysical Journal Supplement Series] {10.3847/1538-4365/acb99a}, 265, 44

\bibitem[\protect\citeauthoryear{White \& Rees}{White \& Rees}{1978}]{whiteCoreCondensationHeavy1978}
White S. D.~M.,  Rees M.~J.,  1978, \mn@doi [Monthly Notices of the Royal Astronomical Society] {10.1093/mnras/183.3.341}, 183, 341

\bibitem[\protect\citeauthoryear{Wiersma, Schaye  \& Smith}{Wiersma et~al.}{2009}]{wiersmaEffectPhotoionizationCooling2009}
Wiersma R. P.~C.,  Schaye J.,   Smith B.~D.,  2009, \mn@doi [Monthly Notices of the Royal Astronomical Society] {10.1111/j.1365-2966.2008.14191.x}, 393, 99

\bibitem[\protect\citeauthoryear{Wijers \& Schaye}{Wijers \& Schaye}{2022}]{wijersWarmhotCircumgalacticMedium2022}
Wijers N.~A.,  Schaye J.,  2022, \mn@doi [Monthly Notices of the Royal Astronomical Society] {10.1093/mnras/stac1580}, 514, 5214

\bibitem[\protect\citeauthoryear{Wijers, Schaye  \& Oppenheimer}{Wijers et~al.}{2020}]{wijersWarmhotCircumgalacticMedium2020}
Wijers N.~A.,  Schaye J.,   Oppenheimer B.~D.,  2020, \mn@doi [Monthly Notices of the Royal Astronomical Society] {10.1093/mnras/staa2456}, 498, 574

\bibitem[\protect\citeauthoryear{Wijers, {Faucher-Gigu{\`e}re}, Stern, Byrne  \& Sultan}{Wijers et~al.}{2024}]{wijersNeVIIIWarmhot2024}
Wijers N.~A.,  {Faucher-Gigu{\`e}re} C.-A.,  Stern J.,  Byrne L.,   Sultan I.,  2024, \mn@doi [The Astrophysical Journal] {10.3847/1538-4357/ad63a0}, 973, 99

\bibitem[\protect\citeauthoryear{Zhang et~al.,}{Zhang et~al.}{2024}]{zhangHotCircumgalacticMedium2024a}
Zhang Y.,  et~al., 2024, \mn@doi [Astronomy \& Astrophysics] {10.1051/0004-6361/202449412}, 690, A267

\bibitem[\protect\citeauthoryear{{van de Voort}, Quataert, Hopkins, {Faucher-Gigu{\`e}re}, Feldmann, Kere{\v s}, Chan  \& Hafen}{{van de Voort} et~al.}{2016}]{vandevoortImpactStellarFeedback2016}
{van de Voort} F.,  Quataert E.,  Hopkins P.~F.,  {Faucher-Gigu{\`e}re} C.-A.,  Feldmann R.,  Kere{\v s} D.,  Chan T.~K.,   Hafen Z.,  2016, \mn@doi [Monthly Notices of the Royal Astronomical Society] {10.1093/mnras/stw2322}, 463, 4533

\bibitem[\protect\citeauthoryear{{van de Voort}, {Springel}, {Mandelker}, {van den Bosch}  \& {Pakmor}}{{van de Voort} et~al.}{2019}]{2019MNRAS.482L..85V}
{van de Voort} F.,  {Springel} V.,  {Mandelker} N.,  {van den Bosch} F.~C.,   {Pakmor} R.,  2019, \mn@doi [\mnras] {10.1093/mnrasl/sly190}, \href {https://ui.adsabs.harvard.edu/abs/2019MNRAS.482L..85V} {482, L85}

\makeatother
\end{thebibliography}
\end{document}